\documentclass{iopart}
\pdfoutput=1 
\usepackage{aasmacros}
\usepackage[square,sort&compress]{natbib}
\usepackage{iopams}
\usepackage{enumitem}
\usepackage{xspace}
\usepackage{subcaption}
\usepackage{xcolor}
\usepackage[skins]{tcolorbox}
\usepackage{rotating}
\usepackage{color}
\usepackage[backref, breaklinks, plainpages=false, colorlinks=true, anchorcolor=cyan, linkcolor=red, citecolor=cyan, filecolor=magenta, urlcolor=magenta]{hyperref}
\hypersetup{plainpages=false, colorlinks=true, anchorcolor=blue, linkcolor=red, citecolor=cyan, filecolor=cyan, urlcolor=magenta}
\newcommand{\eg}{e.\,g.\ }

\usepackage{cleveref}[2012/02/15]
\crefformat{footnote}{#2\footnotemark[#1]#3}

\newcommand{\revision}[1]{{#1}}

\newcommand{\msun}{M_\odot}

\bibpunct[; ]{(}{)}{;}{a}{}{,}

\begin{document}

\title[The Astrophysics of Nanohertz Gravitational Waves]{The Astrophysics of Nanohertz Gravitational Waves}

\author{Sarah Burke-Spolaor\footnote{\label{cauth}Corresponding authors}}
\address{Department of Physics and Astronomy, West Virginia University, P.O. Box 6315, Morgantown, WV 26506, USA}
\address{Center for Gravitational Waves and Cosmology, West Virginia University, Chestnut Ridge Research Building, Morgantown, WV 26505, USA}
\address{Canadian Institute for Advanced Research,
CIFAR Azrieli Global Scholar,
MaRS Centre West Tower,
661 University Ave. Suite 505,
Toronto ON M5G 1M1, Canada}
\author{Stephen~R.~Taylor\cref{cauth}}
\address{TAPIR, California Institute of Technology, 1200 E. California Blvd., Pasadena, CA 91125, USA}
\address{Jet Propulsion Laboratory, 4800 Oak Grove Drive, Pasadena, CA 91109, USA}
\author{Maria Charisi}
\address{TAPIR, California Institute of Technology, 1200 E. California Blvd., Pasadena, CA 91125, USA}
\author{Timothy Dolch}
\address{Department of Physics, Hillsdale College, 33 E. College St., Hillsdale, MI 49242, USA}
\author{Jeffrey S.~Hazboun}
\address{Physical Sciences Division, University of Washington Bothell, 18115 Campus Way NE, Bothell, WA 98011-8246}
\author{A.~Miguel Holgado}
\address{Department of Astronomy and National Center for Supercomputing Applications, University of Illinois at Urbana-Champaign, 1002 W. Green St., Urbana, IL 61801, USA}
\author{Luke Zoltan Kelley}
\address{Harvard--Smithsonian Center for Astrophysics, 60 Garden St., Cambridge, MA 02138, USA}
\address{Department of Physics \& Astronomy and CIERA, Northwestern University, 2145 Sheridan Road, Evanston, IL 60208, USA}
\author{T.~Joseph~W.~Lazio}
\address{Jet Propulsion Laboratory, California Institute of Technology, 4800 Oak Grove Drive, Pasadena, CA 91109, USA}
\author{Dustin R. Madison}
\address{Department of Physics and Astronomy, West Virginia University, P.O. Box 6315, Morgantown, WV 26506, USA}
\address{The National Radio Astronomy Observatory, 520 Edgemont Rd., Charlottesville, VA 22903, USA}
\author{Natasha McMann}
\address{Department of Life and Physical Sciences, Fisk University, 1000 17th Ave N., Nashville, TN 37208, USA}
\address{Department of Physics and Astronomy, Vanderbilt University, PMB 401807, Nashville, TN 37206, USA}
\author{Chiara M. F. Mingarelli}
\address{Center for Computational Astrophysics, Flatiron Institute, 162 Fifth Ave, New York, NY 10010, USA}
\author{Alexander Rasskazov}
\address{E\"otv\"os University, Institute of Physics, P\'azm\'any P. s. 1/A, Budapest, Hungary 1117}
\author{Xavier Siemens}
\address{Center for Gravitation, Cosmology and Astrophysics, Department of Physics, University of Wisconsin-Milwaukee, P.O. Box 413, Milwaukee, WI 53201, USA}
\author{Joseph J.~Simon}
\address{Jet Propulsion Laboratory, California Institute of Technology, 4800 Oak Grove Drive, Pasadena, CA 91109, USA}
\author{Tristan L.~Smith}
\address{Department of Physics \& Astronomy, Swarthmore College, Swarthmore, PA 19081 USA}

\begin{abstract}
Pulsar timing array (PTA) collaborations in North America, Australia, and Europe, have been exploiting the exquisite timing precision of millisecond pulsars over decades of observations to search for correlated timing deviations induced by gravitational waves (GWs). 
\revision{PTAs are sensitive to the frequency band ranging just below 1\,nanohertz to a few tens of microhertz.
The discovery space of this band} is potentially rich with populations of inspiraling supermassive black-holes binaries, decaying cosmic string networks, relic post-inflation GWs, and even non-GW imprints of axionic dark matter. 

This article aims to provide an understanding of the exciting open science questions in cosmology, galaxy evolution, and fundamental physics that will be addressed by the detection and study of GWs \revision{through PTAs}. The focus of the article is on providing an understanding of the mechanisms by which PTAs can address specific questions in these fields, and to outline some of the subtleties and difficulties in each case.
The material included is weighted most heavily towards the questions which we expect will be answered \revision{in the near-term} with PTAs; however, we have made efforts to include most currently anticipated applications of nanohertz GWs.

\end{abstract}


\tableofcontents

\title[The Astrophysics of Nanohertz Gravitational Waves]
\maketitle

\section{Introduction}
\label{sec:intro}
The first direct detection of Gravitational Waves (GWs) in 2015 
 started a new era in astrophysics, in which we can now use gravity itself as a unique messenger from the cosmos 
\citep{GW150914,GW151226,GW170104}. The subsequent detection of a neutron star merger with associated electromagnetic counterparts in 2017 marked the beginning of \revision{gravitational waves' contribution to the era of} multi-messenger astronomy \citep{GW170817}.
The ground-based laser interferometers that made those detections \revision{(LIGO/VIRGO Collaboration)} are sensitive to sources that emit gravitational radiation between about ten and a few thousand Hz. In the coming decades, we will open up additional bands in the GW spectrum, which will allow us to probe entirely new astrophysical sources and physics. 

The detection of nanohertz GWs by Pulsar Timing Arrays (PTAs) is expected to be the next major milestone in GW astrophysics.
In the future, PTAs and ground-based laser interferometry experiments will be complemented by space-based laser interferometers \citep{2017arXiv170200786A} and observations of primordial GWs, imprinted in the polarization of the cosmic microwave background \citep[\eg][]{planck+bicep}, providing comprehensive access to the GW Universe.
Current PTA efforts are spearheaded by a number of groups worldwide, including \revision{the European Pulsar Timing Array \citep[\hbox{EPTA},][]{2016MNRAS.458.3341D,2015MNRAS.453.2576L,2016MNRAS.455.1665B},} the North American Nanohertz Observatory for Gravitational Waves \citep[NANOGrav,][]{nano-11yr-data}, and the Parkes Pulsar Timing Array \citep[\hbox{PPTA}, ][]{ppta-manchester13,2015Sci...349.1522S,2016PhRvX...6a1035L}.  The individual groups are also the constituents of an international collaboration, known as the International Pulsar Timing Array \citep[\hbox{IPTA}, ][]{2010CQGra..27h4013H, VerbiestEtAl:2016}.




This paper presents a comprehensive 
background in astrophysical theory that can be addressed observationally by PTAs, and thus the science that will be extracted from the detection of GWs at nanohertz frequencies. \revision{The immediate focus of PTAs has been a stochastic GW background, hypothesized to result from the ensemble of in-spiraling supermassive black hole binaries. However, the astrophysics resulting from detection and study of GWs by PTAs is much richer, and some of it has been developed alongside steady PTA sensitivity improvements over the past decade.} We limit this paper to describe the astrophysics that is related to GW detection in the PTA band, and in \S\ref{sec:beyond} to gravitational effects on PTAs not due to the pulsars or their companions. Throughout the present work, the ``PTA band'' refers to GW frequencies of approximately 1 -- 1000\,nHz.

We do \emph{not} aim to cover the rich (non-GW-related) astrophysics accessible by pulsar timing.
The prolific ancillary science from a PTA as a whole includes, but it not limited to: neutron star population dynamics \citep{1998ApJ...505..315C,matthews+16}, the formation histories of compact objects (\citealt{bassa+16}, \citealt{fonseca+16}, \citealt{kaplan+16}), and the characterization of the ionized interstellar medium \revision{\citep{keith-ISM,jones+17,lam+16,mckee-crab}}, including plasma lensing events \revision{\citep{coles-ESE,lam+18}}, tests of general relativity \citep{1982ApJ...253..908T,ksm+06,zhu+15} and the physics of nuclear matter \citep{dpr+10}.


For readers seeking a brief summary, each section is led by an outline of the most salient overview points from that section. The layout of the remainder of this paper is as follows: \S\ref{sec:ptas} provides a backdrop of concepts in pulsar timing that are relevant to the understanding of this review. In~\S\ref{sec:smbh}, we discuss PTA applications to supermassive black hole binaries; 
in~\S\ref{sec:strings}, we consider cosmic strings; in \S\ref{sec:grtests}, we assess whether nanohertz GWs can present unique tests of General Relativity; in \S\ref{sec:relic} we consider topics in cosmology, and in \S\ref{sec:beyond} we consider other (potentially more exotic) possibilities. In \S\ref{sec:lisa} and \S\ref{sec:em}, we describe potential synergistic science in multi-band GW studies (in particular with the Laser Interferometer Space Antenna, LISA), and in multi-messenger studies (in particular with electromagnetic observations of binary supermassive black holes), respectively. In \S\ref{sec:conclude}, we summarize the current and the expected near-future developments in this field.


\section{Pulsar Timing in Brief}
\label{sec:ptas}
\vspace{-3mm}
\begin{tcolorbox}[enhanced,colback=gray!10!white,colframe=black,drop shadow]
Here we introduce critical concepts for understanding how PTAs can access their target science.

\vspace{-10pt}\paragraph{\textbf{Timing residuals:}}
Evidence of GWs can be seen by the influence they have on the arrival time of pulsar signals at Earth. The measured versus predicted arrival time of pulses, as a function of time, are referred to as the timing residuals.

\vspace{-10pt}\paragraph{\textbf{Pulsar versus\ Earth term:}} A propagating GW will pass both the pulsar and Earth, affecting their local space-time at different times. Pulsar timing can detect a GW's passage through an individual pulsar (``pulsar term''), and can detect a wave's passage through Earth (``Earth term'') as a signal correlated between pulsars.

\vspace{-10pt}\paragraph{\textbf{Correlation analysis:}} While we can detect the Earth or pulsar term in one pulsar, a GW can only be confidently detected by observing the correlated influence of the GW on multiple pulsars, demonstrating \revision{a dominantly quadrupolar signature.}

\vspace{-10pt}\paragraph{\textbf{GW signals:}}
\begin{itemize}[noitemsep,topsep=0pt]
\item \emph{Continuous waves} from orbiting binary black holes.
\item \emph{GW bursts} from single-encounter \revision{supermassive black hole (SMBH)} pairs and cosmic strings.
\item \emph{Bursts with memory}, singular, rapid, and permanent step changes in space time that can accompany \revision{SMBH binary (SMBHB)} coalescence and cosmic strings.
\item \emph{GW background}, the combined sum from all sources of GW emission.
\end{itemize}
\end{tcolorbox}

\subsection{Pulsar Timing and Timing Residuals}


We time a pulsar by building a ``timing model", which is a mathematical description of anything we know about what will affect the arrival times of its pulses at the Earth (for details on how precision pulse arrival times are measured, see \eg~\citealt{2012hpa..book.....L}). Effects we know about and model include (but are not limited to) the time-dependent position of Earth in the Solar System, the natural slowing of a pulsar's period due to rotational energy loss, and any orbital motion of the pulsar, if it is in a binary. The parameters of the timing model are iteratively refined to minimize the root-mean-square (RMS) of the ``timing residuals", which are the difference between the observed pulse arrival time and the arrival time expected based on the timing model.

As a GW moves between the Earth and a pulsar, it alters the local space-time, and thus changes the effective path length light must travel. By this process, the pulses will arrive slightly earlier or later than expected. GWs and any other processes influencing pulse arrival times that are unaccounted for in the timing model will manifest as structure in the pulsar's timing residuals. Since a pulsar's timing model is modified over time to remove as much structure as possible from the timing residuals (forming so-called ``post-fit" timing residuals), some of the residual structure induced by a GW will be effectively ``absorbed'' by the timing model.    

\begin{figure*}
\begin{subfigure}{0.5\textwidth}
\includegraphics[width=1.0\textwidth,trim=40mm 4mm 18mm 10.3cm,clip]{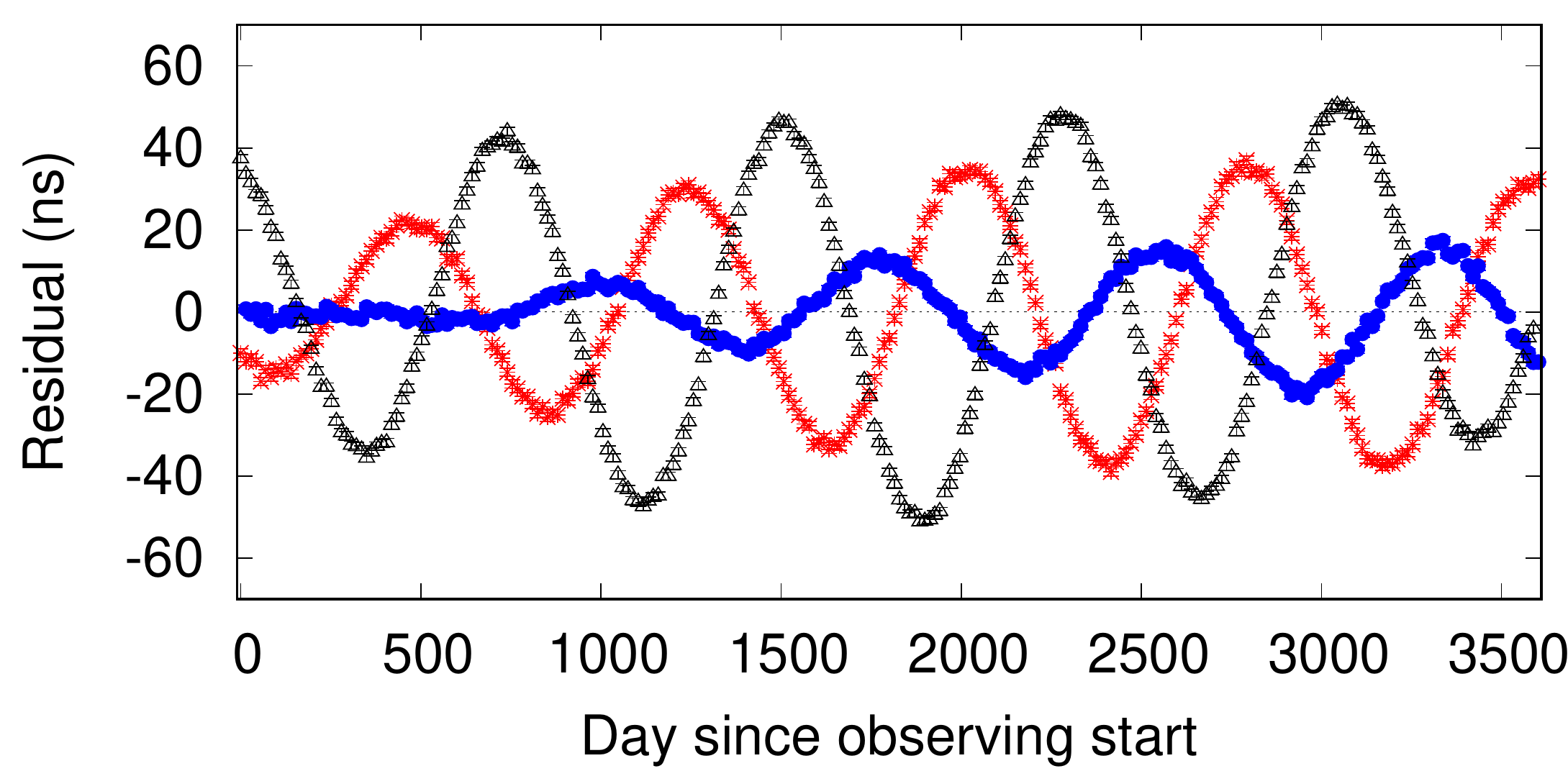}
\vspace{-5mm}
\caption{Continuous wave}\label{sub:cw}
\end{subfigure}
\begin{subfigure}{0.5\textwidth}
\includegraphics[width=1.0\textwidth,trim=40mm 4mm 18mm 10.3cm,clip]{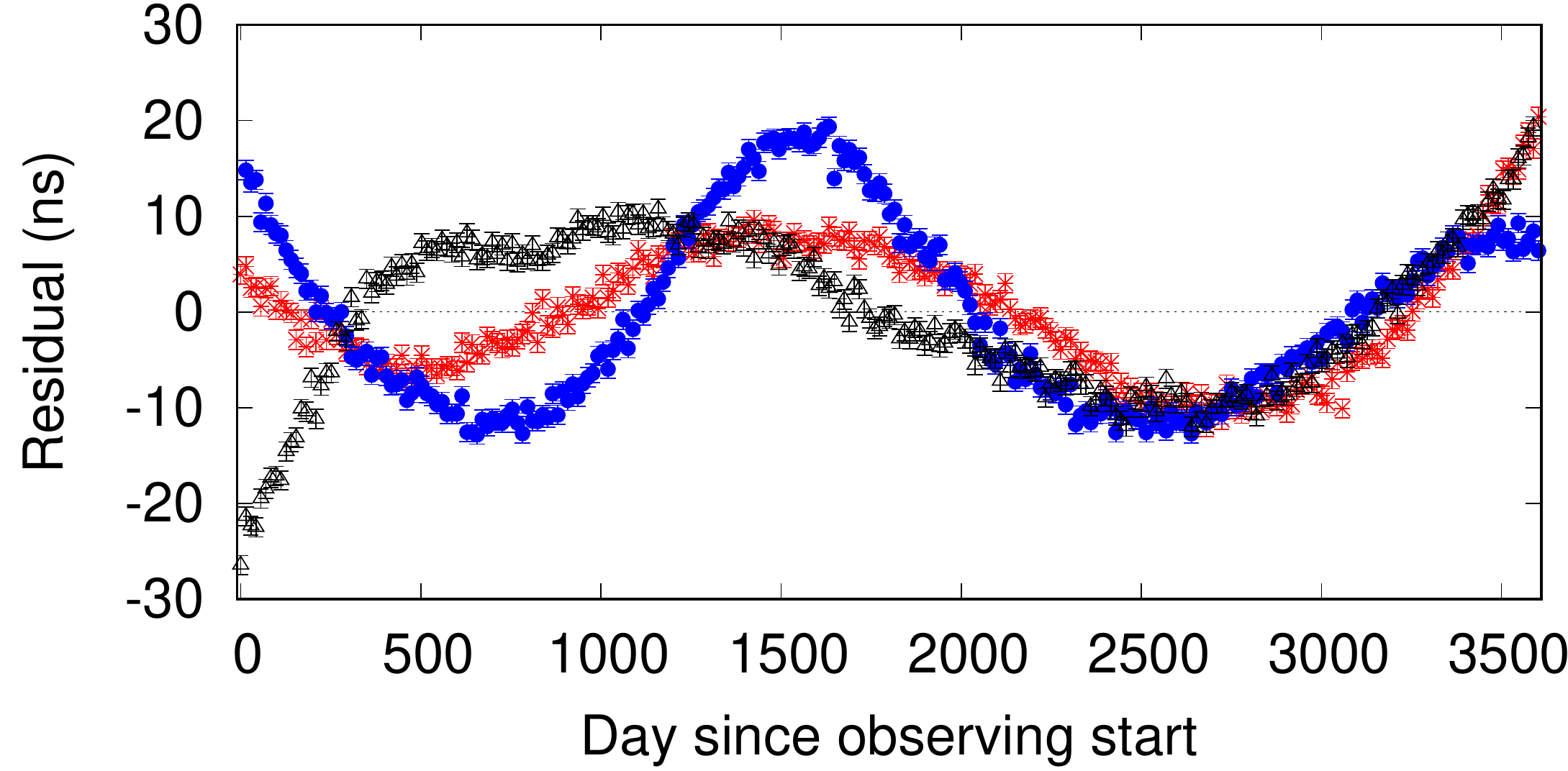}
\vspace{-5mm}
\caption{Background}\label{sub:gwb}
\end{subfigure}

\vspace{2mm}
\begin{subfigure}{0.5\textwidth}
\includegraphics[width=1.0\textwidth,trim=40mm 1mm 18mm 10.3cm,clip]{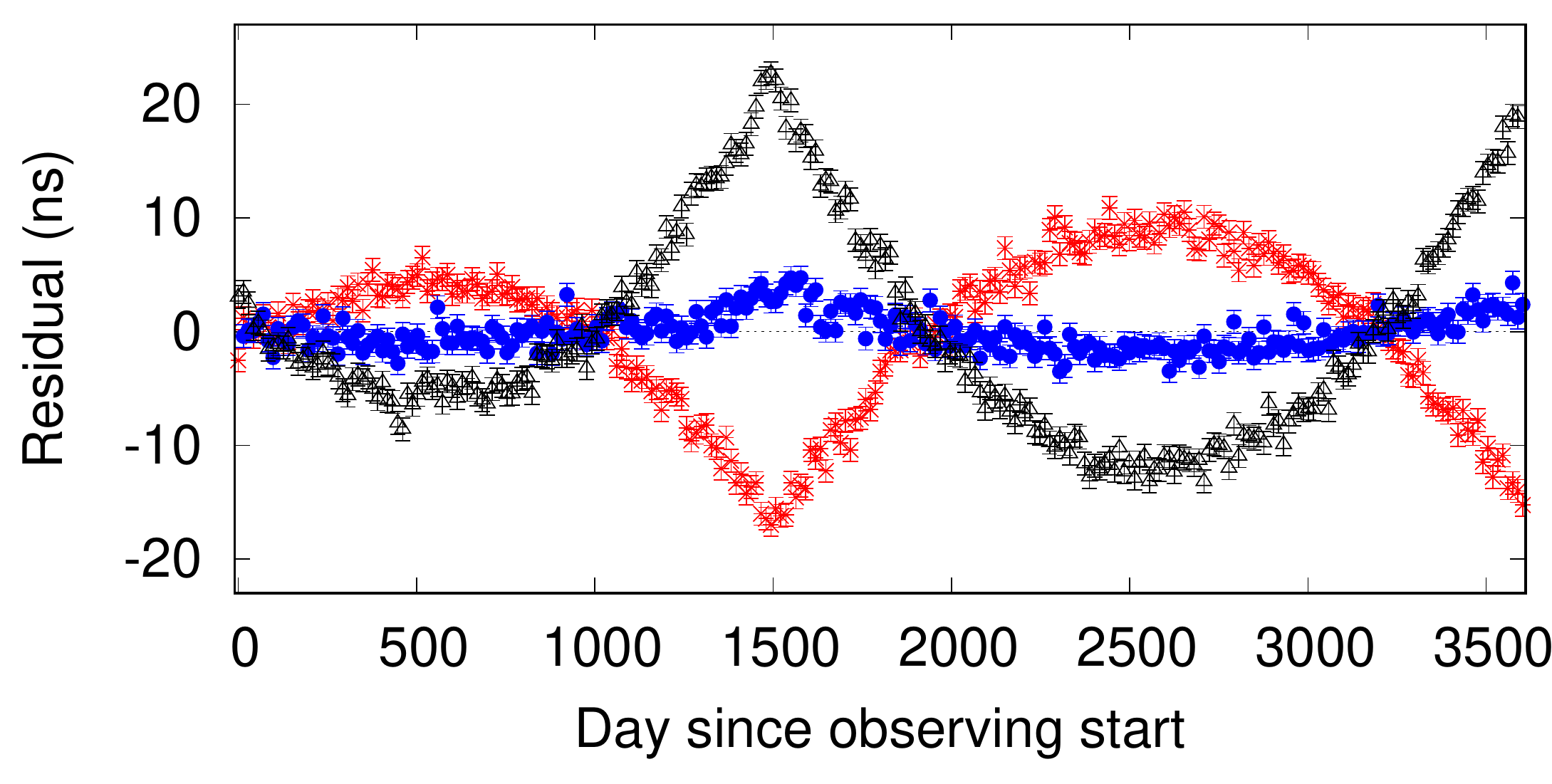}
\vspace{-6mm}
\caption{Burst with Memory}\label{sub:memory}
\end{subfigure}
\begin{subfigure}{0.5\textwidth}
\includegraphics[width=1.0\textwidth,trim=40mm 1mm 18mm 10.3cm,clip]{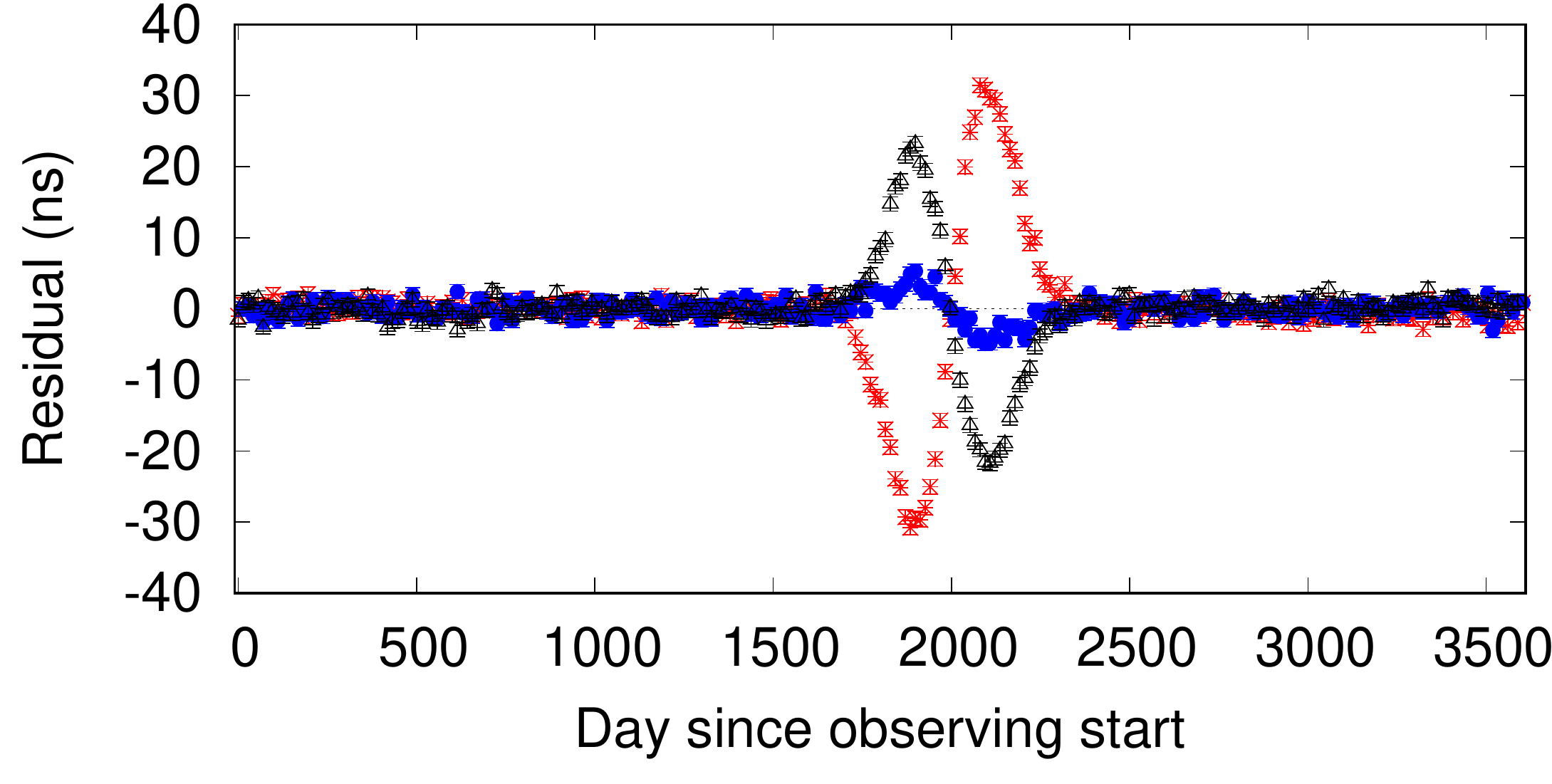}
\vspace{-6mm}
\caption{Burst}\label{sub:burst}
\end{subfigure}
\vspace{-2mm}
\caption{Each panel shows pulsar timing residuals for three pulsars (black triangles, red stars, blue circles) simulated with weekly observing cadence and 1\,ns of white noise in their arrival times. The pulsar-to-pulsar variations demonstrate how the quadrupolar signature of GWs will manifest as correlated timing residuals in distinct pulsars. Note that 1\,ns is not a noise level yet achieved for any pulsar, however here it allows us to demonstrate each observable signal type with a high signal-to-noise ratio. 
Panels are: \emph{(a)} Continuous waves from an equal-mass $10^9\,\msun$ SMBHB at redshift $z=0.01$. The distortion from a perfect sinusoid is caused by self-interference from the pulsar term (Sec.~\ref{sec:ptaterms}). In this case, the pulsar term has a lower frequency because we are seeing the effects on the pulsar from an earlier phase in the SMBHB's inspiral evolution. This interferes with the Earth term, which takes a direct path from the source to Earth and therefore is a view of a more advanced stage of evolution. \emph{(b)}A GW background with $h_c=10^{-15}$ and $\alpha=-2/3$; \emph{(c)} A memory event of $h=5\times10^{-15}$, whose wavefront passes the Earth on day 1500. \emph{(d)} A burst source with an arbitrary waveform.}
\label{fig:residuals}
\end{figure*}

Simulated post-fit residuals influenced by a variety of GW signals, that PTAs are poised to detect, are illustrated in Fig\,\ref{fig:residuals}. Residuals for three different pulsars are shown to demonstrate how the GW signal can vary from pulsar to pulsar.
As can be easily seen, PTAs are sensitive to effects that last on time scales from $\sim$weeks, which is the approximate cadence of pulsar observations, to decades, which is how long PTA experiments have been running. We note that the scale of these GW effects is realistic given the properties of SMBHBs, but the noise level is optimistically small by a factor of 20 or more depending on the pulsar.
Since realistic signals will not have such a high signal-to-noise ratio, PTAs time dozens of pulsars  to mitigate signal-to-noise limitations in individual pulsars, and search for correlations in their timing residuals.


\subsection{The Pulsar Term, the Earth Term, and Correlation Analysis}\label{sec:ptaterms}
A GW passing through the galaxy perturbs the local space-time at the Earth and at the pulsar, but at different times. 
Pulsar timing can detect a GW's passage through an individual pulsar (pulsar term) and also through the Earth (Earth term). The Earth term signal is correlated between different pulsars, while the pulsar term is not.\footnote{Note, a more precise remark is that the Earth and pulsar terms are not correlated as long as two pulsars are separated by many gravitational wavelengths, that is to say that $fL\gg1$, where $f$ is the GW frequency and $L$ is distance to the pulsar.  This assumption is called the short-wavelength approximation \citep[\eg][]{mm18}.}

Nanohertz GW sources of interest are thought to be tens to hundreds of megaparsecs away and perhaps further \revision{\citep[\eg][]{rsg15}} -- it is thus well-justified to approximate the GW as a plane wave. With this simplifying approximation, we can consider the influence of a GW on the observed pulse arrival time as a Doppler shift between the reference frame of the pulsar we're observing, and the solar system barycenter. 
The Doppler-shifted pulsation frequency as measured by an observer at the quasi-inertial solar system barycenter is given by $f_{\rm obs}=(1+z)f_{\rm emit}$. The observed redshift varies with time, depending on the time-varying influence of the GW on the local spacetime of the pulsar and the solar system barycenter. 
More specifically, at time $t$:
\begin{equation}\label{eq:PTermETerm}
z(t)=\frac{\hat{p}^i\hat{p}^j}{2(1+\hat{\Omega}\cdot\hat{p})}[h_{ij}(t)-h_{ij}(t-t_l)],
\end{equation}
where $h_{ij}$ is the space-time perturbation (typically referred to as the ``strain''), and gives the metric perturbation describing the GW in transverse-traceless gauge. 
With the solar system barycenter as a reference position, the parameter $\hat{p}$ is a vector pointing to the pulsar position, $\hat{\Omega}$ is a vector along the direction in which the GW propagates, $t_l=(l/c)(1+\hat{\Omega}\cdot\hat{p})$, and $l$ is the distance between the pulsar and solar system barycenter. The timing perturbation to pulse arrival times is the integral of the redshift over time \citep{det79,abc+09}, which reduces to the difference between the Earth term (evaluated at time $t$) and the pulsar term (evaluated at time $t-t_l$). 


Note the pulsar term, if observed, always depicts an earlier time in the evolution of a GW signal. This is because the Earth term samples GWs arriving to Earth directly from the source, while the pulsar term is associated with a longer path length, encompassing the GW's trip to the pulsar from the source and then the traversal of light from the pulsar to Earth. This may be an important effect in studying SMBHB evolution, as discussed in more detail in Section \ref{sec:bhspin}.

Figure~\ref{fig:residuals} illustrates simulated post-fit residuals for four types of GW signals. For each signal, the residuals of three separate pulsars are shown. Figures \ref{sub:cw} and \ref{sub:gwb}, representing continuous GWs from an individual SMBHB and a stochastic GW background from an ensemble of SMBHBs, respectively, correspond to long-duration signals, for which both the Earth and pulsar terms are active simultaneously. In these cases, the pulsar term interferes with the Earth term and lessens the extent to which the residuals between different pulsars are correlated or anti-correlated. In Figure \ref{sub:cw}, we have modeled the inspiral phase of a SMBHB, which over the course of its evolution changes its frequency and phase. Because each pulsar has a different position relative to Earth and the GW source, the pulsar terms probe different stages of the SMBHB orbit and the pulsar terms interfere with the Earth-term signal in slightly different ways. For burst-like signals, Figures \ref{sub:memory} and \ref{sub:burst}, the Earth term can be active, while the pulsar term is quiescent or \emph{vice versa}. If the Earth term is active, but all pulsar terms are not, the timing perturbation from the GW will be fully correlated or anti-correlated across all pulsars in a PTA.

It is important to note that a GW can only be confidently detected by PTAs if the correlated influence of the GW on multiple pulsars is observed \citep{tiburzi+16,tlb+17}. Earth-based clock errors will influence all pulsar timing residuals the same way (monopolar signature, \eg\ \citealt{2012MNRAS.427.2780H}), and errors in our solar system models will influence ecliptic pulsars more severely (dipolar signature, \eg\ \citealt{2010ApJ...720L.201C}). GWs are expected to have a quadrupolar signature, the directional correlations of which are expected to depend on the nature of gravity, \revision{the polarization of the GWs, and also the nature of other noise sources affecting the PTA \citep{tiburzi+16,tlb+17}}. Therefore, the relative positions of a GW source and two pulsars will dictate how the residuals of those two pulsars are correlated. How these correlations appear as a function of the angle between pulsars depends on the nature of gravity and the polarization of a GW, and is commonly shown for pulsar timing data as the ``Hellings and Downs'' curve \citep{1983ApJ...265L..39H}. Section~\ref{sec:grtests} discusses correlation analysis and the Hellings and Downs curve in more detail, and describes how various models of gravity will dictate the shape of the correlations observed.

\subsection{Types of Gravitational-wave Signal}
Depending on the origin of the GW signal, the induced residuals might appear as deterministic signals or a stochastic background. Here we simply aim to set up a reference point of what signal modes we expect to detect with PTAs. In the remainder of the document, we further discuss what information can be extracted about the Universe, depending on the type of the detected GW signal.

Cyclic signals (continuous waves; Fig.~\ref{sub:cw}) can arise from objects in an actively orbiting binary system. Bursts (Fig.~\ref{sub:burst}) represent rapid but temporally finite accelerations, e.g., during the pericenter passage in a highly eccentric or unbound orbit of two SMBHs \citep[\eg][]{2010ApJ...718.1400F}. These classes can be detected by PTAs as long as their characteristic timescale is between weeks and a few decades.

Bursts with memory (Fig.~\ref{sub:memory}) represent a rapid and permanent deformation in spacetime. A burst with memory (BWM) event is generally expected to occur on timescales less than 1\,day \citep[\eg][]{f10}. Because the duration of memory's ramp-up time is relatively brief compared to PTA observing cadence, it is commonly modelled as an instantaneous step function in strain. PTAs cannot typically detect the memory event as it occurs, but they can see the resulting difference between the pre- and post-event spacetimes; the BWM creates a sudden, long-term change in the apparent period of a pulsar. This leaves a low-frequency ramp-like signature in the pulsar timing residuals \citep[\eg][]{vhl10,2015ApJ...810..150A,mcc14,whc+15}.
In Figure~\ref{sub:memory}, the memory event at day 1500 causes a characteristic ``$\omega$'' shape to be seen in the timing residuals, indicating a difference in the spacetime before and after the event. The residual shape here is influenced by our fit to the period and period derivative of each pulsar, which is required in the pulsar timing model. See \citet{mcc14} and Section \ref{sec:gwmemory} for more details on this effect.

\begin{figure*}
\centering
\includegraphics[width=0.73\textwidth]{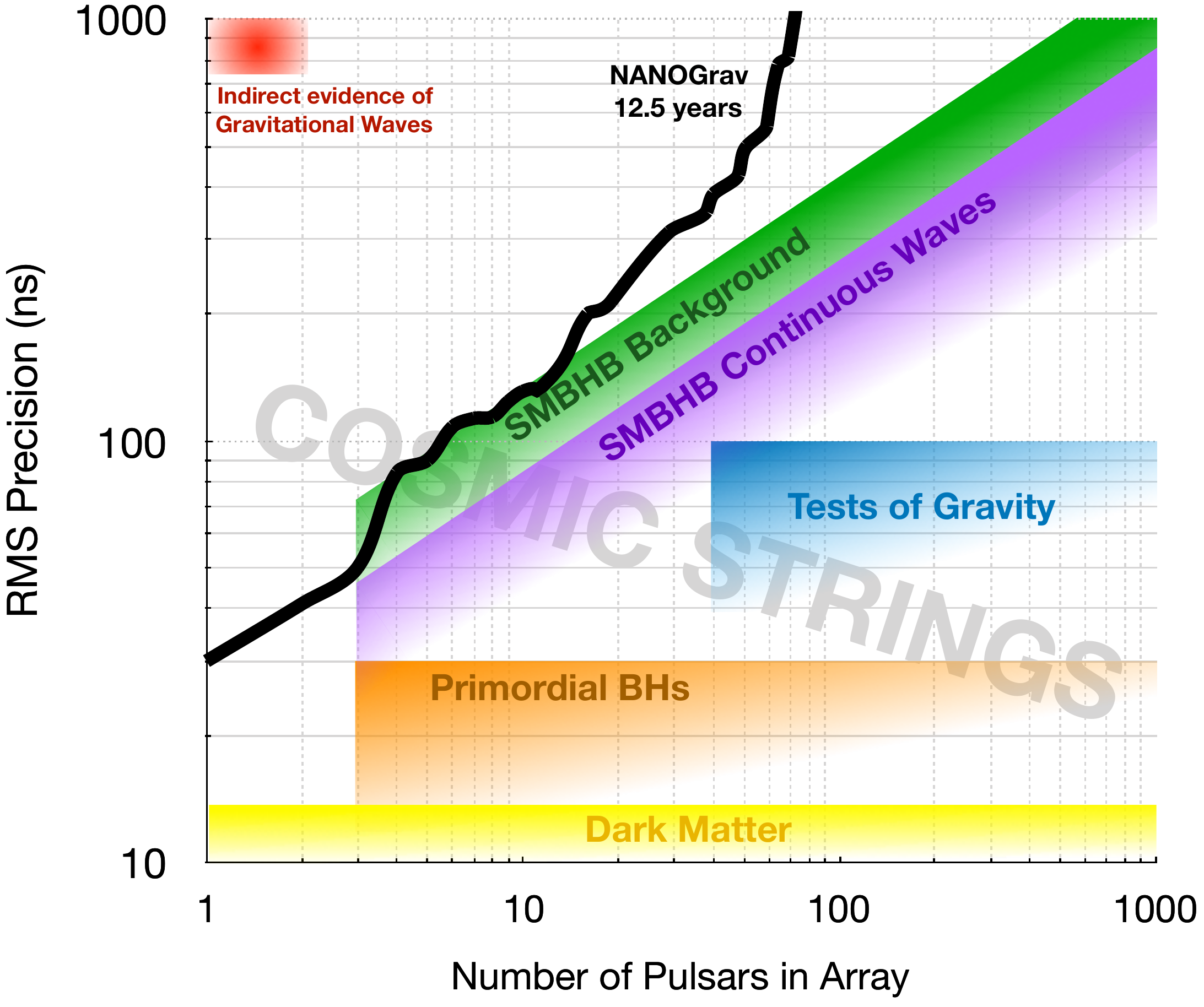}\\
\includegraphics[width=0.73\textwidth]{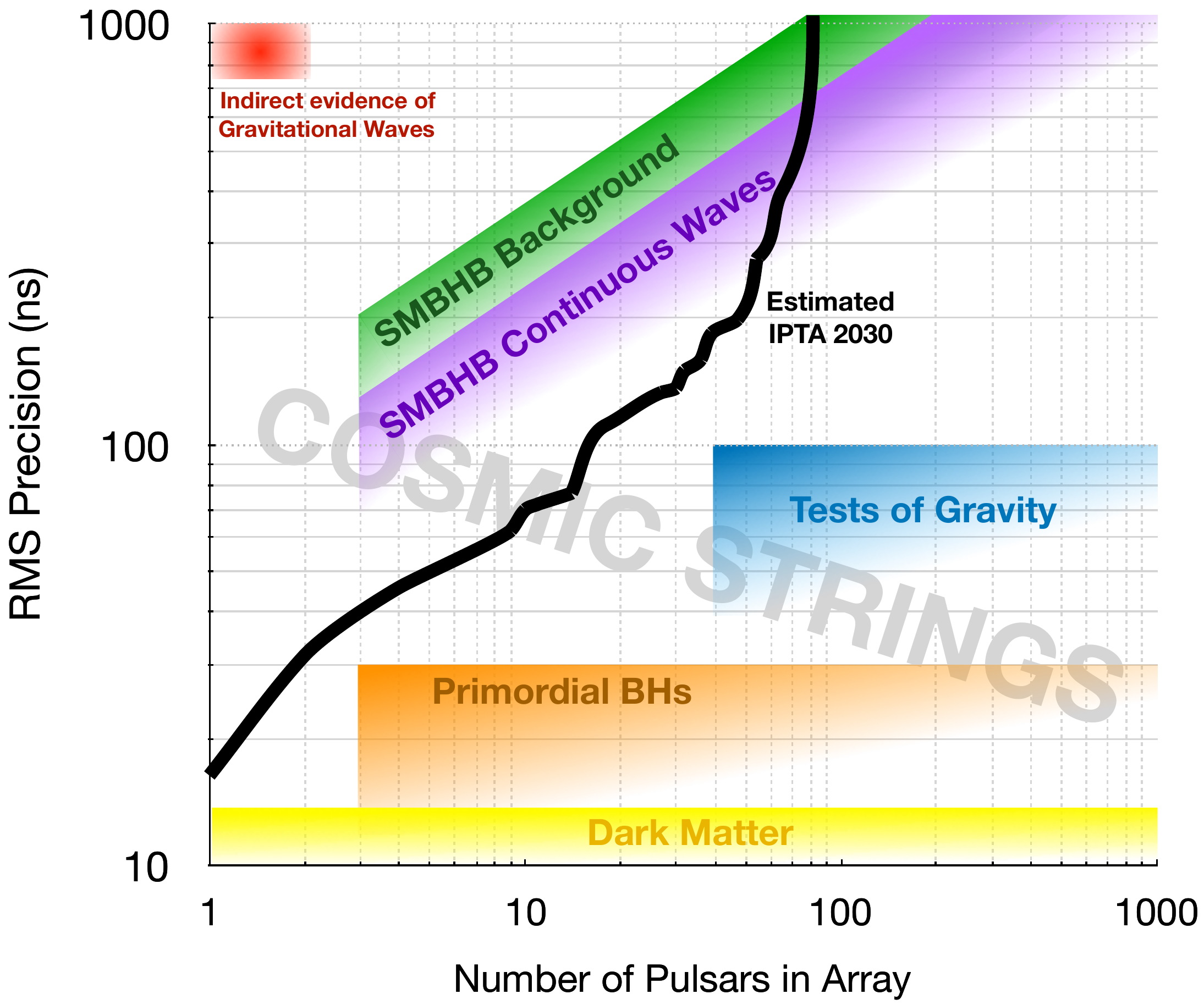}
\vspace{-2mm}
\caption{Here we outline the approximate number of pulsars and timing precision required to access various science, based on current predictions for each signal. The upper and lower panels represent a 10- and 25-year timing array, respectively.
In the top plot, the black curve shows a representative PTA, reflecting the upcoming NANOGrav 12.5-year data release. \revision{That data set contains approximately 70 pulsars, however the timescale over which each pulsar has been timed ranges from $\sim$1--20\,years. The lower plot shows expectations for the future IPTA, assuming approximately 100 pulsars.}
Each curve shows pulsars that are timed to a precision lower than or equal to the indicated RMS timing precision.  The location and shape of the SMBHB regions reflect the scaling relations of \citet{sejr13}. \revision{These assume a detection signal-to-noise ratio of at least five, and a SMBHB background of $h_c\lesssim 1\times 10^{-15},$ which is just below the most recent limit placed independently by several PTAs} on this background source of GWs. A longer-duration PTA requires less precision and fewer pulsars for a detection because the signal-to-noise ratio scales with total observing time.}
\label{fig:summary}
\end{figure*}

Finally, all of nature's deterministic signals can contribute en masse to a stochastic GW background (Fig.\,\ref{sub:gwb}). 
The strain of the background is frequency-dependent, described by the characteristic-strain spectrum, $h_{\rm c}(f)$. This is calculated by integrating the squared GW strain signal 
over the entire emitting population \citep{p01}.
For most GW sources in the nanohertz to microhertz frequency regime, the predicted characteristic strain spectrum can be simplified as a single power-law:
\begin{equation} \label{eq:hc-plaw}
h_{\rm c} (f) = A\left(\frac{f}{f_{\mathrm{yr}}}\right)^{\alpha} \ ,
\end{equation}
or in terms of the energy density of GWs, $\Omega_{\rm GW} \propto f^{2(1+\alpha)}$. 
In this way, the background can be characterized with an amplitude $A$, and a single spectral \mbox{index $\alpha$.} The amplitude $A$ is commonly defined at a frequency $f_{\rm yr}=1\,$year$^{-1} \sim 32\,\mathrm{nHz}$.
As discussed in future sections, the details of physical processes in galaxies, cosmic strings, and inflation can potentially make the spectrum more complex than a single power law. Nevertheless, as the PTA sensitivity to a GW background increases, they will be able to detect the amplitude and spectral shape of the background.
In Figure~\ref{sub:gwb}, the background appears as a red noise process, because the index $\alpha$ is negative, leading to greater variations at longer timescales/lower frequencies. For a good introductory review on methods to detect the stochastic background, we refer readers to \citet{romano_cornish_LRR}.

Figure\,\ref{fig:summary} provides a bird's-eye view of the PTA sensitivities required to successfully breach each area of science that we describe in the remainder of this document. Regardless of the emission source, PTAs will grow in sensitivity by adding pulsars to the array, by decreasing the average RMS residuals, and simply by timing pulsars for a longer duration. Thus, in Fig.\,\ref{fig:summary} we show how the requirements change as a function of these parameters. As seen in the top (``now'') panel, PTAs have already breached the expectations for the background of GWs from SMBHBs formed in galaxy mergers, and are now setting increasingly stringent limits on galaxy/black hole co-evolution. In the coming years to decade, we expect this to become first a detection of the background, and then become an exploration of the physics of discrete SMBHB systems.
Deeper explorations of gravity, dark matter, and other effects should be soon to follow thereafter.

%


\section{The Population and Evolution of Supermassive Black Hole Binaries}
\label{sec:smbh}
\vspace{-3mm}
\begin{tcolorbox}[enhanced,colback=gray!10!white,colframe=black,drop shadow]

SMBHBs are expected to be the brightest sources of nanohertz GWs. They form during major mergers between two galaxies, each of which contain their own SMBH.\vspace{-10pt}
\paragraph{\textbf{Stochastic Background}}
\begin{itemize}[noitemsep,topsep=0pt]
\item In the simplest model, an ensemble of SMBHBs should produce a stochastic GW background with a characteristic strain $h_{\rm c}=A~f^{-2/3}$ ($\Omega_\mathrm{gw}(f)~\propto~f^{2/3}$), with an amplitude $\mathcal{O}[10^{-15}]$ ($\mathcal{O}[10^{-9}]$) at a GW frequency of 1\,year$^{-1}$. 
\item The GW background spectrum might be detected with a shallower slope at frequencies $\lesssim$10\,nHz, if SMBHB evolution is accelerated due to strong interactions with their environments (stars and gas). 
\item 
Galaxy merger rates, SMBH-host co-evolution, dynamical relaxation timescales, and whether SMBHBs might stall at wide separations, can all affect the amplitude scaling of this background. 
\item PTA constraints or measurement of the background's amplitude or spectral shape can give information on all of the above astrophysical uncertainties for the ensemble SMBHB population.
\item A detection of the SMBHB background would confirm the consensus view that SMBHs reside in most or all massive galaxies.
\item Constraints on background anisotropy may indicate local binary clustering or large-scale cosmic features.
\end{itemize} \vspace{-10pt}
\paragraph{\textbf{Discrete Continuous-wave Sources}}
\begin{itemize}[noitemsep,topsep=0pt]
\item Massive or nearby systems may be individually resolvable from the background. Detections will provide the most direct probe of the early-inspiral stage of a SMBHB merger, and can provide measurements of the binary's position, phase, and an entangled estimate of chirp mass and luminosity distance ($\mathcal{M}/D_{\rm L}$). \revision{Chirp mass is defined in the section below.}
\item If detected, the pulsar term can permit 
$\mathcal{M}$ and $D_{\rm L}$ to be disentangled. Evolution of the waveform over PTA experimental durations is unlikely for SMBHBs, however this would also disentangle the $\mathcal{M}/D_{\rm L}$ term.
\end{itemize} \vspace{-10pt}
\paragraph{\textbf{Bursts With Memory}}
\begin{itemize}[noitemsep,topsep=0pt]
\item PTA measurement of a burst with memory can provide the date, approximate sky location, and reduced mass over co-moving distance of a SMBHB coalescence.
\end{itemize}
\end{tcolorbox}

It is now broadly accepted that SMBHs in a mass range around $10^6$--$10^{10}$~M$_\odot$ reside at the centers of most galaxies \citep{kormendy1995, Magorrian1998}, with several scaling relations between the SMBH and galactic-bulge properties \citep[e.g.\ $M_\bullet-\sigma$, $M_\bullet-M_\mathrm{bulge}$,][]{2000ApJ...539L...9F, 2000ApJ...539L..13G} indicating a potential co-evolution between the two.
In the standard paradigm of structure formation, galaxies and SMBHs grow through a continuous process of gas and dark matter accretion, interspersed with major and minor mergers. Major galaxy mergers form binary SMBHs,
and these are currently the primary target for PTAs. In this section, we lay out a detailed picture of what is not known about the SMBHB population, how those unknowns influence GW emission from this population, and what problems PTAs can solve in this area of study.

In Figure~\ref{fig:lifecycle}, we summarize the lifecycle of binary SMBHs. 
SMBHB formation begins with a merger between two massive galaxies, each containing their own SMBH. Through the processes of dynamical friction and mass segregation, the SMBHs will sink to the center of the merger remnant through interactions with the galactic gas, stars and dark matter. Eventually, they will form a gravitationally bound SMBHB \citep{bbr80}.  Through continued interaction with the environment, the binary orbit will tighten, and GW emission will  increasingly dominate their evolution.

Any detection of GWs in the nanohertz regime, either from the GW background or from individual SMBHBs, will be by itself a great scientific accomplishment. Beyond that first detection, however, there are a variety of scientific goals that can be attained from detections of the various types of GW signals. The subsections below discuss these in turn, first setting up GW emission from SMBHB systems, and then detailing the influence of  environmental interactions. Each section describes a different aspect of galaxy evolution that PTAs can access.

Throughout this document we refer to SMBHB parameters using the following conventions: SMBH masses $m_1$ and $m_2$ have a mass ratio $q=m_2/m_1$ defined such that $0\geq q\geq1$. The total mass is $M=m_1+m_2$ and the reduced mass is $\mu=m_1m_2/(m_1+m_2)$. Chirp mass is defined as $\mathcal{M}\equiv (m_1m_2)^{3/5}/(m_1+m_2)^{1/5}$. The binary inclination, eccentricity, and semi-major axis are given by the symbols $i$, $e$, and $a$, respectively.  The parameter $D$ is the radial comoving distance to the binary system.  Other specific parameters will be defined in-line where relevant.

\subsection{GW Emission from Supermassive Black Hole Binaries}

\subsubsection{PTAs and the Binary Lifecycle}~\\\vspace{-3mm}

\begin{figure*}
\centering
\includegraphics[width=1.0\textwidth]{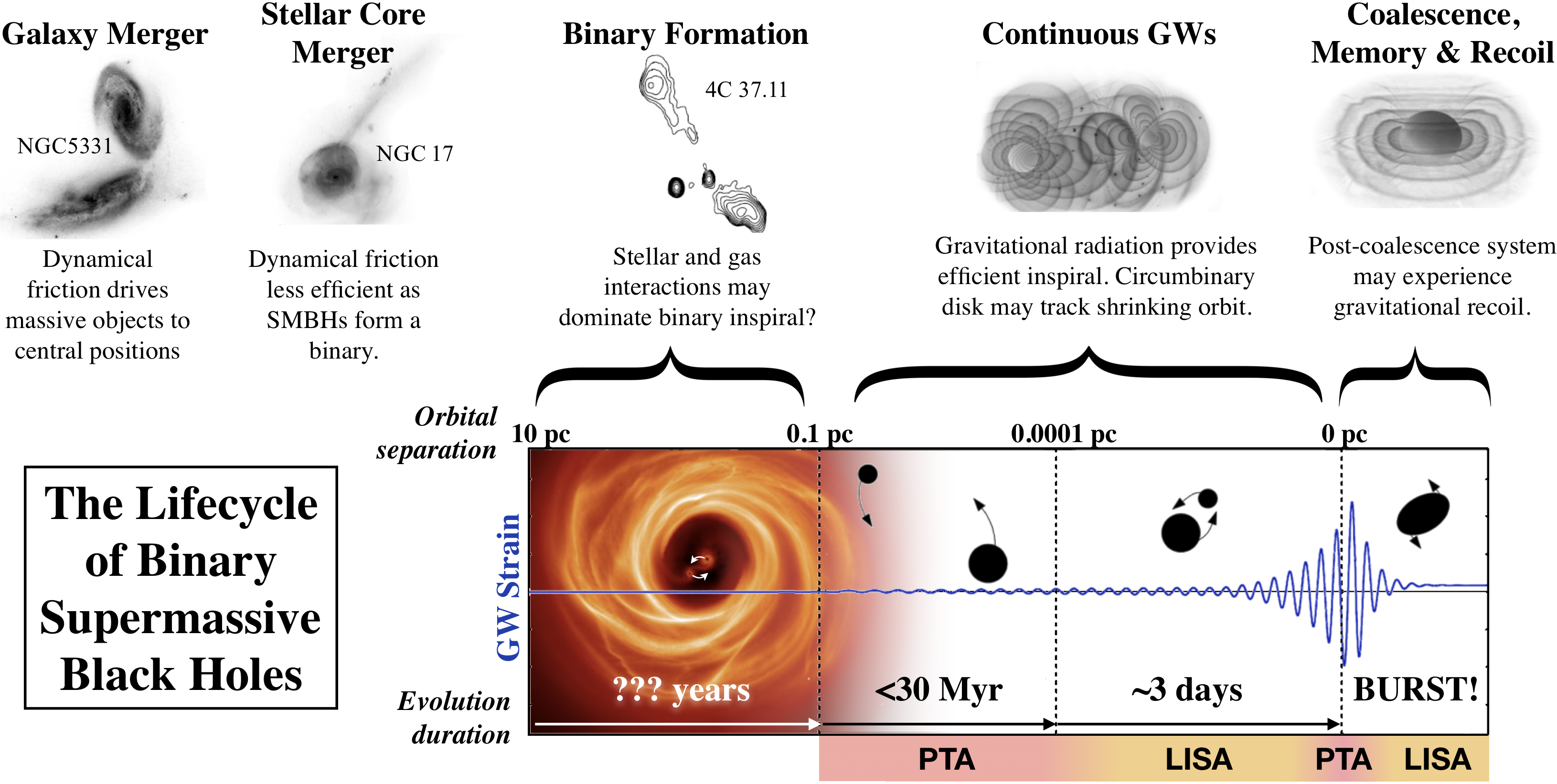}
\caption{\footnotesize 
Binary SMBHs can form during a major merger. Pulsar Timing Arrays' main targets are continuous-wave binaries within $\sim$0.1\,pc separation (second panel in the lower figure; \S\ref{sec:gwinspiral}), although we may on rare occasion detect ``GW memory'' from a binary's coalescence \citep[][Sec.~\ref{sec:gwmemory}]{f10}. Millions of such binaries will contribute to a stochastic GW background, also detectable by PTAs (\S\ref{sec:gwb}).
A major unknown in both binary evolution theory and GW prediction is the means by which a binary progresses from $\sim$10\,pc separations down to $\sim$0.1\,pc, after which the binary can coalesce efficiently due to GWs \citep[\eg][]{bbr80}. If it cannot reach sub-parsec separations, a binary may ``stall'' indefinitely; such occurrences en masse can cause a drastic reduction in the ensemble GWs from this population. Alternately, if the binary interacts excessively with the environment within 0.1\,pc orbital separations, the expected strength and spectrum of the expected GWs will change. 
Image credits: Galaxies, Hubble/STSci; 4C37.11, \citet{rodriguez+06}; Simulation visuals, C. Henze/NASA; Circumbinary accretion disk, C. Cuadra.}
\label{fig:lifecycle}
\end{figure*}

\noindent
As shown in Fig.~\ref{fig:lifecycle}, SMBHBs can emit discrete PTA-detectable GWs in two phases of their lifecycle. PTAs can detect continuous waves during SMBHBs' active inspiral phase, up to a few days before coalescence. PTAs can also detect the moment of coalescence of a SMBHB by detecting its related burst with memory (covered in more detail in \S\ref{sec:gwmemory}).  
As noted earlier (\S\ref{sec:ptaterms}), the ``pulsar term'' of a binary contains information about an earlier stage of binary evolution. For a sufficiently strong GW signal, the pulsar term can be measured in several pulsars, and thus multiple snapshots of the evolutionary progression of the binary can be detected simultaneously \citep{cc10,m+12,ellis2013CQG,teg14}.

Because binary inspiral is slower at larger separations, the number density is much higher for discrete systems at earlier stages of inspiral (that is, at low GW frequency). At these stages, the binary may still be interacting closely with its environment. Here, we review deterministic and stochastic GW emission from binary SMBHs, and in the next sub-sections we develop from this into how environments can influence the GW signals---and how PTAs can uniquely explore these physical processes.

\subsubsection{Continuous Waves: Binary Inspiral}
\label{sec:gwinspiral}~\\\vspace{-3mm}

\noindent
A PTA detection of the correlated signal from a continuous-wave SMBHB (Fig.~\ref{sub:cw}) will produce constraints on a system's orbital parameters \citep{ellis2013CQG,thgm16,2014ApJ...794..141A,2016MNRAS.455.1665B,bs12,z+14,lwk+11}, in much the same way as ground-based instruments can with stellar-mass binaries. PTAs may have potentially poorer parameter estimation; this is because PTAs will typically observe an early portion of the binary inspiral, and only have a glimpse of this phase over the $\sim$1--2 decade observational timespans of PTAs. During this time, we are unlikely to observe frequency evolution of the binary that creates the information-rich ``chirping'' signal seen by ground-based laser interferometers \citep{thgm16}, which can allow GW experiments to derive distances to a binary and a detailed model of the system's evolution. However, if the system is of sufficiently high mass, has initially high eccentricity, or is detected in a late stage of inspiral evolution, then chirping within the observational window may be detectable \citep{kjlee-CWs,thgm16}. 


As a binary evolves and accelerates in its orbit, it has a greater chance at decoupling from the environment. Somewhere below separations of $\sim$1\,pc, this may occur and the binary can be considered as an isolated physical system. In this case, the dissipation of orbital energy will depend only on the constituent SMBH masses, the orbital semi-major axis, and the binary's eccentricity. As explored in the sections below, the timing of this decoupling has a distinct effect on the detectable GW signals from SMBHB systems. Here, we lay out pure orbital evolution due to gravitational radiation.

The leading order equations for GW-driven orbital evolution are \citep{pm63,peters1964}:
\begin{eqnarray} \label{eq:orbit-gw}
\left\langle \frac{{\rm d}a}{{\rm d}t} \right\rangle &= -\frac{64}{5}\frac{m_1m_2(m_1+m_2)}{a^3}\frac{(1+\frac{73}{24}e^2+\frac{37}{96}e^4)}{(1-e^2)^{7/2}}, \nonumber\\ 
\left\langle \frac{{\rm d}e}{{\rm d}t} \right\rangle &= -\frac{304}{15}\frac{m_1m_2(m_1+m_2)}{a^4}e\frac{(1+\frac{121}{304}e^2)}{(1-e^2)^{5/2}}, 
\end{eqnarray}
where the derivatives of the orbital separation and eccentricity are averaged over an orbital period.
One should note from \ref{eq:orbit-gw} that GW emission always causes the eccentricity to decrease, i.e.~the binary will become more circular as it inspirals toward coalescence. 
For purely circular systems, the GW emission frequency will be twice that of the orbital frequency, and will evolve as ${\rm d}f/{\rm d}t\propto f^{11/3}$.

An important concept here is that of \emph{residence times}; because the binary's inspiral evolves more rapidly fast at smaller separations, it spends less time residing in a high-frequency state once its inspiral is GW-dominated (and accordingly, it spends less time residing in a state of small-separation). Thus, we would naturally expect there to be fewer binary systems emitting at high GW frequencies, and many more binary systems emitting at low GW frequencies. As you will see in the next section, this becomes a critical point in assessing the shape of the GW background.

For a population of eccentric binaries, the GW emission will be distributed across a spectrum of harmonics of each binary's orbital frequency. At higher eccentricities, the frequency of peak emitted GW power shifts to higher and higher harmonics. This peak will coincide approximately with the pericenter frequency \citep{krp12}, such that:
\begin{equation}
f_\mathrm{peak} \approx \frac{(1+e)^{1/2}}{(1-e)^{3/2}} \frac{f_{K,r}}{(1+z)}~,
\end{equation}
where
\begin{equation}
f_{K,r} = \frac{f(1+z)}{n}~.
\end{equation}
Here, $f$ is the (Keplerian, observed, Earth-reference-frame) GW frequency, and $z$ is source redshift. The parameter $n$ describes the harmonic of the orbital frequency at which the GWs are emitted; for a circular system, $n=2$. 
Eccentric systems emit at the orbital frequency itself as well as at higher harmonics \citep[\eg][]{2003ApJ...598..419W}.

Binary eccentricity and the Keplerian orbital frequency co-evolve in the following mass-independent way \citep{pm63,peters1964,thgm16};
\begin{equation}\label{eq:ecc1}
\frac{f_{K,r}(e)}{f_{K,r}(e_0)} = \left(\frac{\sigma(e_0)}{\sigma(e)}\right)^{3/2}~,
\end{equation}
where $e_0$ is the eccentricity at some reference epoch, and,
\begin{equation}\label{eq:ecc2}
\sigma(e) = \frac{e^{12/19}}{1-e^2}\left[1 + \frac{121}{304}e^2\right]^{870/2299}~, 
\end{equation}
Hence a binary with (for example) $e_0=0.95$, when its orbital frequency is $1$ nHz, will have $e\approx 0.3$ by the time its orbital frequency has evolved to $100$ nHz. 

Finally, the strain at which GWs are emitted is often quoted as the ``RMS strain'' averaged over orbital orientations:
\begin{equation}\label{eq:hrms}
h(f_{K,r}) = \sqrt{\frac{32}{5}}\,\frac{(G\mathcal{M})^{5/3}}{c^4D}\,(2\pi f_{K,r})^{2/3}~.
\end{equation}
This equation and those above highlight the fact that continuous-wave detection by PTAs will enable a measurement of the system's orbital frequency and eccentricity. However, the strain amplitude is scaled by the degenerate parameters $\mathcal{M}^{5/3}/D$; therefore, chirp mass and source distance cannot be directly measured unless there is orbital frequency evolution observed over the course of the PTA observations, or unless the host galaxy of the continuous-wave source is identified (Sec.~\ref{sec:em}). Some loose constraints on the mass and distance of the continuous wave might be inferred simply based on the fact that statistically, we expect the first few continuous-wave detections to be of the heaviest, relatively equal-mass systems, at low to moderate redshifts ($z\lesssim 1$), as demonstrated originally in \citet{sesanaCW}.

It is worth noting here that, until now in this section, we have ignored the pulsar term (\S\ref{sec:ptaterms}). Because the Earth term is correlated between different pulsars, it will always be discovered at a higher S/N than the pulsar term. If the pulsar term can be measured in multiple pulsars, however, we can map multiple phases of the binary's inspiral history. We can exploit this information through a technique known as temporal aperture synthesis to disentangle chirp mass and distance, as well as improve the precision of other parameters \citep{ellis2013CQG,esc12,cc10,teg14,2016MNRAS.461.1317Z}. Likewise, if we have many pulsars and excellent timing precision then we can potentially place constraints on BH spin terms in the waveform \citep{m+12}. This is discussed further in Section \ref{sec:bhspin}.

\subsubsection{Memory: Binary Coalescence}
\label{sec:gwmemory}~\\\vspace{-3mm}

\noindent
SMBHBs are one of the two leading sources that we expect to produce detectable GW memory events (the other potential source, as noted later in this paper, is cosmic string loops). In the case of SMBHBs, the inspiral and even the coalescence produce the oscillatory waveform that we see as continuous waves. However, the stress tensor after the binary's coalescence will differ from its mean before the coalescence; this is apparent in the ``BURST'' panel in Figure \ref{fig:lifecycle}, where the waveform is offset from zero after the SMBHB coalescence's ring-down period. That offset, which grows over the entire past history of the binary's evolution, but most precipitously during the coalescence, is the non-oscillatory term we call \emph{memory} \citep{first_memory_paper,bt87,c91,t92,f10}. All SMBHBs are expected to produce a GW memory signal.
SMBHBs may produce both \emph{linear} and \emph{non-linear} memory, where the former is related to the SMBHB motion in the final moments of coalescence, whereas the non-linear signal is produced by the GWs themselves \citep[see the discussion in][]{f11}. The memory strain from a coalescing binary depends on the binary parameters and the black hole spin. For a circular binary, the order of the strain can reasonably be approximated as
\begin{equation}
\hspace{-15mm}h \simeq h_+\simeq \frac{(1-\sqrt{8}/3)G\mu}{24 c^2 D}{\rm sin}^2(i)[17+\cos^2(i)][1+\mathcal{O}(\mu^2/M^2)] \simeq 0.02\frac{G\mu}{c^2D} \ ,
\end{equation}
and the cross-hand polarization term $h_\times$ goes to zero \citep{mcc14}.

Note that due to its non-oscillatory nature, this strain deformation is permanent. As previously noted, this leads to a sudden observed change in the observed period of pulsars. After this event, timing residuals will begin to deviate from zero with a linear upwards or downwards trend.
We would observe that ramp as such, if we knew the intrinsic spin period and spin-down rate of pulsars (instead, we measure these from timing data). After fitting the timing residuals for the pulsar's period and period derivative, one finds the signature shown in Figure~\ref{sub:memory}, with the sharp peak of the curve representing the moment of coalescence. Based on the time and the amplitude of this signature in the residuals, PTAs can measure the date of coalescence, and obtain a covariant measurement of the SMBHB's reduced mass and co-moving distance. If the signature is detected in the Earth-term (i.e. correlated between multiple pulsars), a position of the memory event can be loosely inferred, likely to a few thousand square degrees depending on the number of pulsars in the array and how well they are timed. 

Predictive  simulations  have  found  that  PTAs are highly unlikely to detect GW memory from SMBHB mergers due to the extreme rarity of bright events \citep{seto09,cordes+jenet12,ravi+15}. Nonetheless, \citet{2014PhRvD..89d2003C} find memory to be increasingly important for investigating phenomena at high redshift and discuss its potential for discovering unexpected phenomena. Techniques to search for  memory in PTAs have been  developed \citep[\eg][]{2010MNRAS.402..417P,vhl10,mcc14}, and applied to place limits with PTAs \citep{2015ApJ...810..150A,whc+15}. 

\begin{figure*}
\centering
\includegraphics[width=\textwidth,trim=0mm 0mm 0mm 0mm,clip]{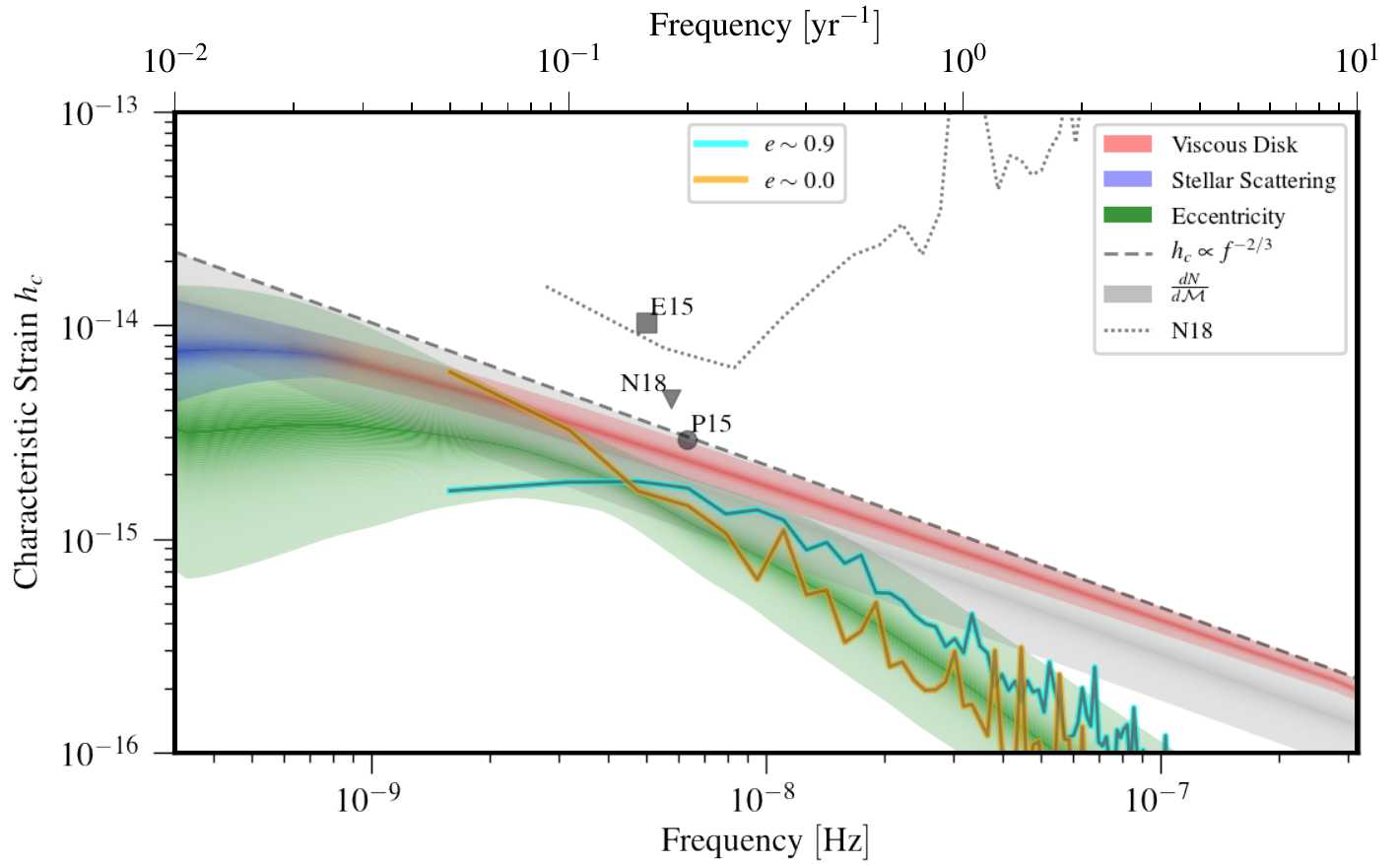}
\caption{The GW spectrum at nanohertz frequencies from supermassive black hole binaries.  We adapted the data from the SMBH binary populations and evolutionary models of \citet{k+17a} and \citet{k+17}, highlighting the effects of variations in particular binary model parameters on the resulting GW spectrum.  The dashed black line is the spectrum using only the population mass distribution and assuming GW-driven evolution, and the grey-shaded region represents the uncertainty in the overall distribution of SMBHB in the Universe.
The cyan (orange) line is the GW background from a particular realization of a SMBHB population using a high (low) eccentricity model.
The time sampling corresponds to a PTA with duration of $20$ yr and a cadence of $0.05$ yr.  The NANOGrav 11yr detection sensitivity and GW background upper-limits \citep{2018ApJ...859...47A} are illustrated with a grey dotted-line and triangle, respectively\revision{, while the EPTA \citep{2015MNRAS.453.2576L} and PPTA \citep{2015Sci...349.1522S} upper-limits are denoted by a square and circle, respectively. We note that the PPTA limit appears to be most constraining, however it is known to be sensitive to the choice of planetary ephemeris; this effect has been accounted for in subsequent analysis of other PTA data and results in less constraining limits \citep{2018ApJ...859...47A}.} \textit{Note:} the shaded regions are schematic.}
\label{fig:gwspectrum}
\end{figure*}

\subsubsection{The Stochastic Binary Gravitational-wave Background}\label{sec:gwb}~\\\vspace{-3mm}

\noindent
The superposition of GWs \revision{from the large population of SMBHBs predicted by hierarchical galaxy formation \citep{vhm03}} will produce a stochastic background. The GW background has greater power at lower frequencies (Fig.~\ref{sub:gwb}); 
we typically visualize this as a plot of characteristic strain, $h_{\rm c}(f)$. Figure~\ref{fig:gwspectrum} demonstrates the effect on the GW spectrum of variations in assumptions about the SMBHB population. This connection between PTA characterization of the background and SMBHB evolution and galaxy dynamical evolution is the focus of the following subsections. Here, we outline how many discrete continuous-wave sources can contribute to a GW background. 

 The calculation reveals the cosmological and astrophysical factors that can influence the spectral amplitude and shape of the GW background \citep{1995ApJ...446..543R, wyithe2003, sesana2004}.  In particular, we typically calculate the characteristic strain spectrum from an astrophysical population of inspiraling compact binaries (e.g.~that shown in Fig.~\ref{fig:gwspectrum}) as

\definecolor{col0}{RGB}{100, 100, 100}   
\definecolor{col1}{RGB}{122, 0, 0}   
\definecolor{col2}{RGB}{0, 80, 15}   
\definecolor{col3}{RGB}{0, 0, 0} 
\definecolor{col4}{RGB}{133, 6, 165} 

\begin{eqnarray} \label{eq:hc-spec}
h_{\rm c}^2(f) = & \int_0^\infty \mathrm{d}z \int_0^\infty \mathrm{d}M \int_0^1 \mathrm{d}q \, 
	{\frac{\mathrm{d}^4N}{\mathrm{d}z\,\mathrm{d}M\,\mathrm{d}q\,\mathrm{d}t_r}} \times \nonumber \\
& \qquad\sum_{n=1}^\infty\left\{
	{\frac{g[n,e(f_{K,r})]}{(n/2)^2}}
    {\frac{\mathrm{d}t_r}{\mathrm{d}\ln f_{K,r}}}
    {h^2(f_{K,r}) } \right\}
\end{eqnarray}
%
where the contributing factors are: 
\begin{enumerate}
{\item $\mathrm{d}^4N / \mathrm{d}z\mathrm{d}M_1\mathrm{d}q\mathrm{d}t_r$} is the comoving occupation function of binaries per redshift, primary mass, and mass-ratio interval, where $t_r$ measures time in the binary's rest frame.  (Uncertainties in this quantity are illustrated as the grey shaded-region in Fig~\ref{fig:gwspectrum}.)
{\item The expression within \{$\cdots$\}} describes the distribution of GW strain over harmonics, $n$, of the binary orbital frequency when the system is eccentric. As previously noted, a circular system will emit at twice the orbital frequency, while eccentric systems emit at the orbital frequency itself as well as higher harmonics. The function $g(n,e)$ is a distribution function whose form is given in \citet{pm63}. Thus, non-zero eccentricity in a binary redistributes the power of the GW spectrum, as shown by green-shaded regions in Fig.~\ref{fig:gwspectrum}.
{\item $\mathrm{d}t_r / \mathrm{d}\ln f_{K,r}$} describes the time each binary spends emitting in a particular logarithmic frequency interval, the ``residence time'', where $f_{K,r}$ is the Keplerian orbital frequency in the binary's rest frame. This is largely controlled by the impact of the direct SMBHB environment, as shown by red- and blue-shaded regions in Fig.~\ref{fig:gwspectrum}. The effects of environment are explored in much greater detail in Section~\ref{sec:binaryinfluence} below.
{\item $h(f_{K,r})$} is the orientation-averaged GW strain amplitude of a single binary as given by Eq.~\ref{eq:hrms}. Note that in Fig.~\ref{fig:gwspectrum}, sharp peaks in the orange and cyan lines indicate contributions from single-sources that may be loud enough to be ``resolved'' from the background.
\end{enumerate}
In the simple case of a population of circular binaries whose orbital evolution is driven entirely by the emission of GWs, the form of $h_{\rm c}(f)$ is easily deduced. In this case, $g(n=2,e=0)=1$ and $g(n \neq 2,e=0)=0$ such that $f=2f_{K,r}/(1+z)$, and the residence time $\mathrm{d}t_r / \mathrm{d}\ln f_{K,r} \propto f^{-8/3}$ is given by the quadrupole radiation formula \citep{peters1964}. Collecting terms in frequency gives $h_{\rm c}(f) \propto f^{-2/3}$, as per Equation~\ref{eq:hc-plaw} (this single power-law spectrum is shown as the black, dashed line in Fig.~\ref{fig:gwspectrum}). 
\footnote{It is important to note that most PTA information on GW backgrounds is derived from the lowest few accessible frequencies in the datasets (the most sensitive point is typically at frequencies $\sim 2/T$). However, published upper limits often quote a standardized constraint on $A_{\alpha}$ within a fiducial single-power-law model, for instance in the SMBH binary case by projecting the limit to a frequency of $f=1$ yr$^{-1}$ using $\alpha=2/3$. If there are spectral turn-overs as described herein, such projections are non-physical, even if practical to use for comparison of two PTAs.}

The $h_{\rm c} \propto f^{-2/3}$ power-law GW background spectrum assumes a continuous distribution of circular SMBHBs evolving purely due to GW emission over an infinite range of frequencies.  As the residence time decreases with frequency, the probability that a binary still exists (i.\,e.\ has not coalesced) also decreases at higher frequencies---that is, there are far fewer binary systems with a high-frequency orbit. At $f \gtrsim 10$ nHz, the Poisson noise in the number of binaries contributing significantly to a given frequency bin becomes important, and realistic GW spectra become `jagged' (e.g.~orange and cyan lines in Fig~\ref{fig:gwspectrum}).  At even higher frequencies, the probability of a given frequency bin containing any binaries approaches zero, and the spectrum steepens sharply relative to the power-law estimate in response \citep[e.g.][]{svc08}.\footnote{The power-law average effectively includes the contribution from \textit{fractional} binary systems.}  At the same time, with fewer sources contributing substantially to the GW spectrum at higher frequencies, the chance of finding a discrete system that outshines the combined background becomes larger; in this case, we say that the binary can be ``resolved'' from the background as a continuous-wave source.

Binaries with non-zero eccentricity emit GW radiation over a spectrum of harmonics of the orbital frequency, rather than just the $2$nd harmonic as in the circular case. For large eccentricities, this can significantly shift energy from lower frequencies to higher ones, and change the numbers of binaries contributing energy in a given \textit{observed} frequency bin.  This can thus substantially change the shape of the spectrum (i.e.~green shaded region in Fig~\ref{fig:gwspectrum}), decreasing $h_{\rm c}$ at low frequencies ($\lesssim 10^{-8}$ Hz) and increasing it at higher frequencies ($\gtrsim 10^{-8}$ Hz) \citep{r+14,enn07,h+15,rm17b,k+17,tss17}.  

Here again it is worth explicitly tying these ideas back to the effect of binary residence times on the GW spectrum; while the strain of an \emph{individual} SMBHB rises with frequency (Eq.\,\ref{eq:hrms}), the number of binary systems contributing to the background falls with frequency, leading to the generally downward-sloped GW spectrum at high frequencies. The turn-over seen at low frequencies for the case of eccentric binaries and strong environmental influence (green, blue, and red curves) can likewise be interpreted in part as due to these effects decreasing the residence time of the binaries at those frequencies: the systems are pushed to evolve much faster through that phase than a circular, purely-GW-driven binary would, therefore fewer systems contribute.

The ``turn-over frequency'' that is seen at low GW frequencies for an eccentric and/or environmentally influenced population, as well as the shape of the spectrum before and after that turn-over in the spectrum, is rich with information about nuclear environments, binary eccentricities, and the influence of gas on binary evolution, as will be explored more fully in Section~\ref{sec:binaryinfluence}.




\subsubsection{Gravitational-Wave Background Anisotropy}
\label{sec:gwanis}~\\\vspace{-3mm}

\noindent
The incoherent superposition of GWs from the cosmic merger history of SMBHBs creates a GW background, as we have discussed. However, some of these SMBHBs may be nearby, but not quite resolvable as continuous waves (\S\ref{sec:gwinspiral}). This can induce departures from isotropy in the GW background \revision{(e.g. \citealt{2017arXiv170803491M,spl+14,rh16})}. Moreover, it may be possible for a galaxy cluster to host more than one inspiralling SMBHB, and thus this sky region may present excess stochastic GW power. 
Indeed, many physical processes can induce GW background anisotropy, which can be characterized and detected using methods developed in numerous works \citep{msmv13, tg13,grt+14,cvh14,ms14, cjm18}.

Importantly, GW background anisotropy will influence the shape of the observed Hellings and Downs curve (\S\ref{sec:ptaterms}; \S\ref{sec:grtests}), leading to different correlation functions between pulsar residuals than would be observed for an isotropic background \revision{\citep{msmv13,grt+14}}. 
The current limit on GW background anisotropy is $\sim 40\%$ of 
the isotropic component \citep{TaylorEtAl:2015}. Indeed, the expected level of anisotropy due to Poisson noise in the GW background is expected to be $\sim 20\%$ of the monopole signal \citep{msmv13,tg13}, in agreement with current astrophysical predictions based on nearby galaxies~\citep{2017arXiv170803491M}. These estimates, however, assume a specific $M_\bullet-M_\mathrm{bulge}$ relation \citep{mm13} for the prediction of anisotropy levels.


The detection of the isotropic GW background may follow after a GW background detection \citep{2016ApJ...819L...6T}. 


\begin{figure*}
\centering
\includegraphics[width=0.48\textwidth]{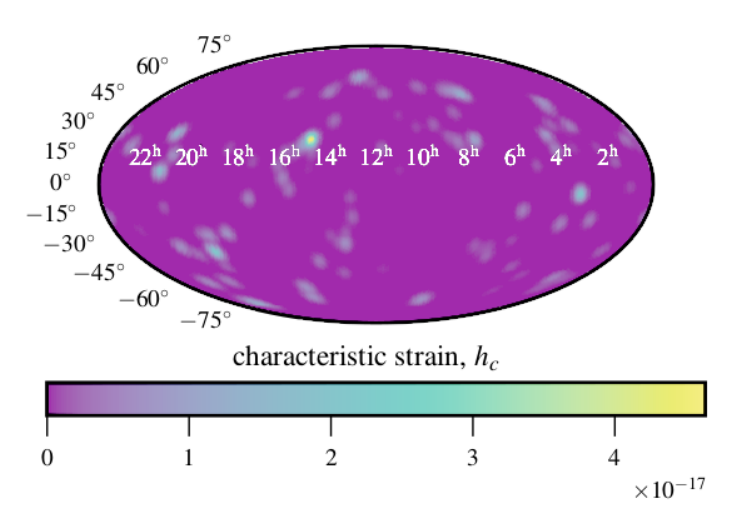}
\includegraphics[width=0.48\textwidth]{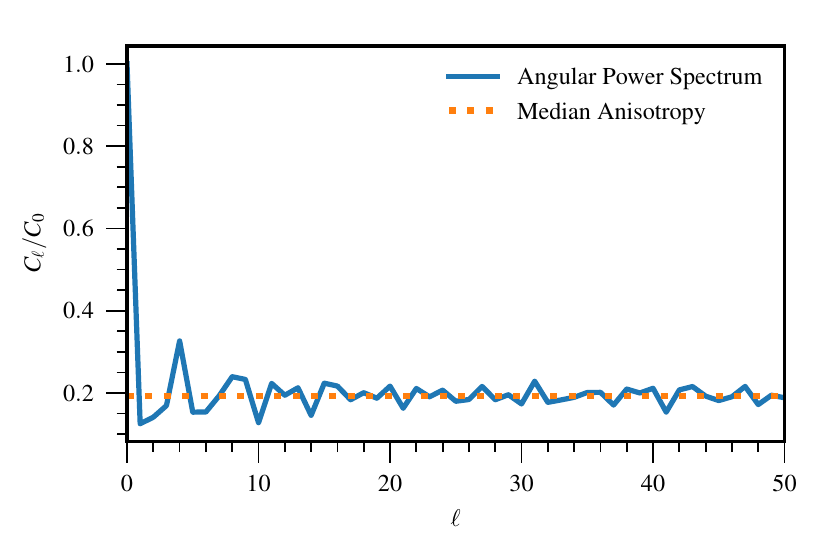}
\caption{A model of anisotropy induced by local nearby SMBHBs. 
Left: A view of the GW \revision{intensity induced by the superposition of many SMBHBs in} a random realization of the local Universe from \cite{2017arXiv170803491M}. Here there are 111 GW sources in the PTA band (at a frequency of 1\,nHz for the sake of this image), with the color scale representing the characteristic strain as a function of sky position.
The level of this anisotropy is determined by taking the angular power spectrum of the background, and normalizing it to the isotropic component, which we have done on the right. Right: angular power spectrum the same GW sky. Though there are fluctuations, the median anisotropy as a fraction of the monopole is $0.19$.}
\label{fig:anisotropy}
\end{figure*}

\subsection{\revision{Supermassive Binaries and Their Environments}}
\label{sec:binaryinfluence}

We now discuss in detail factors which influence both the number and waveforms of continuous-wave sources, and the amplitude and shape of the characteristic strain spectrum for the gravitational wave background from SMBHBs.
PTAs will ultimately measure at least both continuous waves and the amplitude and spectral shape of the GW background at various frequencies. Thus, these measurements have the potential to explore the factors discussed below.

The influence of several of the factors below, such as binary inspiral rate and their average eccentricity in early evolutionary phases, have covariant effects on the expected GW signals (Fig~\ref{fig:gwspectrum}). Thus, the measurements of PTAs for some of the effects discussed below can be enhanced by constraints on these factors from other areas of astrophysics, e.\,g.~through electromagnetic observation of the SMBHB population and through numerical simulations (Section \ref{sec:em}).

This section is laid out as follows. Section \ref{sec:smbh.mdf} discusses how PTAs can directly measure the SMBH mass function via the influence of this parameter on the GW background. 
Except for the SMBH mass, the strain of the GW background at various frequencies depends on the residence time of the binaries, which in turn depends on the physical mechanisms that drive SMBHBs to coalescence.
As illustrated in Figure~\ref{fig:lifecycle}, binaries may be influenced by several external effects, particularly in the phase immediately preceding continuous-wave GW emission in the PTA band. These effects are discussed in Sections \ref{sec:dynamicalfriction}--\ref{sec:CBdisk}. Finally, Sections \ref{sec:eccentricity} and \ref{sec:bhspin} review how PTAs might access information about the eccentricity of binary systems and the spin of individual SMBHs in a binary.


\subsubsection{The Mass Function of Supermassive Black Holes}

\noindent \\

\noindent
The GW amplitude depends strongly on the masses of SMBHB components: $h\propto\mathcal{M}^{5/3}$ (Eq.~\ref{eq:hrms}). Because of that, errors in the assumed SMBH mass distribution may lead to significant errors in GW background level estimates. Unfortunately, there is a tendency for different SMBH mass-estimation techniques (stellar kinematics, gas kinematics, reverberation mapping, AGN emission lines) to systematically disagree with each other, with stellar kinematics usually giving the highest values. For example, the claims of a $\sim2\times10^{10}M_\odot$ SMBH in NGC 1277 \citep{vandenBosch2012} were subsequently found to be too large by factors of 3--5 \citep{Emsellem2013}. Such a bias is unsurprising: beyond the Local Group, only a few galaxies show evidence for a central increase in the RMS stellar velocities (see Figure 2.5 in \citealt{DEGN}) expected in the presence of a SMBH. That implies the SMBH sphere of influence is unresolved and we can only measure an upper limit on its mass. Other methods have their own biases, e.g. different AGN emission lines give different SMBH mass estimates \citep{Shen2008}. These biases were highlighted in the ``bias-corrected'' SMBH mass--host galaxy relations of \citet{shankar}; PTA limits on the SMBH background have also supported the possibility of biased SMBH masses at the upper ($>10^9\msun$) end of the relation, demonstrating that $M_\bullet-M_{\rm bulge}$ relations must be constrained to below certain values, otherwise we should have already detected the GW background \citep{simonBS16}.

Unlike galaxy masses, only a handful of SMBH masses are directly measured. Therefore, when constructing a SMBHB population we have to rely on various SMBH-galaxy scaling relations, such as the relation between SMBH mass and galactic bulge mass: $M_\mathrm{BH} = \beta M_\mathrm{bulge}$. While these types of relations have been thoroughly studied, there may be systematic biases in the SMBH populations they measure \citep[e.g.][]{Shen2008}.  Unsurprisingly, different studies give different values of $\beta$ (the most widely used one is $\beta\approx0.003$); that issue is analyzed in detail in the Section IIC of \citet{rm17b}. Given the observed mass and mass ratio distribution of merging galaxy pairs, $\beta$ is the only free parameter defining the SMBHB mass distribution. \citet{rm17b} came up with the following analytical approximation for the GW background strain spectrum (assuming zero eccentricity and triaxial galaxies):

\begin{eqnarray}
h_{\rm c}(f) &=& A\frac{(f/f_{\rm yr})^{-2/3}}{1+(f_{\rm b}/f)^{53/30}},\\
A &=& 2.77\times10^{-16}\, \left(\frac{\beta}{10^{-3}}\right)^{0.83} 
,\\
f_{\rm b} &=& 1.35\times10^{-9}\,{\rm Hz}\, \left(\frac{\beta}{10^{-3}}\right)^{-0.68}.
\end{eqnarray}
As one can see, lower SMBH masses not only lower the GW background amplitude, but also increase the turnover frequency, since lighter SMBHBs enter the GW-dominated regime at higher orbital frequencies.

\label{sec:smbh.mdf}

\subsubsection{Dynamical friction}~\\

\noindent\label{sec:dynamicalfriction}
When their host galaxies merge together, the resident SMBHs sink to the center of the resulting galactic remnant through \textit{dynamical friction} \citep{antonini2012, merritt2005}. This is the consequence of many weak and long-range gravitational scattering events within the surrounding stellar, gas, and dark matter distributions, creating a drag that causes the SMBHs to decelerate and transfer energy to the ambient media \citep{chandrasekhar1943}. \revision{The most simple treatment of this phase (resulting in a gross over-estimate of the dynamical friction timescale) considers a point mass (the black hole) travelling in a single isothermal sphere (the galaxy). In this case,} the inspiral timescale is on the order of 10 Gyr \citep{BinneyTremaine}:
%

\begin{equation}\label{eq:dynamicalfriction}
T_\mathrm{DF} \approx \frac{19}{\ln\Lambda}\ \left(\frac{R_e} {5\,\mathrm{kpc}}\right)^2\left(\frac{\sigma}{200\,\mathrm{km\ s^{-1}}}\right) \left(\frac{10^8 M_\odot}{M_\bullet}\right)\ \mathrm{Gyr}
\end{equation}
where $R_e$ and $\sigma$ are the galaxy's effective radius and velocity dispersion, $M_\bullet$ is the SMBH's mass and $\ln \Lambda$ is the Coulomb logarithm.\footnote{This is related to the impact factor of galactic material and the binary. For circular binaries to moderate eccentricities, the value should be around $2\lesssim {\rm ln\Lambda}\lesssim 5$, where ${\rm ln}\Lambda \simeq5$ applies to the case of a circular binary \citep{gm08}.} \revision{However, the braking of the individual SMBHs in reality will be much shorter. In a genuine merger system, the $M_\bullet$ in the denominator cannot be modeled with just the SMBHs, as they will initially be surrounded by their constituent galaxies, and later by nuclear stars.} After the galaxies begin to interact, each SMBH is carried by its parent galaxy through the early stages of merger as the galaxies are stripped and mixed into one. \revision{The in-spiral timescale is dominated by the lower-mass black hole (in the case of an unequal mass-ratio merger), which along with a dense core of stars and gas within the SMBH influence radius, will be left to inspiral on its own. 
Extending the above equation to include a} more realistic model of tidal stripping, the resultant timescale can often be shorter than $1\,\mathrm{Gyr}$  \citep{y02,k+17a, Dosopoulou2017}.\footnote{\revision{On the other hand, some simulations have shown that in galaxies that are rapidly stripped or natively lack a dense stellar core, these inspiral timescales may become indefinite \citep{wanderingBH-18}. Note that pessimistic assumptions about these and other evolution uncertainties still give predictions resulting in detectable GW signals \citep[\eg][]{2016ApJ...819L...6T}.}}

By extracting energy from the SMBHs on the kiloparsec separation scale, dynamical friction is a critical initial step towards binary hardening and coalescence.  For systems with extreme mass ratios ($\lesssim 10^{-2}$) or very low total masses, dynamical friction may not be effective at forming a bound binary from the two SMBH within a Hubble time.  In this case the pair might become ``stalled'' at larger separations, with one of the two SMBH left to wander the galaxy at $\sim$kpc separations \citep{y02,mop14,k+15,db17,k+17a}.  It is possible that a non-negligible fraction of galaxies may have such wandering SMBH, some of which may be observable as offset-AGN, discussed in \S\ref{sec:em}.


\subsubsection{Stellar loss-cone scattering}~\\\label{sec:losscone}

\noindent
At parsec separations, dynamical friction becomes an inefficient means of further binary hardening. At this stage, the dominant hardening mechanism results from individual $3$-body scattering events between stars in the galactic core and the SMBH binary \citep{bbr80}. Stars slingshot off the binary which can extract orbital energy from the system \citep{mv92,q96}. The ejection of stars by the binary leads to hardening of the semi-major axis, and eccentricity evolution usually parametrized as \citep{q96}:
\begin{eqnarray} \label{eq:orbit-lc}
\frac{{\rm d}a}{{\rm d}t} &= -\frac{G\rho}{\sigma}H a^2 \ , \label{eq:orbit-lc-sep}\\ 
\frac{{\rm d}e}{{\rm d}t} &= \frac{G\rho}{\sigma} H K a \ , \label{eq:orbit-lc-ecc}
\end{eqnarray}
where $H$ is a dimensionless hardening rate, and $K$ is a dimensionless eccentricity growth rate.  Both of these parameters can be computed from numerical scattering experiments \citep[e.g.][]{shm06}. 

However, only stars in centrophilic orbits with very low angular momentum have trajectories which bring them deep enough into the galactic center to interact with the binary. The region of stellar-orbit phase space that is occupied by these types of stars is known as the ``loss cone" (LC; \citealt{fr76}).  Stars which extract energy from the binary in a scattering event tend to be ejected from the core, depleting the LC.  In general, the steady-state scattering rate of stars is expected to be relatively low as stars are resupplied to the LC at larger radii where relaxation from star--star scatterings is slow.  Like with dynamical friction at larger scales, binaries can also stall here, at parsec scales, due to inefficacy of the LC, which is typically known as the ``final parsec problem" \citep{mm02,mm03}.  Generally, binaries that do not reach sub-parsec separations will be unable to merge via GW emission within a Hubble time \citep{mop14,db17}.  Some simulations suggest, however, that even in the case of a depleted, steady-state LC, the most massive SMBHB, which dominate in GW energy production and also tend to carry the largest stellar masses, may still be able to reach the GW-dominated regime and eventually coalesce \citep{k+17a}.

Various mechanisms have been explored to see whether the LC can be efficiently refilled or populated to ensure continuous hardening of the binary down to milliparsec separations. In general, any form of bulge morphological triaxiality will ensure a continually-refilled LC that can mitigate the final-parsec problem \citep{khb13,vm13,vam14,vam15}. Isolated galaxies often exhibit triaxiality, and given that the SMBH binaries of interest are the result of galactic mergers, triaxiality and general asymmetries can be expected as a natural post-merger by-product. Also, post-merger galaxies often harbor large, dense molecular clouds that can be channeled into the galactic center, acting as a perturber for the stellar distribution that will refill the LC \citep{ys91}, or even directly hardening the binary \citep{gsc17}.   Finally, since binary coalescence times can be of the order of Gyrs, while galaxies can undergo numerous merger events over cosmic time \citep[e.g.][]{r+15}, subsequent mergers can lead to the formation of hierarchical SMBH systems \citep{a+10,Bonetti_2018,Ryu_2018}. In this scenario, a third SMBH can not only stir the stellar distribution for LC refilling, but may also act as a perturber through the Kozai-Lidov mechanism \citep{k62,l62} wherein orbital inclination can be exchanged for eccentricity \citep{bls02,me94,a+10}.  Not only could a third SMBH more effectively refill the LC, but it could also increase the SMBH binary eccentricity which speeds up GW-inspiral (see \S\ref{sec:gwinspiral}).

Even in the absence of a third perturbing SMBH, binary eccentricity can be enhanced through stellar LC scattering. This has been observed in many numerical scattering experiments \citep{q96,shm06}, where the general trend appears to be that equal-mass binaries (most relevant for PTAs) with very low initial eccentricity will maintain this or become slightly more eccentric. For binaries with moderate to large eccentricity (or simply with extreme mass-ratios at any initial eccentricity), the eccentricity can grow significantly such that high values are maintained even into the PTA band \citep{rs12,s10}. 

The rotation of the stellar distribution (when the stars have nonzero total angular momentum) can impact the evolution of the binary's orbital elements. In particular, a stellar distribution that is co-rotating with the binary will tend to circularize its orbit. But if the stellar distribution is counter-rotating with respect to the binary, then interactions with stars in individual scattering events are more efficient at extracting angular momentum from the binary, enhancing the eccentricity to potentially quite high values ($e>0.9$) \citep{sgd11,rm17a,m+17}. However, the binary's orbital inclination (with respect to the stellar rotation axis) also changes: it is always decreasing so that in the end, the initially counter-rotating binaries tend to become co-rotating with the stellar environment \citep{Gualandris2012,rm17a}. In most cases, this joint evolution of eccentricity and inclination leads to the eccentricities at PTA orbital frequencies being much lower than we would expect from a non-rotating stellar environment model \citep{rm17b}. 

Interaction of a binary with its surrounding galactic stellar distribution will lead to attenuation of the characteristic strain spectrum of GWs at low frequencies (i.e.~blue shaded region in Fig.~\ref{fig:gwspectrum}).  This can be separated into two distinct effects: $(1)$ the direct coupling leads to an accelerated binary evolution, such that the amount of time spent by each binary at low frequencies is reduced; $(2)$ extraction of angular momentum by stellar slingshots can excite eccentricity, which leads to faster GW-driven inspiral, and (again) lower residence time at low frequencies. Assuming an isothermal density profile for the stellar population\footnote{stellar density $\rho(r) \propto r^{-1}$, and constant velocity dispersion $\sigma(r) = \sigma_0$.}, we can re-arrange \ref{eq:orbit-lc-sep} to deduce the orbital frequency evolution, and hence the evolution of the emitted GW frequency, such that ${\rm d}f/{\rm d}t \propto f^{1/3}$. Inserting this into \ref{eq:hc-spec} and collecting terms in frequency gives $h_c(f)\propto f$, which is markedly different from the fiducial $\propto f^{-2/3}$ circular GW-driven behavior. The excitation of binary eccentricity by interaction with stars will further attenuate the strain spectrum at low frequencies, leading to an even sharper turnover \citep[e.g.][]{tss17}.

\subsubsection{Viscous circumbinary disk interaction}~\\

\noindent\label{sec:CBdisk}
At centiparsec to milliparsec separations, viscous dissipation of angular momentum to a gaseous circumbinary disk may play an important role in hardening the binary \citep{bbr80, ka11}. This influence will depend on the details of the dissipative physics of the disk, however the simple case of a binary exerting torques on a coplanar prograde disk has a self-consistent non-stationary analytic solution \citep{ipp99}. These studies have assumed a geometrically-thin optically-thick disk whose viscosity is proportional to the sum of thermal and radiation pressure \citep[the so-called $\alpha$-disk,][]{ss73}.
The binary torque will dominate over the viscous torque in the disk, leading to the formation of a cavity in the gas distribution and the accumulation of material at the outer edge of this cavity (i.e.~Type II migration). The excitation of a spiral density wave in the disk torques the binary, and leads to hardening through the following semi-major axis evolution \citep{ipp99,hbm09}:
\begin{equation} \label{eq:orbit-disk}
\frac{{\rm d}a}{{\rm d}t} = -\frac{2\dot{m}_1}{\mu}(aa_0)^{1/2},
\end{equation}
where $\dot{m}_1$ is the mass accretion rate onto the primary BH, and $a_0$ is the semi-major axis, at which the disk mass enclosed is equal to the mass of the secondary BH, given by \citet{ipp99} as
\begin{eqnarray}
a_0 =& 3\times 10^3 \left(\frac{\alpha}{10^{-2}}\right)^{4/7} \left(\frac{m_2}{10^6 M_\odot}\right)^{5/7} \nonumber\\
&\times\left(\frac{m_1}{10^8 M_\odot}\right)^{-11/7} \left(100\frac{\dot{m}_1}{\dot{M}_E}\right)^{-3/7} r_g,
\end{eqnarray}
where $\alpha$ is a disk viscosity parameter, $\dot{M}_E = 4\pi Gm_1 / c\kappa_\mathrm{T}$ is the Eddington accretion rate of the primary BH ($\kappa_\mathrm{T}$ is the Thompson opacity coefficient), and $r_g = 2Gm_1 / c^2$ is the Schwarzchild radius of the primary BH.

The disk--binary dynamics may be much more complicated, for example, the disk may be composed of several physically-distinct regions \citep{st86}, discriminated by the dominant pressure (thermal or radiation) and opacity contributions (Thompson or free-free). Additionally, high-density disks (equivalently: high-accretion rates) may provide rapid hardening, but may also be unstable due to self-gravity.  Furthermore, while the system will initially pass through ``disk-dominated'' Type II migration (where the secondary BH is carried by the disk like a cork floating in a water drain), it will generally transition to ``planet-dominated'' migration (where the secondary BH is dynamically dominant over the disk) which can be significantly slower.

Eccentricity growth may be significant during this disk-coupled phase \citep{an05,c+09}, although there are no generalized prescriptions of the form of \ref{eq:orbit-lc-ecc}. The growth of eccentricity is driven by outer Lindblad resonant interaction of the binary with gas in the disk at large distances \citep{gs03}. However, \citet{r+11} found that binaries with \textit{high} initial eccentricity will experience a reduction in eccentricity, leading to the discovery of a limiting eccentricity for disk--binary interactions that falls in the interval $e_\mathrm{crit} \in [0.6,0.8]$. The emerging picture is that in a low-eccentricity orbit, the density wake excited by the secondary BH will lag behind it at apoapsis, causing deceleration and increasing eccentricity. Whereas in a high-eccentricity orbit, the density wake instead advances ahead of the secondary BH, causing a net acceleration and reduction in eccentricity. All previously mentioned studies considered prograde disks, but if a retrograde disk forms around the binary then the eccentricity can grow rapidly, leading to significantly diminished GW-inspiral time \citep{sk15}.

Coupling of a viscous circumbinary disk with a SMBH binary, like in the stellar LC scattering scenario, will lead to attenuation of the characteristic strain spectrum of GWs through both direct coupling, and excitation of eccentricity (e.g.~red shaded region in Fig.~\ref{fig:gwspectrum}). 
Focusing on direct coupling of a circular binary to its surrounding disks, \citet{ka11} studied scaling relations for the strain spectrum from different disk--binary scenarios, varying from $h_c(f)\propto f^{-1/6}$ for secondary-dominated type-II migration in a radiation-dominated $\alpha$-disk, to $h_c(f)\propto f^{1/2}$ for the \citet{ipp99}-model in \revision{Equation~\ref{eq:orbit-disk}}. Across all models, the characteristic strain spectrum can be flattened or even increasing due to disk coupling (very different from the fiducial $\propto f^{-2/3}$ circular GW-driven behavior), and spectral attenuation is further enhanced through disk excitation of binary eccentricity.  When the disk--binary models are applied to \textit{populations} of SMBHB, the overall GW background spectra tend to more closely resemble the canonical $-2/3$ power-law, because each disk-regime only applies to a fairly narrow frequency range for a given binary mass \citep{ka11, k+17a}.


\subsubsection{Eccentricity}~\\\vspace{-3mm}\label{sec:eccentricity}

\noindent
The influence of an initial eccentricity on SMBH evolution, without the influence of external driving factors as in the previous subsection, was shown in Equations~\ref{eq:ecc1} and \ref{eq:ecc2}
There have been several studies of the influence of non-zero binary eccentricity on the characteristic strain spectrum of nanohertz GWs \citep{h+15,enn07,r+14,k+17,rm17b,tss17}. The exact shape and amplitude of the spectrum will depend on the detailed interplay of direct environmental couplings with eccentricity, and, in the case of the latter, the distribution of eccentricities at binary formation \citep{r+14,Ryu_2018}. Broadly speaking, $(1)$ eccentricity increases the GW luminosity of the binary, meaning it evolves faster and thus spends less time emitting in each frequency resolution bin;
$(2)$ eccentricity distributes the strain preferentially toward higher harmonics of the orbital frequency. These effects lead to the strain being diminished at lower observed GW frequencies, but also somewhat enhanced at higher frequencies---i.e.~the spectrum can exhibit a turnover to a positive slope at low frequencies, but then a small ``bump'' enhancement at the turnover transition.

In simulated SMBHB populations that include eccentricity, non-zero eccentricities tend to reduce the mean occupation number at lower frequencies, thus making the stochastic background appear to have a flatter (or inverted) spectrum than the standard $\alpha=2/3$. However, these eccentricities also work to redistribute the power to higher frequencies, where in a circular population the background would be otherwise dominated by small numbers of binaries
\citep{k+17}.

\begin{figure*}
\begin{center}
\includegraphics[width=0.75\textwidth]{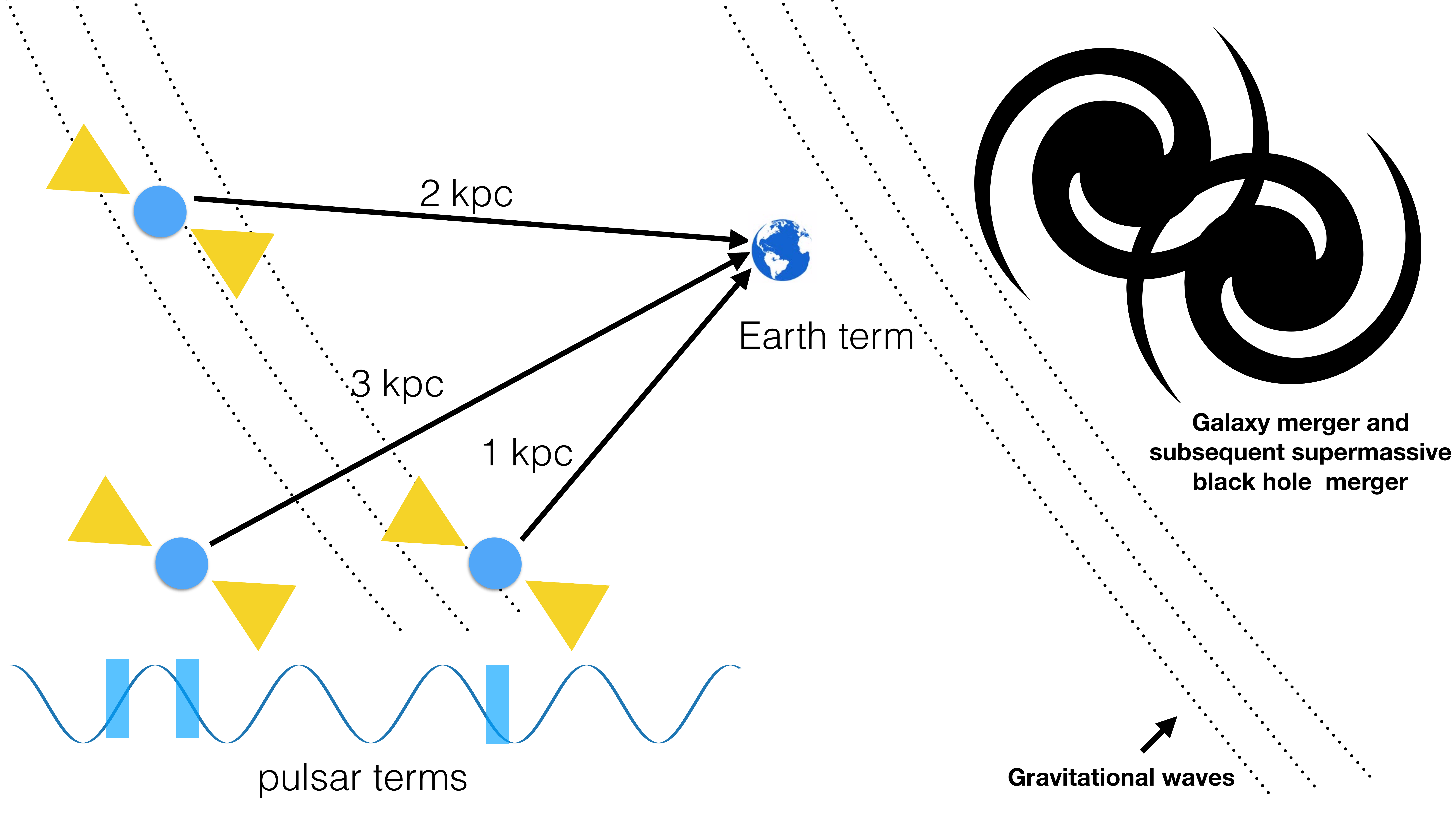}
\caption{Gravitational waves spanning thousands of years in a binary's evolutionary cycle can be detected from a continuous GW source by using the pulsar term.  As an example, we have drawn a few pulsars with line-of-sight path length differences to the Earth. These relative time delays between the pulsar terms can be used to probe the evolution and the dynamics of a SMBHB systems over these many thousands of years. Right: a major galaxy merger leads to the creation of a SMBHB, emitting nanohertz GWs. Left: the pulsar term from each pulsar probes a different part of the SMBHB evolution, since they are all at different distances from the Earth. The blue sinusoid is a cartoon of the GW waveform, and shows that the pulsar terms can be coherently concatenated to probe the binary's evolution, allowing one to measure e.g. the spin of the SMBHB~\citep{m+12}. 
}\label{fig:pulsarterm}
\end{center}
\end{figure*}

\subsubsection{Measuring the Spin of Supermassive Black Hole Binaries}~\\\vspace{-3mm}\label{sec:bhspin}

\noindent When GWs transit our Galaxy, they perturb pulsar signals both at the Earth and at the pulsar, i.e. they affect both the ``Earth term'' and the ``pulsar term''; as a reminder, the pulsar term shows a GW signal that is delayed in time by an amount proportional to the light travel time between the Earth and the pulsar (\S\ref{sec:ptaterms}). That is, 
if a source (such as a SMBHB) is evolving, the pulsar term encodes \emph{information about earlier phases in the SMBH evolution.}
We can use this information to our advantage: when a continuous GW signal is detected, one can look for the perturbation caused by the same source in the pulsars, but thousands of years ago. 
\revision{These pulsar terms can be used to map the evolution of a SMBHB system over many thousands of years: each pulsar term is a snapshot of the binary during a different point in the history of its evolution (Figure \ref{fig:pulsarterm}; \citealt{m+12}), and
the phase evolution of the SMBHB can thus be measured. This is important, since SMBH spins affect the phase evolution of the binary, thus, constraining the phase evolution allows one to constrain the SMBH spins \citep{m+12}.

One estimates the number of expected gravitational wavecycles observed at the Earth via the post-Newtonian expansion \citep{b06}, which is a function of the SMBH mass and spin. For example, over a 10 year observation, an equal-mass $10^9~M_\odot$ SMBHB system with an Earth-term frequency of 100~nHz should produce 32.1 gravitational wavecycles, of which 31.7 are from the leading Newtonian order (or p$^{0}$N), 0.9 wavecycles are from p$^1$N order, and -0.7 are from p$^{1.5}$N order.
This last term is from spin-orbit coupling, and depends on the SMBH spins. Accessing the pulsar term when it arrives at the Earth gives information about the SMBHB system over $\sim 3000$ yrs ago (roughly equivalent to the typical light-travel time between the Earth and the pulsar). Over this time, one expects 4305.1 wavecyles, of which 4267.8 are Newtonian, 77.3 come from p$^1$N order,  $-45.8$ are from p$^{1.5}$N order, etc... One can therefore see at a glance that spinning binaries evolve more quickly, which in turn affects the phase evolution of the waveform. This signal is imprinted in the pulsar terms of the pulsars in the array, and is therefore only accessible via PTA observations of the pulsar terms. }

However, in order to do pulsar-term phase matching, we require that $2\pi f L <1$ to not lose a single wavecycle, where $f$ is the GW frequency and $L$ is the distance to the pulsar. Therefore, it is in principal necessary to measure the pulsar distances to e.g. $\sim 1.5$~pc for a GW signal at 1~nHz.

Many pulsar distances are poorly constrained, since most are estimated via the dispersion measure of the pulse \citep[\eg][]{cl02, ymw17}. However, a number of nearby, well-timed pulsars have accurate position measurements based on parallax measured by very long baseline interferometry \citep[\eg][]{psrpi}. For instance, the well-timed millisecond pular PSR~J0437--4715 has a distance measurement of $156.3 \pm 1.3$~pc~\citep{dvtb08}, and is thus suitable for this measurement. Pulsar timing can also be used to obtain a parallax measurement to the pulsar, as in \cite{lgs12}, but with relatively large errors. Optical surveys such as Gaia \citep{gaia} can be used to measure parallaxes to some pulsars' white dwarf companions (\citealt{jkc+18})
though again with limited accuracy due to the low brightness of the white dwarfs. The independent distance measurements to the pulsar's binary companion can also be used in combination with the pulsar distance measurements to improve this estimate \citep{mab+19}. The most precise way to constrain pulasr distances is through measuring the binary's orbital period derivative -- a {\it dynamical} distance measurement \citep{s70} -- which in the case of PSR~J0437--4715 furthers its distance constraints to $156.79\pm 0.25$ pc \citep{rhc+16}.

Thus, while current pulsar distances are generally not suitable for an extensive study of this effect, future instruments such as the Square Kilometre Array \citep[\eg][]{smits+11} or Next-generation Very Large Array\footnote{http://ngvla.nrao.edu/page/scibook} and NASA's WFIRST telescope \citep{wfirst} hold tremendous promise for enabling precise pulsar (and binary companion) distance measurements, which will also enable more tests of fundamental physics.

\section{Cosmic Strings and Cosmic Superstrings}
\label{sec:strings}
\vspace{-3mm}

\begin{tcolorbox}[enhanced,colback=gray!10!white,colframe=black,drop shadow]
Cosmic strings and superstrings produce GWs in the nanohertz band.
Cosmic strings are topological defects that can form during phase transitions in the early Universe, while cosmic superstrings are fundamental strings stretched to cosmological scales due to the expansion of the Universe. In a cosmological setting, and for the most simple superstring models, both cosmic string and superstring networks evolve in the same way. Cosmic (super)strings can exchange partners when they meet and produce loops when they self-intersect. These loops then oscillate and lose energy to GWs generating a stochastic background, with a power-law spectrum having a spectral index close to that of SMBHBs. Detection of such a background, or GWs from an individual cosmic (super)string loop, would therefore provide a unique window into high-energy physics. Currently, pulsar-timing experiments are producing the most constraining bounds on the energy scale and other model parameters of cosmic strings and superstrings. 
\end{tcolorbox}

\begin{figure*}
\centering
\includegraphics[width=0.99\textwidth]{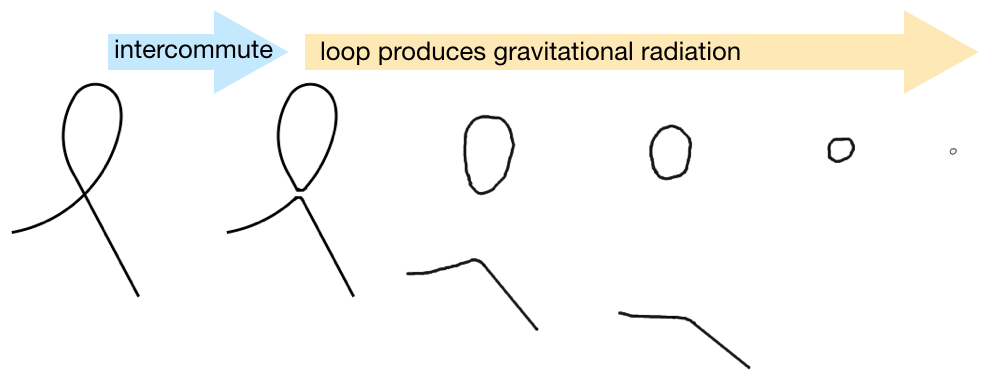}
\caption{A depiction of the production of a cosmic string loop from a self-intersecting string. The loop oscillates and produces gravitational waves slowly decaying away. This process allows the string network to reach the {\it scaling regime}.}\label{fig:sillystring}
\end{figure*}

Topological defects are a generic prediction of grand unified theories. As the Universe expands and cools, the symmetries of the grand unified theory are broken down into the Standard Model in one or more stages called phase transitions. At each of these phase transitions, topological defects generically form, with the type of defect depending on what symmetry is being broken. Cosmic strings are a one-dimensional (or line-like) type of topological defect that can form in the early Universe during one (or more) of these phase transitions. The other common types of topological defects are monopoles and domain walls. Both of these are ruled out, however, because they lead to cosmological disasters (e.g. the monopole problem), and much of the attention of the theoretical cosmology community has focused on strings as the only viable candidate that could lead to potential observational signatures.  The most simple symmetry breaking that leads to the formation of cosmic strings occurs in the Abelian Higgs model, where the symmetry group $U(1)$ breaks  $$U(1) \rightarrow 1$$ at some temperature, or energy scale, $T_s$. They are characterized by their mass per unit length $\mu$, which in natural units is determined by the temperature at which the phase transition takes place, $\mu \sim T_s^2$. The tension of a string is normally given in terms of the dimensionless parameter $G\mu/c^2$ which is just the ratio of the string energy scale to the Planck scale squared. It is worth pointing out that analogous processes abound in condensed matter systems such as superfluid helium, Bose-Einstein condensates, superconductors, and liquid crystals, which can also lead to the production of topological defects.  

\begin{figure*}
\centering
\includegraphics[width=0.6\textwidth]{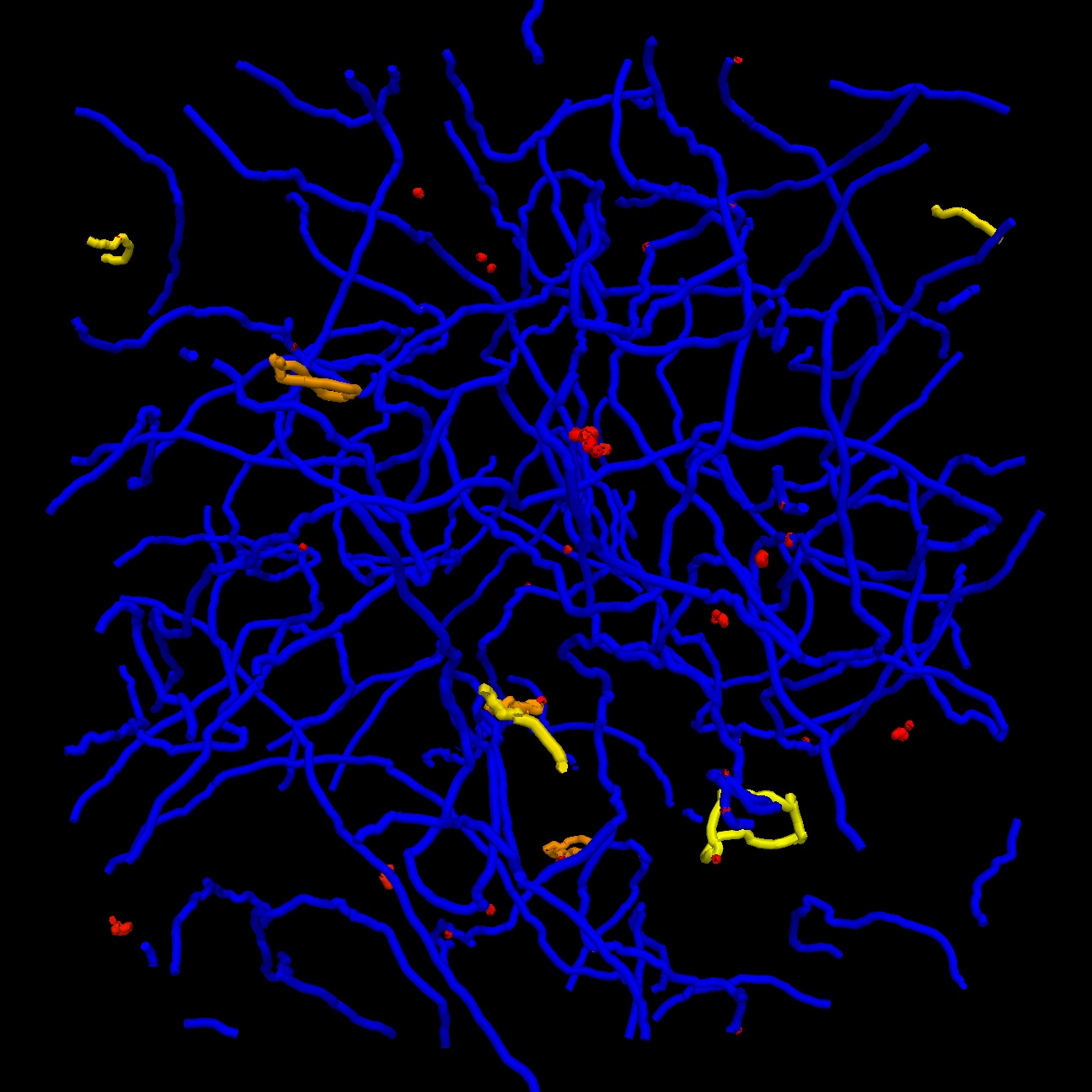}
\caption{Simulation of a cosmic string network in the matter era. Long strings are shown in blue and loops are color coded to show their size with red being the shortest to yellow being the largest. Image credit K.D. Olum.}\label{fig:stringnetwork}
\end{figure*}

Phase transitions in the early Universe can therefore lead to the formation of a network of cosmic strings. Analytic work and numerical simulations show that the network quickly evolves toward an attractor called the {\it scaling solution}. In this regime, the statistical properties of the system---such as the correlation lengths of long strings and the size of loops---scale with the cosmic time, and the energy density of the string network becomes a small constant fraction of the radiation or matter density. This attractor solution is possible due to reconnections: when two strings meet they exchange partners (``intercommute''), and when a string self-intersects it chops off a loop (see Fig.\,\ref{fig:sillystring}). The loops produced by the network oscillate, generate gravitational waves, and shrink, gradually decaying away. This process removes energy from the string network, converting it to gravitational waves, and providing the very signal we seek to detect. The way the {\it scaling solution} works is as follows: if the density of strings in the network becomes large, then strings will meet more often and reconnect, producing extra loops which then decay gravitationally, removing the surplus string from the network. If, on the other hand, the density of strings becomes too low, strings will not meet often enough to produce loops, and their density will start to grow. In this way the cosmic string network finds a stable equilibrium density and a Hubble volume of the Universe with a string network statistically always looks like that shown in Fig.~\ref{fig:stringnetwork}, stretched by the cosmic time.

\begin{figure*}[t]
\centering
\includegraphics[width=0.99\textwidth]{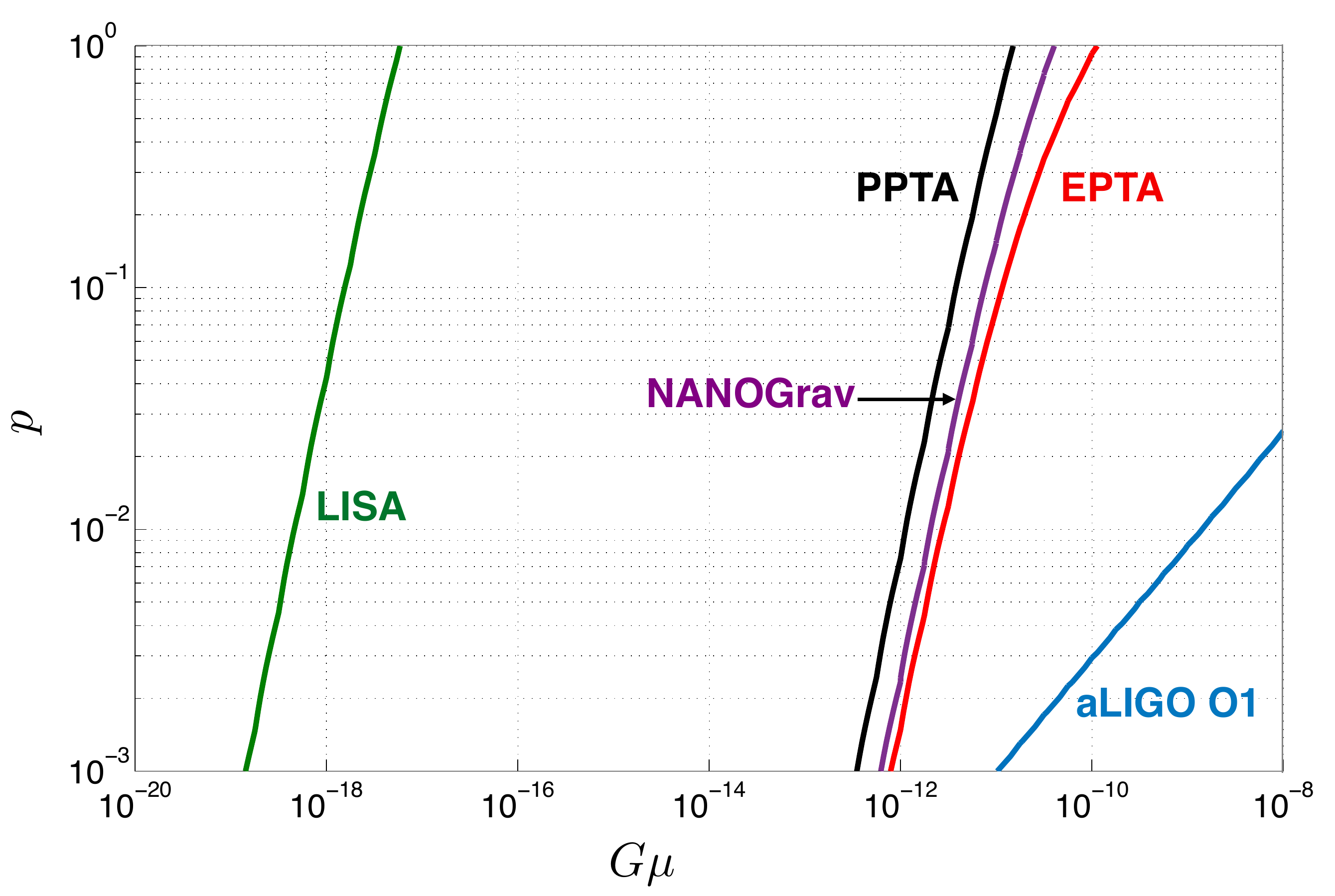}
\caption{Plot of regions of the cosmic (super)string $G\mu/c^2$-$p$ plane excluded by present experiments LIGO and PTAs. PTAs currently place the strongest constraints on cosmic strings. LISA has the potential to improve these limits (or provide a detection). The excluded areas are to the right of each curve. Figure is from \citet{Blanco-Pillado:2017rnf}.}\label{fig:pvsGmu}
\end{figure*}

Superstrings refer to the fundamental strings of string theory that in some models can be stretched to cosmological scales by the expansion of the Universe. Fortunately, much of what we have learned about the evolution of cosmic string networks can be applied to the evolution of cosmic superstrings. Aside from the possibility of forming more than one type of string, the most significant difference is that cosmic superstrings don't always reconnect when they meet. This occurs for two reasons: (i) because these string theory models require more than 3 dimensions, and though strings may appear to meet in 3 dimensions they can miss each other in the extra dimensions, and (ii) because superstrings, being the fundamental strings of string theory, interact with a probability proportional to the string coupling squared.
The net effect is to lower the reconnection probability from $p=1$ for cosmic strings to $p \le 1$ for cosmic superstrings. A network of cosmic superstrings still enters the {\it scaling regime}, albeit at a density higher by a factor inversely proportional to the reconnection probability (because strings need to interact all that more often to produce loops that gravitationally radiate away). The smaller reconnection probability of superstrings therefore actually increases the chances of finding observational signatures because the equilibrium string density of the {\it scaling solution} is higher. Figure \ref{fig:pvsGmu} shows the areas of the cosmic (super)string $G\mu/c^2$-$p$ parameter space excluded by present and potential future experiments, including LIGO and the three leading PTA experiments (PPTA, NANOGrav, and EPTA) as of 2017, as well as the planned LISA space mission. As the reconnection probability decreases, the density of strings in the scaling regime increases, and thus experiments are sensitive to lower and lower string tensions. Following our previous discussion, the limits at $p=1$ represent the limits specifically for cosmic strings. 

In the 1990s, cosmic microwave background data ruled out cosmic (super)strings as the primary source of density perturbations, placing constraints on the string tension at the level of $G\mu/c^2 \lesssim 10^{-7}$, relegating them to at most a secondary role in structure formation. However, cosmic (super)strings are still potential candidates for the generation of a host of interesting astrophysical phenomena: including ultra high energy cosmic rays, fast radio bursts, gamma ray bursts, and of course gravitational radiation. Clearly, any positive detection of an observational signature of cosmic (super)strings would have profound consequences for theoretical physics.

\begin{figure*}[t]
\includegraphics[width=0.99\textwidth]{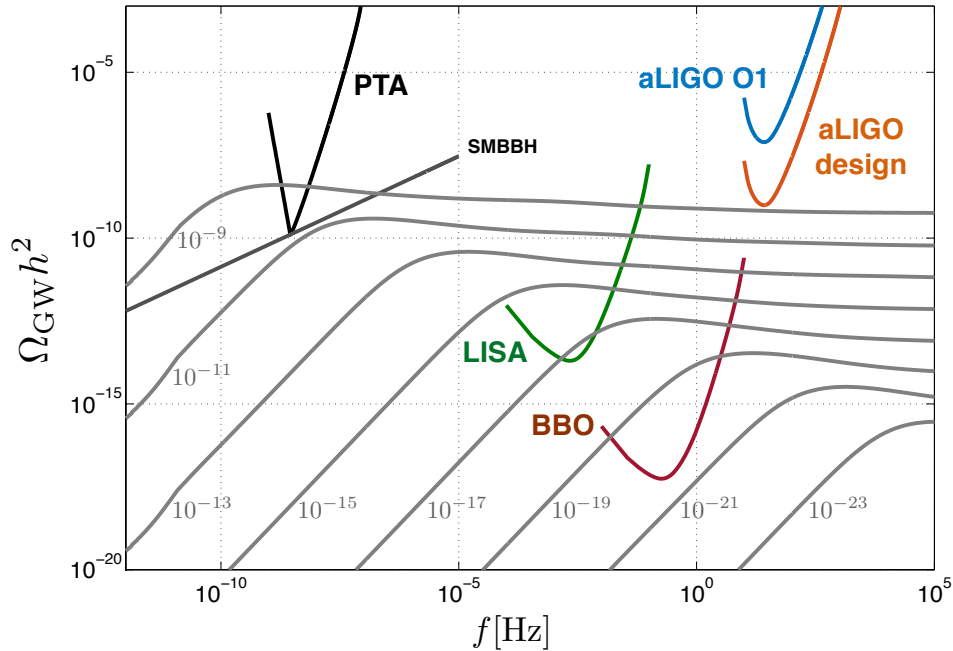}
\caption{Plot of the gravitational wave spectrum in terms of the dimensionless parameter $\Omega$, as a function of frequency in Hz. The figure shows cosmic (super)string spectra for $p=1$ for values of the (dimensionless) string tension  $G\mu/c^2$ in the range of $10^{-23}$-$10^{-9}$, as well as the spectrum produced by supermassive binary black holes (SMBBH), along with current and future experimental constraints. Figure is from \citet{Blanco-Pillado:2017rnf}.}\label{fig:omega}
\end{figure*}

\setcounter{footnote}{0}
PTAs are currently the most sensitive experiment for the detection of cosmic (super)strings and will remain so for more than a decade and a half. Correspondingly, the most constraining upper limits on the energy scale of cosmic (super)strings come from PTA analyses. As of the writing of this paper, \revision{the most constraining upper limit  published by a PTA collaboration (for $p = 1$) is $G\mu/c^2 < 5.3(2) \times 10^{-11}$ from the the NANOGrav collaboration \citep{2018ApJ...859...47A}. Later, Blanco-Pillado et al used results from all PTAs and recalculated upper limits on the string tension; Figure \ref{fig:omega} shows the stochastic background
spectrum produced by cosmic strings in terms of the dimensionless density parameter $\Omega$ versus frequency for dimensionless string tensions $G\mu/c^2$ in the range $10^{-23}$-$10^{-9}$ for $p=1$. Overlaid are current experimental constraints from 
PTAs\footnote{While in \citet{Blanco-Pillado:2017rnf} the PPTA data give the most constraining upper limit, the PPTA result used by Blanco-Pillado et al.\ did not account for uncertainties in the ephemerides which would likely weaken the given constraint \citep{2018ApJ...859...47A}.} and ground-based GW detectors, and future constraints from spaced-based detectors.} PTA sensitivity will not be superseded until the LISA mission which is scheduled for launch in 2034.

\section{The Nature of Gravity}
\label{sec:grtests}
\vspace{-3mm}
\begin{tcolorbox}[enhanced,colback=gray!10!white,colframe=black,drop shadow]
Understanding gravity---and testing general relativity as a theory of gravity---are leading pursuits of most GW detectors, and PTAs are no exception to this. 

\vspace{-10pt}\paragraph{\textbf{Polarization modes:}}
Upon a high-significance detection of gravitational radiation, PTAs will be able to assess its polarization; this is done by correlating the residuals of pairs of pulsars, and constructing an angular correlation function based on the angle between the pair. This can be measured regardless of the type of radiation (continuous, memory, background, etc.). By characterizing the shape of the correlation function, the polarization mode of the signal can be measured. Various classes of gravity models, including General Relativity, will be supported or ruled out. PTAs will explore different targets than ground- and space-based interferometry, permitting competitive and independent constraints on theories of gravity.

\vspace{-10pt}\paragraph{\textbf{Graviton mass and gravity group velocity:}}
Variations in the graviton mass will cause slight variations to the correlation function which may be detected if PTAs acquire sufficient measurement precision. Graviton mass can also be constrained or measured by the temporal offset of any co-detected electromagnetic counterpart to a single GW source. If the graviton mass is minimal ($m_g\ll10^{-22}\,$eV), which it appears to be based on recent LIGO measurements, PTAs will instead produce precise limits on the group velocity of GWs.\\

\vspace{-7pt}For context, we finish this section by summarizing three examples of theories that PTAs can test, which make several predictions that differ from general relativity.
\end{tcolorbox}

General relativity has been an exceptionally successful physical theory, and continues to stand up to over 100 years of tests of its accuracy \citep{Will:2014kxa}. \revision{Pulsar timing of a pulsar-neutron star binary provided the first indirect proof of gravitational radiation \citep{1982ApJ...253..908T}, while subsequent studies of the pulsar-pulsar binary have led to tests of GR to exquisite precision \citep[\eg][]{kramer-review}.}
This includes the discovery of what had been the remaining unobserved prediction of Einstein for general relativity: the detection of GWs from two black holes by LIGO \citep{2016PhRvL.116f1102A}. Over the years there have been many other theories of gravity put forward. Some of these are based on physically aesthetic motivations, for instance a change in symmetry or adding a generalization, such as the scalar field in Brans-Dicke theory \citep{PhysRev.124.925}. Others, especially in recent years, strive to explain an observed phenomenon that is unexplained by current physical theories, such as dark matter or dark energy \citep{Akrami:2012vf}. Whatever the nature of the theory, it must pass all of the observational tests that have made general relativity such a successful theory. Myriad new tests of gravity have become possible with GW observations \citep[\eg][]{Eardley:1973br,Yunes:2009ke,GBM:2017lvd,Chatziioannou:2012rf,Yunes:2013dva} 

\subsection{How Many Gravitational Wave Polarization States Exist?}
\label{sec:grtests.polarization}
General relativity predicts the existence of GWs which travel at the speed of light, are transverse, and have two polarizations, the standard plus and cross.  Other theories of gravity may predict the existence of GWs with different properties. 

For any theory in which a metric encodes the dynamics of spacetime, there are six distinct GW polarizations possible \citep{Eardley:1973br}.  A linear GW is a small deviation from flat spacetime with a metric given by $g_{\mu \nu} = \eta_{\mu \nu} + h_{\mu \nu}$, where $\eta_{\mu \nu}$ is the flat-space Minkowski metric and $|h_{\mu \nu}| \ll 1$.  Since the metric is a symmetric 4$\times$4 matrix, $h_{\mu \nu}$ has ten independent components: using our freedom to choose the coordinate system (i.e., gauge freedom), we can remove four of these components, leaving only six.  These components can be grouped into the way each transforms under rotations, giving us two scalar components, two vector components, and two tensor components \citep{Eardley:1973br,Eardley:1974nw}. The tensor components are those most commonly pursued, and are commonly referred to as the plus and cross polarizations.

There has recently been considerable interest in using observations of gravitational waves to look for evidence of these non-Einsteinian polarizations. For example, \citet{Isi:2015cva} determined that the signal-to-noise of match-filtered searches for gravitational waves with aLIGO can be greatly impacted if they consist of non-Einsteinian polarization modes. Analysis of the detection of GW170814 (the first GW event detected by the two LIGO observatories and Virgo) favored tensor polarization modes over a pure vector or scalar modes by a Bayes-factor of more than 200 and 1000, respectively \citep{Abbott:2017oio,Isi:2017fbj}.  

\begin{figure*}
\begin{center}
\resizebox{!}{7cm}{\includegraphics{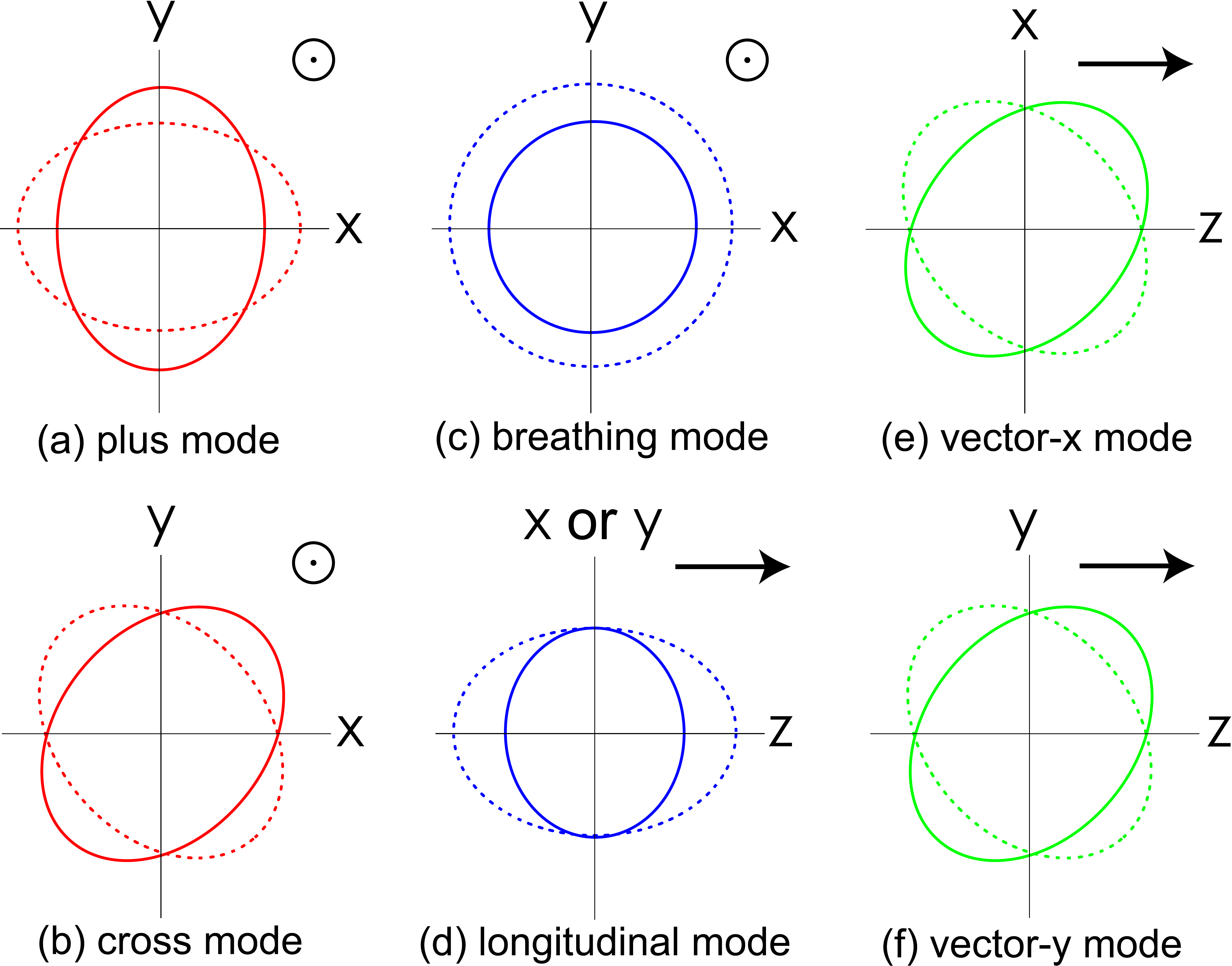}}
\caption{The six possible GW polarizations in synchronous gauge where $h_{\mu 0} = 0$. The solid and dotted line in each case represent maxima and minima in the strain variations induced by an oscillatory GW. General relativity predicts only plus and cross modes, however some theories outlined in Section~\ref{sec:examples.polarizations} give rise to the other modes.
Reproduced from \citet{2009PhRvD..79h2002N}.}
\label{fig:gw-pol-all}
\end{center}
\end{figure*}

We will work in a coordinate system in which $h_{\mu 0}=0$, which is called the synchronous gauge. If we consider a plane wave traveling in the $z$-direction then a generic GW is described by six polarization tensors \citep{Eardley:1973br,Eardley:1974nw}:
\begin{eqnarray}
\label{pluscrosspol}
e_{ij}^+ &=& \left(\begin{array}{ccc}1 & 0 & 0 \\0 & -1 & 0 \\0 & 0 & 0\end{array}\right), \ e_{ij}^{\times} = \left(\begin{array}{ccc}0 & 1 & 0 \\1 & 0 & 0 \\0 & 0 & 0\end{array}\right),\label{eq:pol1}\\
e_{ij}^b &=&  \left(\begin{array}{ccc}1 & 0 & 0 \\0 & 1 & 0 \\0 & 0 & 0\end{array}\right), \ e_{ij}^\ell =  \left(\begin{array}{ccc}0 & 0 & 0 \\0 & 0 & 0 \\0 & 0 & 1\end{array}\right), \label{breathinglongpol} \\
e^x_{ij} &=& \left(\begin{array}{ccc}0 & 0 & 1 \\0 & 0 & 0 \\1 & 0 & 0\end{array}\right), \ e^y_{ij} = \left(\begin{array}{ccc}0 & 0 & 0 \\0 & 0 & 1 \\0 & 1 & 0\end{array}\right), \label{eq:pol3}
\end{eqnarray}
where we have normalized these polarization tensors to satisfy $e^P_{ij} e_{P'}^{ij} = \delta_{P,P'}$.  Three of these polarization tensors ($+$/$\times$, and $b$) are transverse; the other scalar mode and the two vector components are longitudinal, as shown in Fig.~\ref{fig:gw-pol-all}.

\begin{figure*}
\begin{center}
\resizebox{!}{5cm}{\includegraphics{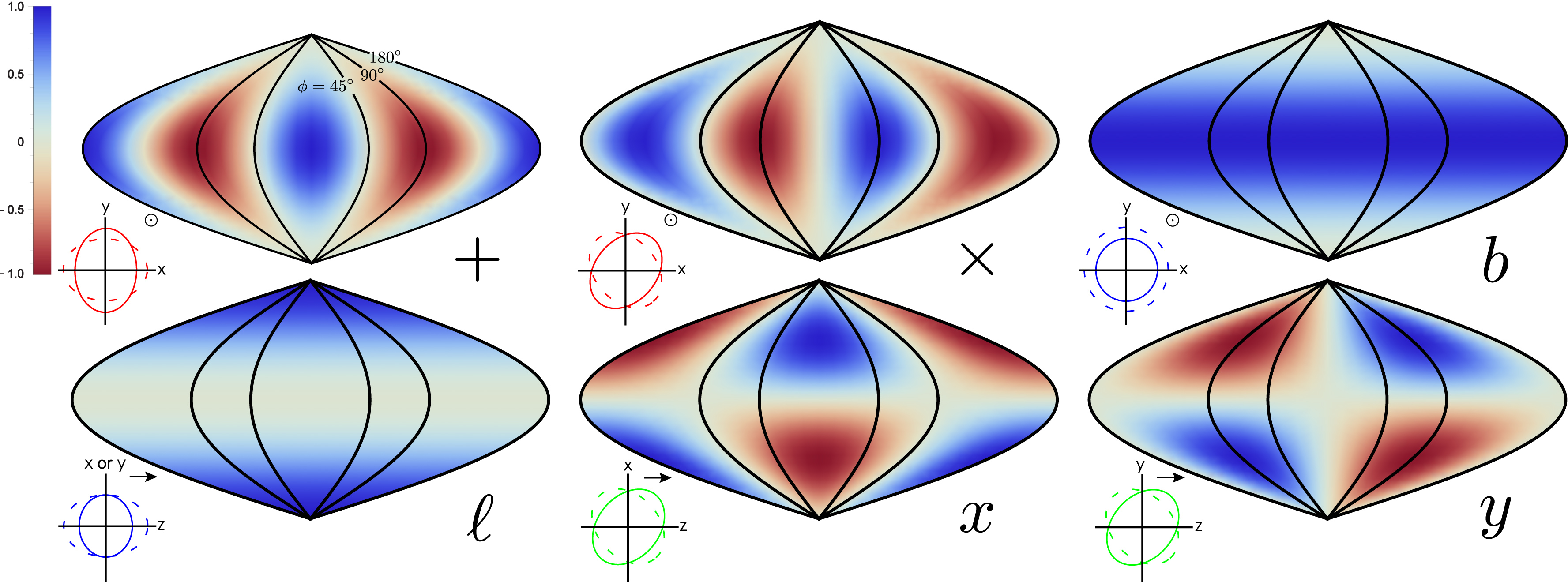}}
\caption{The induced variations in observed pulse phase as a function of pulsar sky-location for a long-wavelength GW traveling along the positive $z$-axis. The scalar/vector/tensor and longitudinal/transverse nature of each polarization mode is apparent.}
\label{fig:v_response}
\end{center}
\end{figure*}

PTAs are sensitive to the polarization of GWs of any sufficiently bright source, including single sources, the stochastic background, etc.~\citep{Chatziioannou:2012rf}. For illustration, let us imagine a single, high-intensity plane GW traveling along the positive $z$-axis as we observe a PTA with pulsar pairs scattered across the sky, with any two pulsars separated by some arbitrary sky angle $\theta$. 
For our single wave with various polarization modes induced, we show the predicted modulation of the observed pulse phase at various sky positions in Fig.~\ref{fig:v_response}. Here we can see the quadrupolar response to the usual plus and cross polarization, the dipolar response to the vector polarizations, and the monopolar response to the scalar polarizations. 

\begin{figure*}
\begin{center}
\resizebox{!}{7cm}{\includegraphics{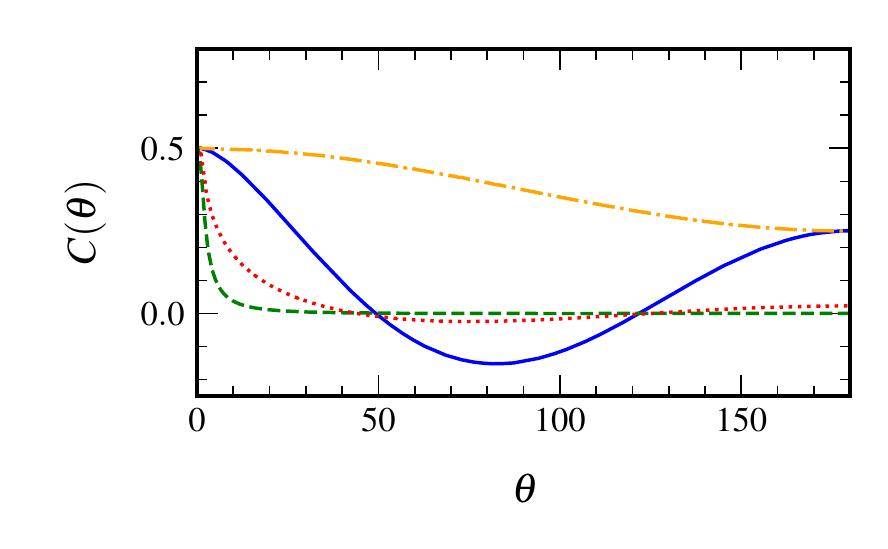}}
\caption{The angular correlation functions (Hellings and Downs curves; Section~\ref{sec:grtests.polarization}) as a function of the sky angle between two pulsars. We show the six possible polarizations: plus/cross (solid blue), vector $x/y$ (dotted red), scalar breathing (dot-dashed yellow), scalar longitudinal (dashed green). Note that there are only 4 curves because the plus/cross and vector $x/y$ modes have been polarization averaged, respectively. See \citet{Maggiore:1900zz} for more details. \citet{2008ApJ...685.1304L} found that a PTA can discriminate between the tensor and non-tensor modes with a PTA with bi-weekly observations of 500 pulsars for five years with an RMS timing accuracy of 100 ns.}
\label{fig:HDcurves}
\end{center}
\end{figure*}

For each pulsar pair, we can measure the correlated response of their timing residuals, and plot this response as a function of a pairs' angular separation, $\theta$. The shape of this angular correlation function of pulse residuals, $C(\theta)$, will be different depending on the type of polarization in the GW being observed.
In Fig.~\ref{fig:HDcurves}, we show the correlation of residuals for a collection of pulsar pairs separated by an angle $\theta$ (i.e., the Hellings and Downs curve, first formulated by \citealt{1983ApJ...265L..39H}).
We can see that each of the different polarization states generates a distinct correlation structure which can, given sufficient PTA sensitivity, be measured \citep{Gair:2015hra}. As discussed in \citet{2008ApJ...685.1304L}, biweekly observations for five years with RMS timing accuracy of 100\,ns can detect non-Einsteinian polarization (i.e. other than the $+$ and $\times$ modes) with 60 pulsars for the longitudinal scalar mode and the vector modes; 40 pulsars for the transverse scalar mode.  To discriminate non-Einsteinian modes from Einsteinian modes, the PTA needs to monitor 40 pulsars for the transverse scalar mode, 100 pulsars for the longitudinal scalar mode, and 500 pulsars for the vector modes. As such, currently the detection of such modes is believed to be at least in the intermediate future; currently, up to $\sim$10 pulsars have the required residual RMS levels. However, this will change rapidly with the timing programs beginning on next-generation radio facilities. 

As an example, \citet{Cornish:2017oic} established the first PTA upper limits on non-Einsteinian polarizations. Currently these limits are derived from the auto-correlation of the residuals of each pulsar with itself. This analysis assumed a gravitational wave background produced by a large collection of unresolved binaries and took into account the fact that dipole and quadrupole radiation will be emitted at different rates from a binary system. Given that the scalar longitudinal mode produces the largest autocorrelation (see Fig.~\ref{fig:HDcurves}) it has the most stringent PTA upper limit.

\subsection{Characterizing the Dispersion Relation of Gravitational Waves}
\label{sec:grtests.dispersion}
Theories which predict these novel polarization states generically also predict non-standard GW dispersion relations which can result in a number of effects measurable by PTAs. We can model a range of modified dispersion relations using the parameterization in geometric units \citep{Blas:2016qmn}:
\begin{equation}
\omega^2 = m_{\rm g}^2 + c_{\rm g}^2 k^2,
\end{equation}
where $m_{\rm g}$ is the mass of the graviton and $c_{\rm g}$ gives a group velocity different from the speed of light. Theories of massive gravity \citep{deRham:2010kj} allow for a non-zero graviton mass and some quantum gravity theories predict a modified group velocity and dispersive effects \citep{Blas:2014aca,Liberati:2013xla}. 

A number of experiments have already placed limits on the graviton mass. In the presence of a graviton mass, the Newtonian potential has a length-dependent suppression from which $m_{\rm g}$ can be constrained using a variety of observations.  The most constraining limit comes from weak lensing observations which yield $m_{\rm g}< 6.9 \times 10^{-32}\ {\rm eV}$ and  can be translated into a frequency $f_{\rm g} = m_{\rm g}c^2/h > 1.7\times 10^{-17}\ {\rm Hz}$ \citep{Choudhury:2002pu}\revision{. Recent independent analyses have focussed on using galaxy clusters and the Sunyaev-Zeldovich effect to limit the mass to $m_{\rm g}< 1.27 \times 10^{-30}\ {\rm eV}$ \citep{2018AnPhy.399...85G}}. A massive graviton changes the functional form of the gravitational potential from a shape of $\sim1/r$ to $\sim e^{-km_{\rm g}r}/r$; this is the so-called Yukawa potential. This change in the potential leads to a change in density fluctuations of matter, which in turn affects the power spectrum of fluctuations measured in weak gravitational lensing surveys. It should be noted that these constraints are subject to model-dependent uncertainties in the exact distribution of dark matter, and so should be used with caution. Other, less model-dependent, constraints from the dynamics of objects in the solar system yield a limit $m_{\rm g}< 4.4 \times 10^{-22}\ {\rm eV}$ or  $f_{\rm g} > 10^{-7}\ {\rm Hz}$ \citep{Talmadge:1988qz,Will:1997bb}.

With a non-zero graviton mass, GWs acquire a frequency-dependent phase velocity leading to an additional phase term in GW signals that would be detected by LIGO \citep{Will:1997bb}. Again, fixing $c_{\rm g} = 1$, constraints from GW170104 provide the best dynamical constraint to the graviton mass and yield $m_{\rm g}< 7.7 \times 10^{-23}\ {\rm eV}$ or $f_{\rm g}>1.8 \times 10^{-8}\ {\rm Hz}$ \citep{Abbott:2017vtc}. 

\begin{figure*}
\begin{center}
\includegraphics[width=0.7\textwidth]{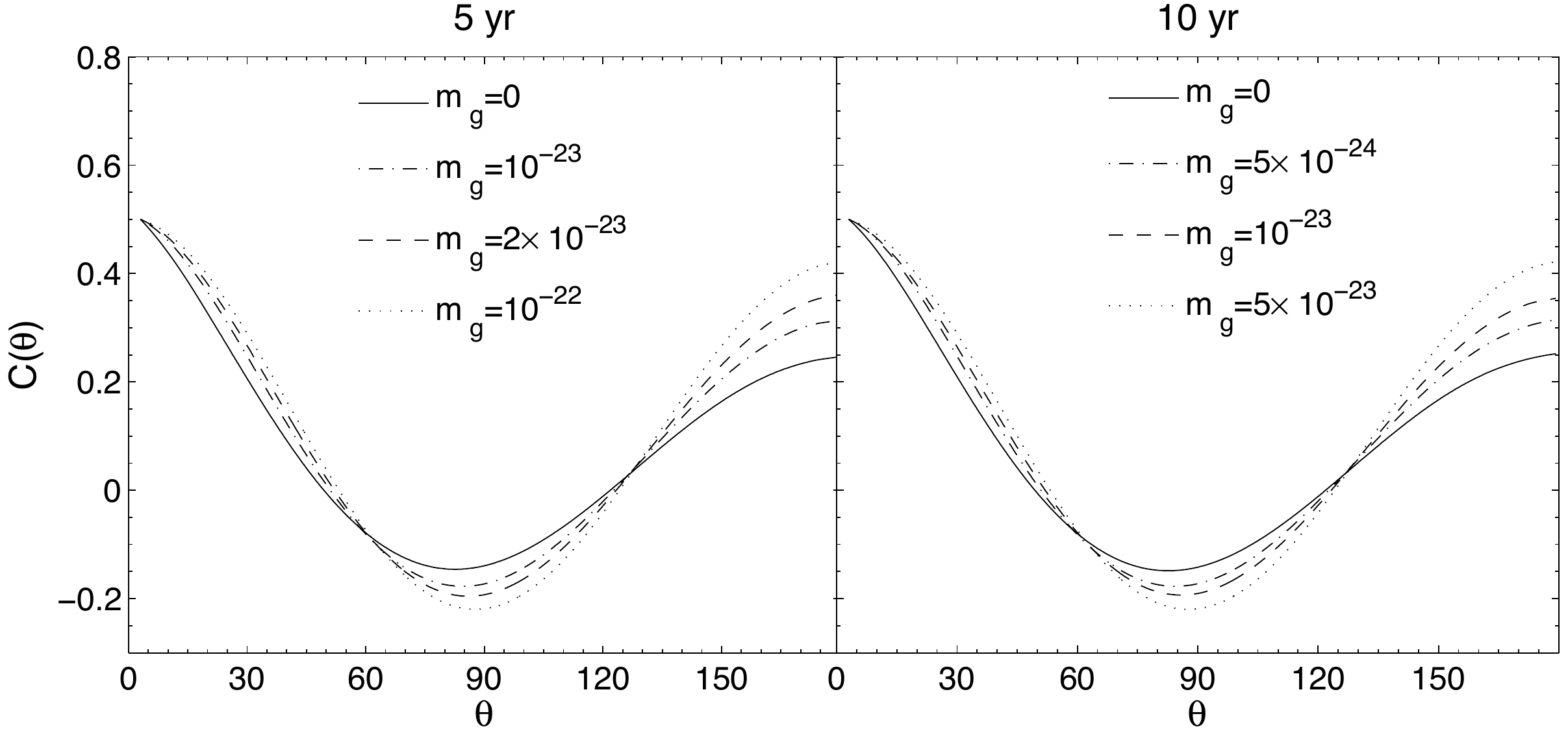}
\caption{The Hellings and Downs curves for the standard polarization modes (plus/cross) but with different values of the graviton mass. Reproduced from \citet{Lee:2010cg}. A PTA with bi-weekly observations of 300 pulsars for 10 years with a RMS timing accuracy of 100~ns can probe graviton masses as low as $m_g = 3 \times 10^{-23}$\,eV. This would be competitive with the current LIGO limit and thus could possibly provide independent confirmation of a LIGO measurement. However, given that we currently only have $\sim$10 pulsars with RMS accuracy of 100~ns, PTA measurements are not yet competitive with other experiments for probing graviton mass.}
\label{fig:f2}
\end{center}
\end{figure*}
Pulsars can also be sensitive to the graviton mass, and given that the best dynamical limit to the graviton mass lies in the middle of the PTA frequency band it is unsurprising that PTAs may complement the limits set by LIGO. A massive graviton would be detectable through slight variations in the Hellings and Downs curve, and an example of this is shown in Fig.~\ref{fig:f2}. 
As discussed in \citet{2010ApJ...722.1589L}, a PTA with bi-weekly observations of 60 pulsars for 5 years with a RMS timing accuracy of 100 ns, massless gravitons can be distinguished from gravitons heavier than $m_g = 3 \times 10^{-22}$\,eV with 90\% confidence level. A 10-year observation with the same RMS timing accuracy and cadence but with 300 pulsars would probe graviton masses down to $m_g = 3 \times 10^{-23}$\,eV. 

In addition to using the correlated residuals between pulsars in response to a stochastic GW background, if a PTA detects a single SMBH binary merger which has a time-tagged electromagnetic counterpart, a comparison between the arrival time of the two signals allows for a complementary probe of the graviton mass.  We discuss this possibility further in Sec.~\ref{sec:gr.propagation}.

Pulsar timing can also place a limit on the group velocity of GWs.  Assuming that the graviton mass $m_{\rm g}\ll 10^{-22}\ {\rm eV}$ so that $f_{g} \gg {\rm year}^{-1}$, in the PTA band the gravitational dispersion relation becomes 
$\omega^2 \simeq c_g^2 k^2$.
If $c_g<1$, the pulse train from a pulsar travels faster than the group velocity of GWs, thus ``surfing'' on the wave. These interactions can lead to phase changes in the waves that significantly amplify the observed timing residuals.
Assuming a reasonable model for the SMBH binary background and taking the timing residuals for PSR B1937+21,
\citet{Baskaran:2008za} find that $1-c_g<10^{-2}$. This limit will improve significantly as PTAs become more sensitive. 

In other GW bands, the GW group velocity can be estimated through measurements of the propagation time between the two LIGO detectors. \revision{Multi-messenger detections at Hz-frequencies can also provide constraints \citep{lombriser+taylor};} an analysis of GW150914 allowed the first direct constraint on $c_g$ giving $c_g < 1.7$ \citep{Blas:2016qmn}. An analysis of the neutron star binary merger GW170817 \citep{TheLIGOScientific:2017qsa} with its electromagnetic counterpart placed the most restrictive constraint on the relative propagation speed of gravitational waves and electromagnetic waves of $-3\times {10}^{-15}<c_g-1<7\times {10}^{-16}$ \citep{Monitor:2017mdv}.


\subsection{Examples of Gravity Theories That Predict Additional Polarizations and Modified Dispersion Relations}
\label{sec:examples.polarizations}
There are several gravity theories which predict additional polarizations and modified GW dispersion relations.  Here we will briefly describe three of them: Einstein Aether \citep{Jacobson:2000xp}, massive gravity \citep{deRham:2010kj}, and $f(R)$ gravity \citep{Carroll:2003wy}. We note that these three theories have been chosen for having a complete and consistent theoretical description and that each face some challenges when confronted with observations. See \citet{Isi:2018miq} for a detailed discussion of how to probe alternative gravity theories with gravitational wave measurements. We also note that the binary neutron star merger observed by \revision{LIGO/VIRGO} \citep{TheLIGOScientific:2017qsa} places strong constraints on the properties of alternative theories of gravity \citep{lombriser+taylor,Sakstein:2017xjx,Baker:2017hug,2018PhRvD..97d1501B}. 

\subsubsection{Einstein Aether}~\\\vspace{-3mm}

\noindent Einstein Aether posits the existence of a Lorentz symmetry-violating, dynamical, time-like vector field. This is in addition to the gravitational metric, a 2-tensor field. This arrangement preserves three-dimensional rotational symmetry while generating a preferred rest-frame. 

The dynamical time-like vector field introduces a preferred frame arising from local physics leading to a generally covariant theory. Einstein Aether is an example of the more general vector-tensor theories \citep{1972ApJ...177..757W,1972ApJ...177..775N} and can be thought of as a simplified model of spontaneous Lorentz symmetry breaking which may occur in string theory \citep{Kostelecky:1988zi}. 

Considering a theory that includes all covariant couplings which is quadratic in all derivatives the theory can be specified by the value of four new constants, $\{\mathcal{C}_1,\mathcal{C}_2,\mathcal{C}_3,\mathcal{C}_4\}$ \citep{Jacobson:2004ts}. In the limit where these constants vanish this theory is indistinguishable from general relativity.  As shown in \citet{Jacobson:2004ts}, there are five GW polarizations in this theory.  Transforming to a gauge where $h_{\mu 0}=0$ and taking the $\mathcal{C}_i \rightarrow 0$ limit the theory predicts: plus/cross tensor polarizations traveling at the speed of light; two longitudinal vector modes traveling at a speed possibly different from the speed of light; and a linear combination of the two scalar modes 
\begin{equation}
h_{ij}^s=v_0\left[\frac{3}{c_s^2}e_{ij}^\ell-(\mathcal{C}_1+\mathcal{C}_4)e_{ij}^b\right],
\end{equation}
where $v_0$ is the linear perturbation to the time-component of the Aether field, $c_s$ is the speed of the scalar GWs, and $e_{ij}$ is as given in Equations~\ref{eq:pol1}--\ref{eq:pol3}. 

\subsubsection{Massive Gravity}~\\\vspace{-3mm}

\noindent 
Building a well-behaved model of gravity in which the graviton has a non-zero mass has been an issue for decades; however, recent attempts at understanding the nature of dark energy have reinvigorated the conversation.  In such models, a massive graviton might introduce a scale that could explain the observed long-range behavior of the gravitational field.  Since the early 70s it has been known that adding a mass in the absence of non-linear interactions gives rise to an observationally ruled-out discontinuity, the dDVZ discontinuity \citep{vanDam:1970vg,Zakharov:1970cc}.  The addition of non-linear interactions can cure this, but at the cost of a degree of freedom.  This is known as the Boulware-Deser ghost \citep{Boulware:1973my}, proof that only five propagating degrees of freedom can exist in any massive, interacting extension of general relativity.

Only very recently was there a solution to this issue; the imposition of a symmetry, the Gallileon symmetry \citep{deRham:2010kj,Hassan:2011hr}, restricts the form of the non-linear interactions and projects out the Boulware-Deser ghost. Later it was shown that bimetric theories can be derived from these new massive gravity theories, and so the theories have been connected \citep{Hassan:2011zd}.   

Just as in the case of Einstein-Aether theory the stable propagating mode is a superposition of the 6 degrees of freedom identified in Sec.~\ref{sec:grtests.polarization}. Transforming the metric far away from the source (where the non-linear terms in the field equations vanish) into synchronous gauge, massive gravity generates the standard plus/cross tensor polarizations which travel at the speed of light and a scalar wave which travels at a speed $c_s$ with a polarization
\begin{equation}
    h^{s}_{ij} = \frac{\pi}{6M_{\rm pl}}\left[e^b_{ij} + \left(1-\frac{1}{c_s^2}\right)e^\ell_{ij}\right]
\end{equation}
which is a linear combination of the breathing and longitudinal modes.

\subsubsection{$f(R)$ Gravity}~\\
\vspace{-3mm}

\noindent  Another commonly considered alternative theory of gravity imagines a gravitational Lagrangian which is a function of the Ricci curvature scalar, $f(R)$ where $f(R)=R$ gives the Lagrangian for GR.  The original motivation for $f(R)$ gravity was to provide an alternative gravity theory that could account for the observed accelerated expansion of the Universe \citep{Carroll:2003wy}. In its original formulation the gravitational Lagrangian was modified to include a term inversely proportional to the Ricci scalar: $R+\mu/R$. As the Universe expands the homogeneous value of the Ricci scalar decreases until, at a late enough time, this new term becomes dynamically important.  Although it was shown that this specific modification was at odds with measurements of the deflection of light around the Sun \citep{Chiba:2003ir,Erickcek:2006vf}, there are specific forms of the function $f(R)$ which pass Solar System tests and lead to a late-period of accelerated expansion \citep{Chiba:2006jp,Hu:2007nk}. 

$f(R)$ gravity is an example of a scalar-tensor theory (see, e.g., \citealt{Will:1993ns}): by performing a conformal transformation the theory can be described as a scalar-tensor theory with a scalar mass 
\begin{equation}
m^2 = \frac{1}{3} \left[\frac{1}{f^{\prime \prime}(R_0)} - \frac{R_0}{f^{\prime}(R_0)} \right],
\end{equation}
where $R_0$ is the Ricci scalar for the background over which the GWs propagate. 
An analysis of the polarization modes shows that this theory predicts the standard plus/cross tensor polarizations propagating at the speed of light and again, a linear combination of the breathing and longitudinal scalar-modes \citep[Eqs.~\ref{breathinglongpol} and \ref{eq:pol3};][]{PhysRevD.95.104034}
\begin{equation}
h^s_{ij}= -\frac{1}{2}f^{\prime}(R_0)\left(e^b_{ij}+\frac{m^2}{\omega^2} e^\ell_{ij}\right),
\end{equation}
and follow the dispersion relation $m^2 = \omega^2 - k^2$.

\section{Relic Gravitational Waves and Early-Universe Cosmology}
\vspace{-3mm}
\begin{tcolorbox}[enhanced,colback=gray!10!white,colframe=black,drop shadow]
Leading theories for the evolution of the early Universe invoke an inflationary epoch to explain the isotropy and other broadly observed properties of the Universe.\\\vspace{-2mm}

If this epoch occurred, it would have produced GWs by rapidly expanding quantum fluctuations present in the pre-inflation epoch. The nature of inflationary expansion will be encoded in the strength and spectrum of the produced broad-band GW background. For standard 
inflation models, the inflationary background will likely be fainter than that of SMBHBs, however some inflationary modes may have a spectrum that rises into the PTA/LISA/LIGO bands. PTAs have already begun to place stringent limits on those models. PTAs will most sharply probe the spectrum (scalar spectral index), while cosmic microwave background (CMB) experiments will probe the relative strength of these ``relic'' GWs and the more standard density waves (the latter have already been detected and characterized by CMB probes).\\\vspace{-2mm}

If primordial GWs can be detected, PTAs may also be sensitive to phase transitions in the early Universe, allowing (amongst other possibilities) an independent constraint on the existence of the cosmological constant.
\end{tcolorbox}

\subsection{Relic Gravitational Waves}\label{sec:relic}
\begin{figure*}
\centering
\includegraphics[width=0.8\textwidth,trim=2cm 1.6cm 2cm 2cm,clip]{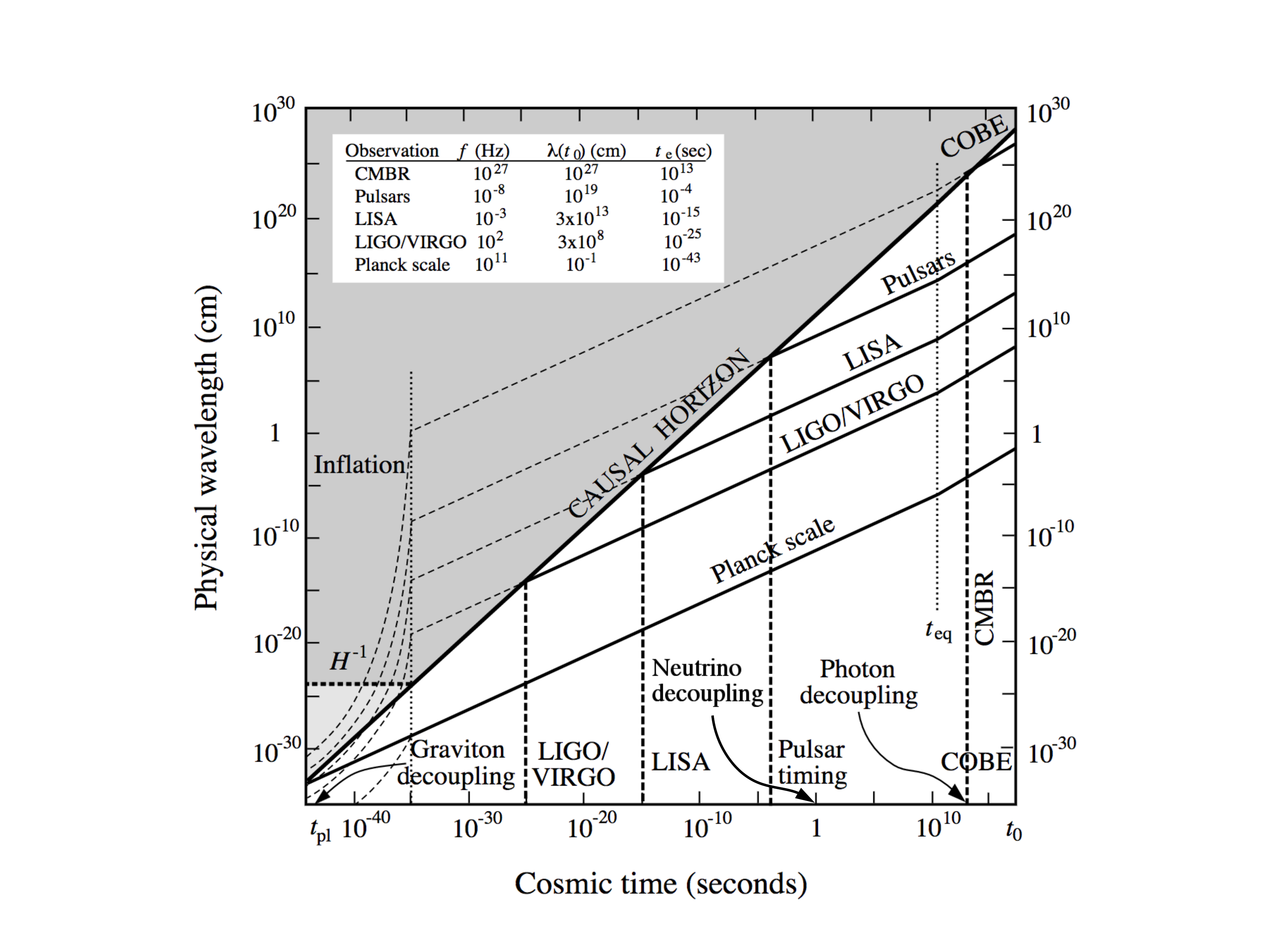}
\caption{The quantum fluctuations in the very early Universe, if rapidly amplified by inflation, could cause a highly broad-band GW background signal in addition to affecting the cosmic microwave background. While cosmic microwave background experiments can constrain the scalar-to-tensor ratio, higher-frequency experiments (as PTAs, LISA, LIGO/VIRGO shown here) will most sharply probe the spectrum of these early fluctuations.  
Figure reproduced from \citep{battye-shellard-astroph}.}
\label{fig:inflation}
\end{figure*}

GW signals can arise from the early Universe if a period of rapid inflation occurred (Fig.~\ref{fig:inflation}). Quantum fluctuations from early in the Universe would have been amplified by inflation, and like strings, would produce a broad-band signal detectable by multiple instruments. 
These ``relic'' GWs are a long-standing target for current and future CMB experiments, looking for the tensor modes induced by these waves \citep[][and references therein]{2016ARA&A..54..227K}. CMB experiments will constrain only the long-wavelength (low-frequency) portion of the spectrum of inflationary GW signals, and can most effectively constrain the scalar-to-tensor ratio. However, higher-frequency experiments, like PTAs and space- and ground-based laser interferometry, are able to more effectively constrain the spectrum of these early fluctuations by looking at the scalar spectral index (the spectral index of the detected background, which reflects the spectrum of the fluctuation scales and the mode of their amplification). Various models of inflation predict differing values for the observed scalar spectral index.

Only weak---and often highly model-dependent constraints---exist on the shape of the inflationary GW spectrum.
Following the recognition that primordial black holes could contribute to the LIGO detections, there has been a renewed interest in examining what kind of constraints could be placed on the spectrum of inflationary perturbations at shorter wavelengths, where PTAs can contribute broad-band limits. In fact, the most stringent limit on relic GWs, before LIGO's first observation, had been set at $\Omega_{\rm gw}(f)<2.3\times 10^{-10}$ by the combination of PPTA, LIGO, and CMB limits \citep{2016PhRvX...6a1035L}. This marked the first time the most optimistic predictions of various models were impinged upon by observations.

Given the rapid progress across the GW spectrum (from ground-based interferometers to PTAs to CMB experiments), our assessment is that this area is likely to remain observationally driven.  That is, while models can always be constructed to evade the observational limits, the extent to which a blue inflationary GW spectrum (i.\,e.\ one where the energy density rises with GW frequency) remains viable will be determined by the observations across the GW spectrum.

While we can currently place constraints on inflationary GWs, we state with some confidence that PTAs will not be the primary driver for the study of inflationary GWs. This is because it is highly unlikely that the inflationary epoch signal will dominate over the much brighter GW background from inspiraling SMBHBs. \revision{For instance, the most widely accepted ``slow-roll inflation'' model has an expected $\Omega_{\rm gw}$ of $10^{-15}$, while the most pessimistic backgrounds from SMBHBs are expected to be around two orders of magnitude above that level \citep{2016PhRvX...6a1035L}.

Certain axion inflation models also predict that a large bump in the GW background spectrum may be produced by the production of stellar-mass primordial BHs in the early Universe \citep[\eg][]{pbh_inflation_twofield_bumpprediction}. If the primordial BHs are produced within a narrow mass range with a peak at a mass of on the order of a few to on the order of 100\,$\msun$, this GW background bump may peak in the pulsar timing band, significantly enhancing the inflationary signal in a limited gravitational bandwidth that is defined by the primordial BH mass distribution \citep{2017JCAP...09..013G}.}

\subsection{Cosmological Measurements and the Cosmological Constant}\label{sec:other.cosmology}

As the early Universe cooled, successive constituents decoupled and began to evolve (more or less) independently. The most notable such example is the decoupling of photons, which led to the formation of the CMB.  Prior to photon decoupling, it is expected that neutrinos underwent the same process, leading to a relic neutrino background.  
\citet{2010CQGra..27s4008L} and \citet{2011GReGr..43..945B} note that cosmic neutrino decoupling should have happened a few seconds after the Big Bang, at a redshift $z \sim 10^{10}$.  A GW entering the horizon at this time, currently has frequency of order 1~nHz (e.g., see \ Fig.~\ref{fig:inflation}), suggesting that the neutrino decoupling time can be probed by nanohertz GWs. Specifically, they predict that the effective viscosity from the neutrinos will suppress such GWs.
They also conclude that a multi-decadal dataset would be required to detect this effect.  With current PTAs beginning to surpass the decadal observational duration, it seems that this effect still remains out of reach.

The early Universe may have also experienced phase transitions as it cooled, likely before the epoch of neutrino decoupling.  \citet{2010PhRvD..82f3511C} consider a quantum chromodynamics phase transition (which they take to occur at $z > 10^{17}$), and the GWs produced during that epoch.  Such a phase transition could produce GWs with $f \sim 1$~Hz.  Their assessment was that the NANOGrav sensitivity, at the time, was insufficient to detect GWs from this phase transition, but predicted that PTAs might be able to detect these GWs on the time scale of the year 2020.  However, their prediction was based on an idealized \hbox{PTA} (with higher precision timing residuals, and on a smaller number of pulsars than currently timed). For a realistic estimate, this calculation would need to be reformulated to encompass the actual sensitivity of current PTAs, i.e. taking into account the true number of pulsars being timed by the various PTAs, and realistic timing residuals.

The standard cosmological model includes a ``dark energy'' component~$\Lambda$ \citep[and references therein]{2016A&A...594A...1P}, which may be equivalent to the ``cosmological constant.''  The expectation is that dark energy becomes relevant on large scales, and significant international efforts are devoted to placing constraints on the properties of dark energy.  \citet{2011PhRvD..84f3523B} and \citet{2013ApJ...764..163E} describe an approach in which measurements of pulsar timing residuals due to GWs emitted in the nearby Universe ($z \ll 1$) could place a complementary constraint on~$\Lambda$.  One (acknowledged) caution about their results is that their estimates are based on a somewhat idealized PTA with a relatively short observation duration (3~yr).  Given that current PTAs have durations in excess of a decade, and substantially more pulsars, it is likely that different (and probably more strigent) constraints could be obtained by repeating their analysis for current PTAs.

\section{Additional Science from ``Hidden Planets'' to Dark Matter}
\label{sec:beyond}
\vspace{-3mm}
\begin{tcolorbox}[enhanced,colback=gray!10!white,colframe=black,drop shadow]
There are numerous scientific endeavors that \emph{may} be attainable with PTAs, including those \revision{for} which: 1) our theoretical understanding is still under active development or is speculative in nature, or 2) the pursuit/detection of the science is likely much further in the future than the topics described in previous sections.  This includes GW studies of stellar convection and dark matter, as well as the \revision{effect of primordial black holes to produce a GW background or direct gravitational perturbations on Earth or the pulsar, or direct perturbations on Earth caused by} our own solar system bodies.

\vspace{2mm} The discussion here is structured in terms of specific sources and predictions. We also recommend reading \citet{2014PhRvD..89d2003C}, which considers the blind discovery space of PTAs. That work showed that, whatever the source, GW memory at high redshift is a potential discovery area.
\end{tcolorbox}




\subsection{Stellar Convection}\label{sec:other.stars}

Although the common assumption for nanohertz GWs is that they are produced by extragalactic or cosmological sources, \citet{2014ApJ...792...55B} evaluate the mass quadrupole produced by turbulence within a convective star, such as the Sun.  They show that the ensemble of stars should produce an irreducible background.  While their primary focus is \hbox{LISA}, they note that the amplitude of the GW signal is amplified in the near-field zone, which is certainly the case for the Earth term for frequencies relevant to PTAs.  A simple extrapolation of their results into the nanohertz GW frequency band suggests that the Sun may be a contributor to the PTA noise budget, and thus potentially detectable. Such an extrapolation relies upon assumptions regarding the nature of turbulence within a star, and that the actual magnitude of any GW signal may be substantially lower; thus, with increasingly stringent limits, PTAs may place constraints on solar and stellar convection.

\subsection{Dark Matter}\label{sec:other.dark}

Several authors have considered various classes of dark matter candidates that may produce observable signatures in PTA data.
\revision{
\subsubsection{Cold Dark Matter}~\\\vspace{-3mm}

\noindent
 The concordance $\Lambda$-Cold-Dark-Matter ($\Lambda$CDM) model for the evolution of the Universe predicts that the mass function of dark matter halos may extend to very small masses ($\ll\msun$), with the mass function cutoff related to the nature of dark matter. Thus, detection of small-scale dark matter clumps can both provide support for the cold dark matter scenario and constrain the mass of dark matter particles. There are two possible influences on pulsar timing \citep{DMclumps,drtz19,baz11,shf07,2010ApJ...723L.195I}: $(i)$ Shapiro delay of the radio pulses propagating through a distribution of dark-matter sub-structure (i.e. the integrated Sachs-Wolfe effect), and $(ii)$ Doppler delay due to an acceleration of the Earth or pulsar caused by the nearby passage of a dark matter clump. 
 
 The Shapiro effect is line-of-sight dependent, leading to delays in pulsars that are related only by the statistical properties of dark-matter sub-structure. This is expected to be challenging to disentangle from intrinsic pulsar noise processes, but could give access to clumps in the mass range $10^{-4}-10^{-3}M_\odot$. In the Doppler delay, a clump passing by a pulsar produces a timing-residual influence only in that pulsar, whereas a clump passing by Earth produces a dipolar-correlated timing delay in all pulsars in the array (similar to unmodeled Solar-system ephemeris systematics; \citealt{baz11}). The Doppler delay effect is expected to dominate over the Shapiro delay from sub-structure \citep{drtz19}, potentially constraining clumps in the mass range $10^{-9}-10^{-8}M_\odot$. 
 

}
\subsubsection{Scalar-field Dark Matter}~\\\vspace{-3mm}

\noindent
\revision{Scalar-field dark matter models (sometimes referred to as ``fuzzy dark matter``) address some of the issues that $\Lambda$CDM has in reproducing the observed number density of sub-galactic scale structures in the Universe \citep{hu+2000,hotw17}.} Structure formation is suppressed below the de~Broglie wavelength, 
\begin{eqnarray}
\lambda_{dB}\approx 600~{\rm pc}\left(\frac{10^{-23}~{\rm eV}}{m}\right)\left(\frac{10^{-3}}{v}\right),
\end{eqnarray}
where $m$ is the mass of the dark matter particles and~$v$ is their characteristic velocity in units of~$c$. Such a population of particles will produce an oscillating pressure that, though averaging to zero, will cause sinusoidal variations in the local gravitational potential with a frequency 
\begin{eqnarray}
f\approx 5\cdot10^{-9}~{\rm Hz}\left(\frac{m}{10^{-23}~{\rm eV}}\right)
\end{eqnarray}
and an amplitude
\begin{eqnarray}
\Psi({\bf x})=\pi\frac{G\rho({\bf x})}{m^2},
\end{eqnarray}
where $\rho({\bf x})$ is the density of dark matter particles at position ${\bf x}$. While it is not a GW effect, pulsar signals propagating through such a time-variable gravitational potential will have their pulsation frequencies sinusoidally modulated. The perturbation to the observed times of arrival could be as large as several hundred nanoseconds, which would be a signal that is accessible to a number of currently timed pulsars \citep{kr14}.

\citet{pp14} searched for this signature of \revision{scalar-field dark matter} and placed the first observational constraints on the mass of such particles using the NANOGrav Five-Year Data Release \citep{d+13}.  Their upper limits on the local oscillating gravitational potential were an order of magnitude above the current predictions. In follow-up work, \citet{p+18} used $26$ PPTA pulsars observed over $12$ years to compute rigorous Bayesian and frequentist constraints on the local density of ultralight scalar-field dark matter. They found that $\rho_\mathrm{SF}\leq 6$ Gev cm$^{-3}$ at $95\%$ confidence for ultralight bosons with $m\leq 10^{-23}$ eV. This improves upon previous constraints by a factor of $2-5$. 

The ability to measure this oscillating gravitational potential improves as more pulsars are added to a \hbox{PTA}. \revision{The current IPTA contains more than twice the number of pulsars than were used by \citet{p+18}, and more continue to be added to the constituent PTA programs on an annual basis. The prospects for near-future constraints from the IPTA on ultralight scalar-field dark matter are thus very exciting.}

\subsection{Primordial Black Holes}\label{sec:other.primordialbhs}
The first detection of GWs from merging stellar-mass BHs by LIGO/Virgo \citep[][]{2016PhRvL.116f1102A}, as well as the subsequent mergers detected, raised the question of the formation channel that produced these BHs.
While a full discussion of this topic is beyond the scope of this paper, it has been proposed these BHs are \emph{primordial}, produced during the inflationary period in the early Universe. \revision{As previously discussed in the last paragraph of Section~\ref{sec:relic}, some of these inflation models predict solar mass and larger primordial BHs that may be produced as a by-product of inflation. Inflation itself, in addition to the production of primordial BHs, both contribute to a potential GW background signal \citep[\eg][]{pbh_axion_inflation,pbh_inflation}. The peak frequency of the bump that can be produced in the GW background spectrum depends on the primordial BH mass, and the 10--100 solar masses range precisely maps into the nanohertz GW band \citep{2017JCAP...09..013G}. Thus, PTA searches of the GW background also serve as probes of primordial BH mass distributions in the critical range that LIGO is detecting.}

\revision{Much smaller primordial BHs have also} been proposed as relevant to PTAs, by \citet{2007ApJ...659L..33S} and \citet{2012MNRAS.426.1369K}, due to the perturbations they can cause when passing close to the Earth.  In contrast to the BHs responsible for the LIGO events, the primordial BHs considered by these authors have much smaller masses ($< 10^{-10}\,M_\odot$).  Nonetheless, from the perspective of an observationally driven constraint on primordial BHs, these constraints remain of potential interest.  These authors show that a primordial BH may produce perturbations of order 20~ns over a duration of order 15~yr.  With the various PTA data sets now exceeding a decade, this signal may become feasible to detect in the next decade, if other contributions to pulsar timing residuals at similar levels (tens of nanoseconds) can be sufficiently controlled or modeled.

\subsection{Solar System Ephemerides and Wandering Planets}\label{sec:other.ssb}

A fundamental term in pulsar timing is an astrometric term, designed to transfer the pulsar TOAs into the frame of the solar system barycenter, which is assumed to be (quasi-)inertial. The time delay between the Earth and Barycentric reference frames is given by \citep{2012hpa..book.....L}:
\begin{equation}
\Delta t_{\mathrm{SSB}} = \frac{\mathbf{R}_{\mathrm{SSB}}\cdot\hat{\mathbf{n}}}{c},
\label{eqn:ssb}
\end{equation}
where $\mathbf{R}_{\mathrm{SSB}}$ is the vector connecting the Earth to the solar system barycenter, $\hat{\mathbf{n}}$ is the unit vector in the direction of the pulsar, and~$c$ is the speed of light. The barycenter position is the center of mass of the solar system, involving weighted contributions from all planets and important dynamical objects. Any uncertainties in the position or masses of solar system bodies will create a corresponding uncertainty in the barycenter position.

In Equation~\ref{eqn:ssb}, it is clear that uncertainties in the knowledge of the position of the solar system barycenter will affect the accuracy to which pulsars can be timed.  As the position of the solar system barycenter~$\mathbf{R}_{\mathrm{SSB}}$ is determined from the orbits and masses of the solar system planets (and minor bodies), there is a long history of assessing whether, and to what extent, precision pulsar timing can be used to constrain properties of the solar system \citep[e.g.,][]{1971ApJ...165..105M,2016RAA....16...58L}.  These efforts generally find that the position of the solar system barycenter is not known to better than about 100\,m, a value that is consistent with expectations from the spacecraft data used to construct the solar system ephemeris (W.~M.~Folkner~2017, private communication).  There have also been efforts \revision{using PPTA and IPTA data to measure the masses of the solar system planets by using precision pulsar timing (\citealt{2010ApJ...720L.201C} and \citealt{IPTA_solar_system}, respectively).}

Recent efforts to constrain the nanohertz stochastic GW background with the NANOGrav $11$-yr dataset uncovered differences in upper-limits and detection statistics caused by systematic errors in published solar system ephemerides \citep{2018ApJ...859...47A}. Upper limit variations were small, but Bayes factors for a long-timescale red noise process shared by all pulsars (a potential first signature of GWs) varied by over an order of magnitude, from compelling to insignificant evidence. To mitigate these systematic errors, \citet{2018ApJ...859...47A} added gas-giant mass perturbations and Jupiter orbital-element perturbations to the global PTA signal and noise model. This led to consistent constraints and detection statistics, regardless of the assumed baseline ephemeris version. Work is ongoing to expand this model into a full Bayesian solar system ephemeris. 

There is also a long history of assessing whether precision pulsar timing could reveal currently unknown members of the solar system or a distant companion star to the Sun \citep[e.g.,][]{1977Natur.270..324H,2005AJ....130.1939Z,glc18}.
Though some of the early efforts may have been affected by severe selection effects \citep{1978Natur.273..132H}, the existence of distant members of the solar system is a question that has recently attracted considerably more attention, given the suggestion of a previously unknown ``Planet Nine'' \citep{2016AJ....151...22B}.
While we are unaware of any large-scale effort to constrain the existence of this body with the modern PTAs, \citet{glc18} estimate that PTAs could constrain the mass to be $\lesssim 10^{-4} M_\odot$, and \citet{2005AJ....130.1939Z} show that measuring the acceleration of the solar system's barycenter with precision pulsar timing may be highly competitive with other methods for constraining the existence of distant companions ($> 300$~au).

\section{Gravitational-Wave Synergies: Multi-band GW Science}
\label{sec:lisa}
\vspace{-3mm}
\begin{tcolorbox}[enhanced,colback=gray!10!white,colframe=black,drop shadow]
PTAs have some target science in common with other GW detectors at higher frequencies (LIGO, LISA) and lower frequencies (CMB experiments). In some cases, this allows for synergistic or even direct multi-band GW studies.

\vspace{2mm}PTAs and higher-frequency millihertz experiments like LISA uniquely share SMBHBs as a direct common target. PTAs and LISA will make synergistic measurements of galaxy and SMBHB evolution.
However, LISA will probe lower-mass and higher-redshift systems, while PTAs will preferentially detect the largest systems in the nearest few Gpc.
These experiments will work together to fully characterize black hole growth over cosmic time. For a small subset of discrete systems, \emph{both} experiments may co-detect the selfsame system at different phases of evolution. 

\vspace{2mm} Cosmic strings and relic GW signals are also intrinsically broad-band GW emitters, and where relevant we have summarized the links between experiments sensitive to those phenomena in Sections \ref{sec:strings} and \ref{sec:relic}, respectively. Complementary constraints from multiple wave bands on the nature of gravity are described in Section \ref{sec:grtests}.
\end{tcolorbox}

The LIGO/Virgo GW detections \citep{GW150914, GW151226, GW170104} in the high-frequency GW regime ($\gtrsim$Hz) have ushered us into the era of GW astrophysics.  In addition to PTAs, 
there are many other instruments and techniques currently in the design, planning and prototyping phases for future GW observatories. For instance, LISA is a planned space-based GW detector, sensitive to intermediate frequencies ($\sim$ mHz) expected to be launched in the 2030s \citep{2012CQGra..29l4016A}.  

\begin{figure*}
\begin{center}
\resizebox{!}{8cm}{\includegraphics{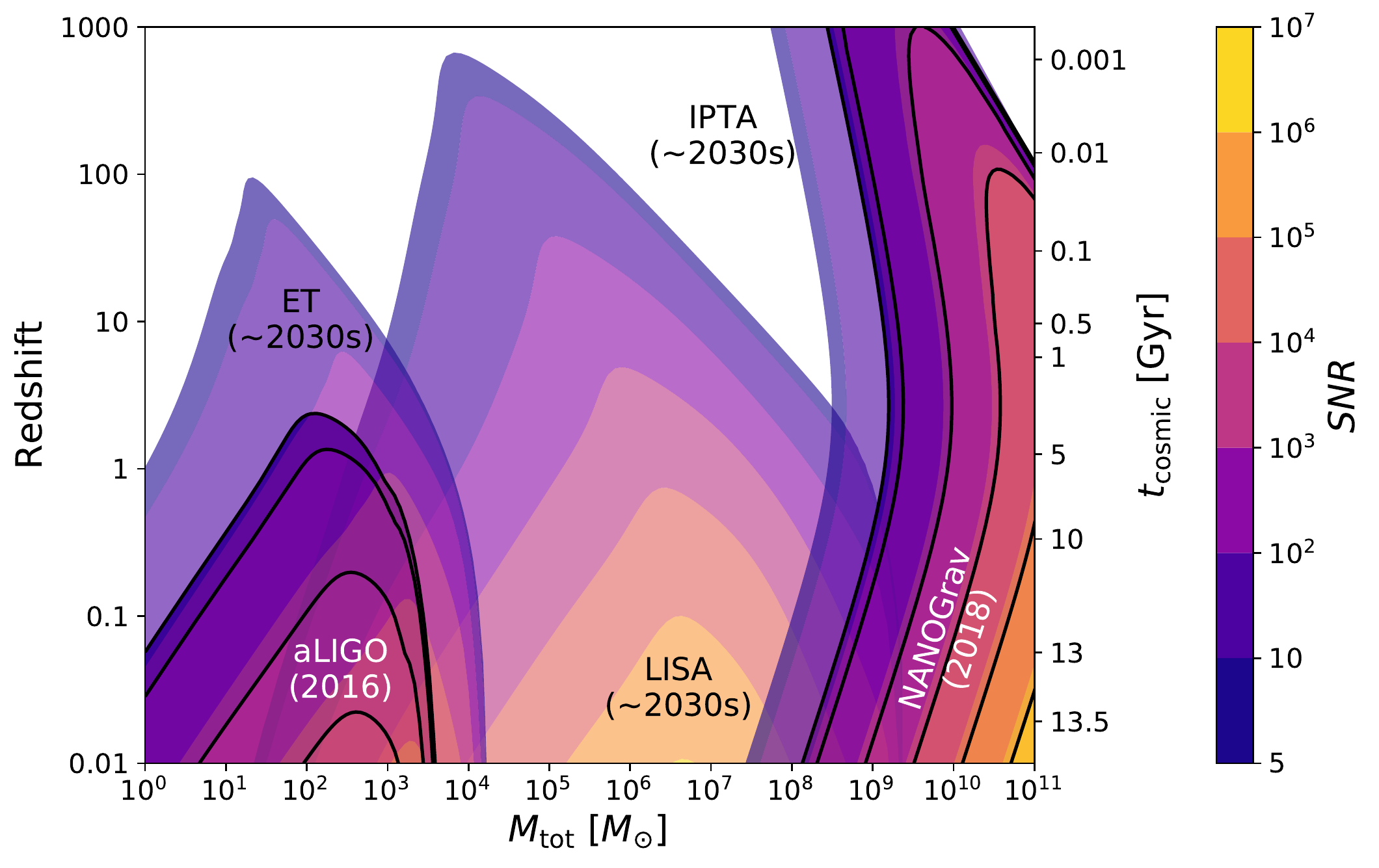}}
\caption{\revision{LIGO, LISA, and PTAs have complementary coverage to study the full
range of black hole masses at various stages of the Universe. Here we
show the approximate signal-to-noise ratio for the complementary
wavebands of these three instruments as they are currently (darker
shading/black contours) and in the early- to mid-2030's era (lighter
shading). This plot focuses only on \emph{individual} (rather than stochastic) black hole detections. All curves assume
instrument-limited sensitivity, without an astrophysical background.
Individual inspiral/coalescence events at high redshift will be
detectable by LISA, while systems in the extended inspiral phase at
higher masses and lower redshift are detectable by PTAs as continuous
gravitational waves. The source classes of LISA and PTAs are
particularly linked through the evolution of MBHs across cosmic time.
Understanding the growth of MBHs will require the contributions of
both PTA and LISA data. Figure produced by Andrew Kaiser and Sean
McWilliams (WVU); a more rigorous version will be published in Kaiser
\& McWilliams (in prep).}}
\label{fig:pta-lisa}
\end{center}
\end{figure*}


SMBHBs are among the primary targets of both PTAs and LISA. However, the two experiments probe different stages of SMBHB evolution and they are sensitive to SMBHBs in different mass ranges. PTAs are most sensitive to the early inspiral (orbital periods of years or longer) of nearby sources with $\mathcal{M} \sim 10^9 \, M_\odot$ \citep{2017arXiv170803491M}. In contrast, LISA is sensitive to the inspiral, merger, and ringdown of SMBHBs with masses from $10^4 - 10^7~M_\odot$ at a wide range of redshifts \citep{2012CQGra..29l4016A}. The two populations of SMBHBs probed by PTAs and LISA are linked via the growth and evolution of SMBH across cosmic time, as shown in Figure \ref{fig:pta-lisa}. Given that the same fundamental physics is required to produce and evolve both populations of BH binaries, there exists a strong link between the methodology of evolutionary models used to study them, and the insights that observations of their GW signatures will provide. In particular PTA observations of the GW background, and measurement of its spectral index, will provide valuable constraints on the mass-function and eccentricity-distribution of SMBH \citep[e.g.][]{k+17} which will better constrain the detection rates for LISA.

Under the right circumstances, an individual source could be observed by PTAs and LISA at different stages of its evolution \citep{pch+2008,Spallicci2013}. For a PTA with high-frequency cutoff of $4\times10^{-7}$~Hz,
a SMBHB will transition from being observable by PTAs to being observable by LISA in 1 - 50 years. This transition time can be reduced by increasing the pulsar observing cadence, which improves the sensitivity of PTAs at high frequencies. However, astrophysical rate estimates suggest the probability of a sequential detection is extremely low ($4.7\times10^{-4} - 3.3\times10^{-6}$ per year to merger per year of survey) due to the small number of individual sources observable by both detectors \citep{Spallicci2013}.

It is also possible to 
use ringdown observations made by LISA as triggers to search PTA data for past continuous waves, or for memory-inducing SMBHB coalescence events (Sec.~\ref{sec:gwmemory}). LISA can observe the ring-down of higher mass sources, even when the inspiral and merger happen outside of the LISA band, meaning there is better overlap for direct observations of these sources with both PTAs and LISA. Parameter constraints from observing the ring-down can be used to improve the search for the inspiral, extrapolating a SMBHB model back in time to predict the expected gravitational waveform throughout the previous years of pulsar observations. Currently, the planned launch date for LISA is 2034, at which point PTAs will have accumulated over 30 years of data that can be used for such a search.

Galactic sources of GWs that LISA will be able to study may, under certain circumstances, be possible to investigate with pulsar timing. Globular clusters (GCs) likely host GW sources detectable by LISA \citep{kcb+18}. GCs are also known to host large populations of pulsars \citep{rhs+05,frk+17}. Pulsars in a GC will be within a few parsecs of GW sources in that cluster and could act as sensitive probes of those GWs \citep{jcl05,mcc17}. \revision{Pulsars in GCs have some limitations in sensitivity due to accelerations from intra-cluster dynamics, which tend to cause low-frequency structure in the timing residuals of those pulsars, possibly masking or mimicking low-frequency GWs. Still, several GC pulsars are currently timed as part of the PPTA and EPTA, and could serve as probes of Galactic LISA targets.}

Alternatively, a pulsar behind a GC, significantly more distant than the cluster, would be extremely sensitive to GWs from sources near its line-of-sight. However, such an arrangement is unlikely. In the absence of multi-messenger information, several such pulsars would also be necessary in order to distinguish GWs from unmodeled effects in an individual pulsar. Additionally, for pulsars within or behind a GC, the conceptually straightforward division of the timing signal into an Earth term and pulsar term does not apply because the GWs cannot be approximated as plane waves. This fact will require PTAs to adopt currently undeveloped detection techniques. Finally, several pulsar-timing GW limits at high GW frequencies ($10^{-6}~$Hz -- $10^{-3}~$Hz) overlapping with the LISA band have been published \citep{yardley+10,yi+14,Dolch+16,perera+18}. This is possible because of high-cadence observations of selected pulsars. These limits are many orders of magnitude above the Doppler tracking GW limits from the Cassini spacecraft and the expected GW sensitivity of LISA \citep{armstrong+03}. However, both Cassini and LISA are entirely solar-system bound; one possible advantage of pulsar timing in the $\mu$Hz to mHz band is its sensitivity to a Galactic GW source close to a pulsar line-of-sight. Studying Galactic sources is an area of investigation that needs to be further developed by the PTA community, but it could yield exciting overlap with LISA science.

Recently, there has also been work on the possibility of astrometric GW detection \citep{schutz2010, book2011}, for example using an instrument like GAIA that can accurately measure the positions of over a billion stars in the Milky Way \citep{gaia2016}.  This technique uses correlated deviations in the observed positions of astrophysical objects (most likely stars, or possibly quasars) to measure passing GWs.  PTA use a similar procedure based on time delays which are determined by integrating a GW induced redshift over the path of each photon.  This integration introduces a frequency dependence in which the characteristic-strain sensitivity of the array is linearly related to the GW frequency, and thus is worse at high frequencies.  Astrometric detection, on the other hand, performs an effectively instantaneous measurement of each photon's source direction.  Within the Nyquist band, determined by the observational cadence and duration of the observatory, astrometric detection has a sensitivity relatively independent of frequency.  This detection method is still in its early days of study, however astrometric GW-detection may be able to complement PTAs at the high-frequency end of the PTA sensitivity range \citep[$f \gtrsim 3\times10^{-8}~$Hz;][]{moore2017}, and could potentially be used in a coherent multi-messenger PTA+Gaia analysis to validate the detection of nanohertz GWs, or even constrain beyond-GR GW polarizations \citep{m+18,oc18}.

\section{Electromagnetic Synergies: Multi-messenger Science}
\label{sec:em}
\vspace{-3mm}
\begin{tcolorbox}[enhanced,colback=gray!10!white,colframe=black,drop shadow]
Almost all of the science cases described in the previous sections can be significantly augmented with electromagnetic studies. However, for the PTA GW band, the science that would benefit most clearly from a concerted multi-messenger and multi-wavelength effort is the identification of the inspiral and/or coalescence from discrete SMBHBs. Here, we outline a selection of studies related to SMBHs that can be uniquely achieved with a host galaxy identification and the direct measurement and tracking of a SMBHB electromagnetic counterpart (note, the latter implies the host galaxy has been identified).
\vspace{3mm}

We also summarize a few practical aspects of multi-messenger detection in the nanohertz wave-band, which may include both new and archival data.
\end{tcolorbox}

\subsection{Identifying and Tracking a SMBHB Host}\label{sec:localize}
\begin{figure*}
\begin{center}
\includegraphics[width=0.75\textwidth]{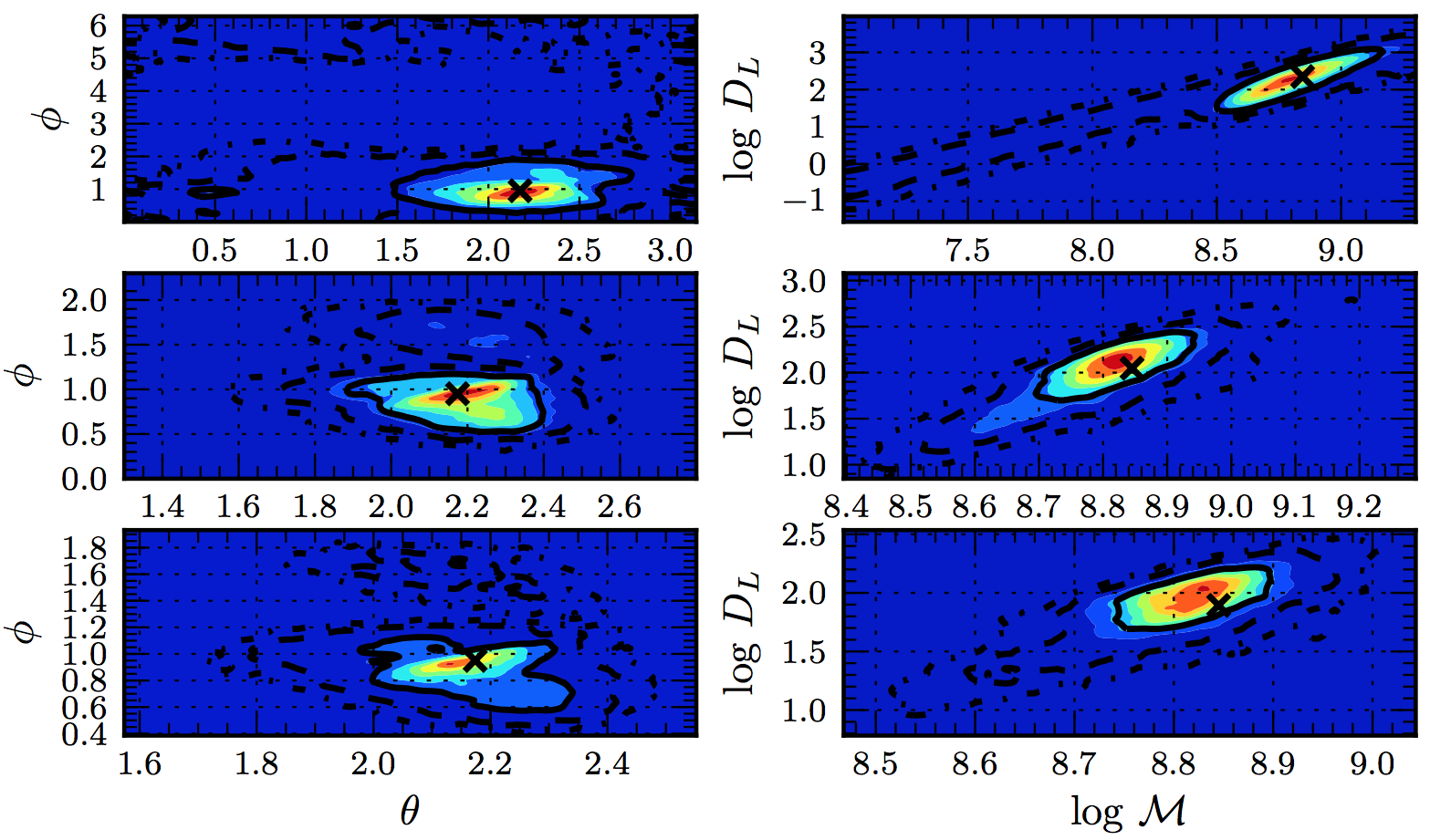}
\caption{Representative localization and parameter measurements for a continuous-wave detection. These images show the marginalized 2-D posterior probability density functions in the sky coordinates ($\theta,\phi$) and the log of the chirp mass and distance for injected signal-to-noise ratios of 7, 14, and 20 shown from top to bottom. The $\times$ symbol indicates the injected parameters and the solid, dashed and dot-dashed lines represent the 1, 2, and 3 sigma credible regions, respectively. Figure from \citet{ellis2013CQG}.
}\label{fig:skyloc}
\end{center}
\end{figure*}

To identify the host of a discrete (continuous wave/memory/burst) GW source, we need to be able to localize the object on the sky. The size of a PTA's localization error depends on the signal-to-noise ratio of the GW detection, and the relative position of pulsars with respect to the GW source \citep[\eg][]{ellis2013CQG}. However, typical localizations for at least the near future (detection significance of $\sim$7$\sigma$) will be up to a few times 1000~deg$^2$ (Fig.\,\ref{fig:skyloc}). Although the chirp mass of and distance to a SMBH binary are covariant in a PTA detection (Eq.\,\ref{eq:hrms}), we can estimate an upper limit on the distance by assuming a maximum chirp mass ($\sim$10$^{10}\,\msun$). Still, the large number of resulting galaxies in this error region will require us to turn to electromagnetic information to identify the most likely host.

There are two main goals of multi-messenger astronomy in the nanohertz regime:
(1)~We aim to simply identify the likely galactic host of the GW signal, and determine its distance. This information will allow us to determine the GW source's chirp mass, by breaking the chirp-mass/luminosity-distance degeneracy, mentioned above.
(2)~We aim to perform true ``multi-messenger'' monitoring of a target, by detecting electromagnetic signals from the vicinity of a SMBH binary, during the inspiral or coalescence. This will be possible, if the SMBHs are illuminated as AGN, or if the binary leaves a distinct observable signature on the stellar or gas dynamics, the photometry, or the morphology of their host. 
We summarize the methods (and efforts) to electromagnetically identify SMBHBs in section \ref{sec:smbhb_em}.

SMBHBs are formed in galaxy mergers, which for much of their duration are ostentatious events; they exhibit large-scale asymmetries, tidal tails, sudden bursts of star-formation (e.g., ULIRG stage), and potentially an abundance of tidal disruption events \citep{2013CQGra..30v4013B}. However, since, as discussed extensively in Section~\ref{sec:smbh}, the efficiency of SMBHB formation and inspiral is highly uncertain, it is also unclear whether these easily-observable features would be present by the time a binary SMBH enters the GW-dominated phase.
If the final-parsec problem is resolved efficiently, it is possible that these large-scale indicators of a galaxy merger persist, until the binary reaches the inspiral phase. Therefore, we can easily survey for this type of signature in the GW error region, and potentially narrow down the list of candidate hosts.

However, if the archival data prove inadequate, new dedicated observing campaigns will also be required to systematically catalog and identify outstanding features of galaxies in the error region of PTAs.

Before we move on to discuss direct SMBHB detection and multi-messenger science, it is worth noting an important practical point about electromagnetic SMBHB emission.
The evolution of SMBHBs within the PTA band is slow, and the inpsiral phase may last for long periods of time (of order years to millennia). Additionally, SMBHBs at these stages may have long-enduring electromagnetic signatures. Therefore, search efforts for multi-messenger signals in the nanohertz regime have a tremendous advantage (e.g., compared to LIGOs electromagnetic counterparts); for the majority of objects, tracking of a GW target will not necessarily be time-critical. Electromagnetic data collected years before the GW detection might reveal key information for a binary SMBH, its host, and its evolution. Therefore, the vast availability of archival images, and spectra, along with time series from the ongoing time-domain surveys, as well as data from the emergent highly sensitive instruments will allow great potential for discovery. They may identify any electromagnetic signatures on the large scale (i.e. unique identifiers in the properties of the host galaxy), or they may allow the tracking of any multi-messenger counterparts that mark the GW-emitting binary itself.

\subsection{Direct Searches for SMBHBs in Electromagnetic Bands}\label{sec:smbhb_em}

\noindent 
Given the significance of SMBHBs in galaxy evolution and the prospects for GW astrophysics, over the past few decades, several groups have searched for compact sub-pc binaries using electromagnetic data across the electromagnetic spectrum. In principle, direct imaging of material near the SMBHB would be the most direct identification method. However, only radio interferometry can provide sufficient resolution to probe sub-parsec scales \citep{radiocensus,kharb-recent}; 
although this provides a convenient observation method to test some binary SMBHB candidates, it can only reach those up to moderate redshifts at the earlier stages of binary evolution.

Therefore, many recent searches have relied on the effects of the binary on its environment (e.g., Doppler-shifted broad emission lines, periodic variability, distorted morphology of radio jets, etc; see \citealt{2013CQGra..30v4013B} for a discussion of multi-wavelength counterparts to SMBHBs). However, since these signatures are indirect, other physical processes can also explain the observed features. 

For instance, in the standard picture of AGN, the broad emission lines arise in regions close to the central SMBH, whereas the narrow emission lines are associated with large-scale features of the host galaxy. In the presence of a compact binary, the orbital motion of the SMBHB will be imprinted in the broad emission lines, producing noticeable shifts compared to the narrow lines. Alternatively, if both SMBHs have their own broad-emission-line region, the broad lines should be double-peaked, reflecting the motion of the binary.  Several candidates have been identified, especially from systematic searches in the spectroscopic database of SDSS \citep{eracleous+12,jgr13}. However, similar features can be produced by inflows/outflows, the geometry of the broad-emission-line region, as well as by a special type of AGN, the so-called double-peaked emitters.  Therefore, only long-term monitoring can prove/disprove the binary nature of these candidates, if coherent changes in the spectra are observed, as expected due to the orbit of a binary.

Periodic variability in AGN and quasars can also signify the presence of SMBHBs. The binary is expected to periodically perturb the circumbinary disk and several hydrodynamical simulations have found that SMBHBs can produce bright quasar-like luminosities, with the mass accretion rate onto the BHs periodically modulated at the orbital period of the binary. Several candidates ($\sim$150) have emerged \revision{from systematic searches in analysis of time-resolved photometry \citep{graham2015a,graham2015b,Charisi_ptf,liu2015,Zheng2016}.} However, the identified periods are relatively long, and only a few cycles have been observed. This, combined with the stochastic underlying variability of quasars, can lead to false detections \citep{Vaughan2016}. Again long-term monitoring can test whether the detected periodicity is persistent and thus attributed to a SMBHB. \revision{For instance, follow-up observations showed that the periodicity of quasar PSO J334.2028+01.4075 is not persistent \citep{liu2016}, whereas for quasar PG1302-102, the periodic model is preferred compared to pure stochastic noise \citep{liu2018}.} Additionally, several independent signatures have been proposed, which, if observed along with the periodicity, will increase our confidence in the binary nature of the candidates. These include multi-wavelength observations of Doppler-boosted emission \citep{Dorazio_Doppler,Dorazio_ir,Charisi2018}, periodicity with a characteristic frequency pattern \citep{Charisi2015,Dorazio2015}, and periodic self-lensing flares \citep{Dorazio_selflensing}.

Last but not least, the orbit of the binary can affect the morphology of radio jets. If one of the BHs is emitting a radio jet, the orbital motion of the base of the jet will result in a jet with helical structure \citep{hellicaljets}. Furthermore, the presence of a secondary BH can lead to jet precession, producing a jet with conical or helical morphology \citep[\eg][]{britzen+17}. As before, these signatures are not unique to SMBHBs, as instabilities in the jet or the misalagnment of the accretion disk with the jet axis can also produce similar features. Independent lines of evidence are required to test whether these galaxies host SMBHBs. We note that the above signatures, albeit indirect, will be significant in searches for the host galaxy, once PTAs detect a GW source.

\subsection{Topics in Binary Supermassive Black Hole Multi-messenger Science}

\subsubsection{Accretion Dynamics; Active Nucleus Geometry}~\\\vspace{-3mm}

\noindent
In gas-rich mergers, copious amounts of gas will be funneled into the central regions of the post-merger galaxy. Once the two SMBHs become gravitationally bound, the gas will settle in a circumbinary accretion disk. 
The disk can dissipate the SMBHB energy and angular momentum, driving the binary to the GW regime. Since the existence of ambient gas can catalyze the binary evolution, this has been suggested as one solution to the final-parsec problem (Sec.~\ref{sec:smbh}). 
Additionally, torques from a circumbinary disk could influence the inspiral evolution of the binary significantly enough for this effect to be visible in the gravitational waveform.
Theoretical work on circumbinary disks covers a variety of topics, including the timescales of SMBHB evolution, whether the disks will be retrograde or prograde with respect to the binary orbit \citep{sk15}, whether such disks will support accretion onto one SMBH or both SMBHs \citep{c+09}, the effect of the gaseous disk on the binary eccentricity, etc.\ (see also Sections \ref{sec:gwinspiral} and \ref{sec:CBdisk}).

More importantly, the presence of gas is crucial in providing bright electromagnetic counterparts, which will allow us to pinpoint the tentative SMBHBs. This is an area of intense investigation, with several remaining open questions regarding the expected emission signatures. For instance:
\begin{itemize}
\item\emph{Will the binary have one or two radio jets?}
\item\emph{Will the circumbinary disk itself support a stable or unstable jet?}
\item\emph{Will a secondary black hole clear a gap or a cavity \citep{uv-disk-gaps}, and if so, can we use this as an observable to infer its mass ratio?} 
\end{itemize}

The GW signal will provide an independent measurement of the chirp mass $\mathcal{M}_c$ (provided that the distance is constrained electromagnetically from the redshift of the source). It will also allow for the measurement of the separation of the two black holes and time-tagged tracking of the orbital motion of the binary. These, in combination with any observed electromagnetic signatures, will allow us to explore in detail the interplay between the disk and the binary orbit, the accretion rate onto the individual BHs and the overall disk geometry, as well as the processes involved in jet generation, thus providing unprecedented insight into the physics of active galactic nuclei.

\subsubsection{Constraining Black Hole Mass--Host Galaxy Relations}~\\\vspace{-3mm}

\noindent The mass of the central SMBH $M_\bullet$ is strongly correlated with several properties of the host galaxy, e.g., bulge mass \citep{Magorrian1998}, galactic velocity dispersion \citep{2000ApJ...539L...9F,2000ApJ...539L..13G}, and S\'ersic index \citep{sersicBH}, amongst other properties. This suggests that SMBHs co-evolve with their host galaxies. These relations have been critical throughout astronomy because of their ability to predict SMBH masses where there is no direct measurement. However, the reliability of the SMBH masses derived from these relations has recently come into question, and it has been suggested that SMBH masses might be biased upwards \citep{shankar}.
As mentioned above, PTAs can only measure a degenerate chirp mass and distance to the source. However, with the identification of the galactic host, this degeneracy can be broken, permitting a measurement of $M_c$. 
As we write this, PTAs are sensitive to SMBHBs of $M_c>10^9\,\msun$ \revision{out to approximately 200\,Mpc, or a redshift of $z\simeq 0.05$ \citep{nanograv_11yr_cw,2016MNRAS.455.1665B}.} 
As the PTA horizon grows and more distant sources are detected, PTAs will have some limited capability to probe the redshift dependence of these relations, particularly at the highest-mass end.

Note that, while $M_c$ does not reflect the total black hole mass directly, numerous simulations have demonstrated that all SMBHBs detected by PTAs will likely have mass ratios $q\gtrsim0.1$, implying an error when inferring $M$ of $\sim$0.5\,dex. Even in the more conservative limit of $q\gtrsim0.01$, the error in the total mass inferred from PTA measurement of $\mathcal{M}_c$ would be only $\sim1\,$dex.

As discussed here, PTAs can probe the SMBH mass--host relations through the detection of continuous waves, however \citet{simonBS16} demonstrated that they can also probe these relations by looking at the amplitude of the GW background. Because SMBH masses are currently linked to GW background predictions through modelling of host galaxy populations, that study was able to use limits on the GW background to place constraints on the high-mass slope, scatter, and intercept of the relationship between SMBH mass and host bulge mass.

\subsubsection{Active Nuclei Preceding and Following SMBHB Coalescence}~\\\vspace{-3mm}

\noindent
In the case of a GW memory event, identification of the host will allow us to ``wait and watch'' for a post-coalescence ignition of an active galactic nucleus. If there is a circumbinary disk, it will closely track the shrinking orbit of the SMBHB. If circumbinary disks are highly viscous, at some point it is possible that the binary and circumbinary disk will ``decouple'' if the GW-induced inspiral becomes so rapid that the viscosity of the disk no longer allows it to track the binary.
After the coalescence, there may be a delay in any nuclear activation while the disk settles; this delay depends both on the disk viscosity, which governs the infall rate of the disk, and the SMBHB masses, which govern the rapidity of the final break-away inspiral. The delay between coalescence and subsequent turn-on of a luminous X-ray source can be modeled as:
\begin{equation}
t_{\rm X-ray} \simeq 7(1+z)\bigg(\frac{M}{10^6\,\msun}\bigg)^{1.32}\,{\rm yr}
\end{equation}
\citep{phinney}. For the high-mass systems that PTAs are expected to discover ($M\gtrsim10^8\,\msun$), this implies timescales that are impractically long, on the order of hundreds to thousands of years. However, \citet{tm+10} proposed a much more rapid temporal evolution of the circumbinary disk, finding that there may be a rapid brightening to super-Eddington luminosities in the few years following a merger. 
These signatures would be detectable in X-rays, with potential signatures in optical and infrared bands. Subsequent ignition of a radio jet might also arise once the accretion disk is settled.

Another promising prospect is to search through archival data to uncover the evolution of the system leading up to and during the merger in electromagnetic bands. In practice, the extent of such studies will be highly dependent on the position of the source on the sky, and the prevalence of historical data in that direction. However, given the broad coverage of current and upcoming synoptic sky surveys, like the Catalina Sky Survey \citep[CRTS,][]{crts}, the Panoramic Survey Telescope and Rapid Response System \citep[Pan-STARRS,][]{panstarrs}, the Zwicky Transient Facility \citep[ZTF,][]{ztf}, the Large Synoptic Survey Telescope \citep[LSST,][]{LSST}, and the Extended Roentgen Survey with an Imaging Telescope Array \citep[eROSITA,][]{eROSITA}, we expect PTAs to be detecting SMBHBs in an era rich with time-domain data over broad areas of the sky.

\revision{
\subsubsection{SMBHB Effects on Galaxy Central Morphology and Dynamics}~\\\vspace{-3mm}

\noindent Under certain circumstances (\eg\ dry/low-gass-fraction mergers), the stellar scattering described in Section \ref{sec:losscone} can work to flatten the density profile of a galaxy away from a central stellar cusp, leading to a ``core'' in a post-merger system. Such cores are commonly observed in the most luminous galaxies, and are marked by a flat stellar photometric profile of the central $\sim$kpc of a galaxy \citep{faber97}, as opposed to a much steeper profile external to the core. Simulations have shown direct links between the core phenomenon and SMBHB stellar scattering \citep{ebisuzaki91, makino97, milomerritt01}, and some work has indicated these cores may even survive subsequent mergers \citep{volonteri-cores}. However, direct observational proof of these links does not yet exist.

SMBHBs discovered by PTAs will have already undergone the stellar scattering phase. With the identification of multiple galactic hosts of continuous-wave sources, the presence of cores in those hosts could indicate direct links between the dynamical actions of SMBHBs and the dynamics and morphology of the larger-scale core.

Taking a more general view the central dynamics of galaxies containing a continuous-wave emitter, compared to those without a binary or at earlier stages of orbital inspiral, may serve to reveal unanticipated stellar or gas dynamical interactions during a merger.
}

\subsubsection{Tidal Disruption Events}~\\\vspace{-3mm}

\noindent
PTAs will not directly detect tidal disruption events, however a number of studies have postulated that tidal disruption events may occur at much higher rates in SMBHB systems. This is largely caused by the chaotic three-body interaction of the SMBHB with stars in the stellar core, which can cause an increase of a few to several thousand over the rate of disruption events around an isolated black hole \citep{tde-largefactor,tde-smallfactor}. 
It has also been postulated that tidal disruption events might encode electromagnetic information about the binary's orbit in their time-resolved signatures \citep{tde-binary}.

LSST is expected to detect thousands of tidal disruption events in the coming decade \citep{LSST-TDEs}. This may highlight individual galaxies with excess event rates. Such galaxies could signify the host of a potential continuous-wave GW source, which would assist in host identification for any future continuous-wave detections by PTAs.

Similarly, if we can otherwise identify the host of a GW, this detection can be cross-matched with transient catalogs to test for excess events, or to allow direct multi-messenger studies of any tidal disruption event caused by the continuous-wave emitter.
Any binary modelling extracted from a tidal disruption event will provide PTAs greater sensitivity to the continuous waves from that event.

Extreme-mass-ratio inspiral events (commonly termed EMRIs; in this case, the inspiral of the object before it is disrupted) may also be detected by LISA. The coordinated detection of GWs from a binary SMBH by PTAs, an EMRI signature from the low-mass object, and the electromagnetic observation of a tidal disruption event could provide a fascinating set of studies of distinct processes in a common object.

%

\subsubsection{Testing Electromagnetic SMBHB Emission Models}~\\\vspace{-3mm}

Several binary SMBHB candidates have emerged, but it has proven difficult to find a smoking-gun signature that can conclusively confirm the binary nature of these sources. PTAs, on the other hand, can provide a straight-forward way to test the binary hypothesis for such candidate systems identified in electromagnetic bands. Given the current PTA capabilities, this test can be applied for candidates that are sufficiently nearby, massive, and at the right orbital phase to be emitting GWs in nanohertz frequencies, i.e., they need to be at sub-parsec orbital separations. 

Even without a direct detection of nanohertz GWs, the PTA upper limits already provide astrophysically relevant constraints on the SMBHB population. For instance, recently, \citet{sesana_testing_2018} and \citet{holgado_pulsar_2018} examined the GW background from the population of SMBHBs inferred from the samples of periodic quasars and blazars, respectively. They found that, even if the individual candidates are below the sensitivity of PTAs, the inferred population is in moderate tension with the current limits on the GW background, which suggests that the samples may be contaminated by false detections. 

Additionally, in some cases, electromagnetic constraints on a binary system may raise the PTA sensitivity for the same source by constraining the searchable parameter space.
%
There are already several examples of this type of multi-messenger astrophysical application. Most famously, a $\mathcal{M}_c\simeq 10^{10}\,\msun$ SMBHB was hypothesized to exist in galaxy 3C66B due to an elliptical motion detected through long-baseline interferometry; it was suggested that this implied a binary-modulated jet precession \citep{sudou3C66B}. However, \citet{jenet3C66B} used only seven years of data on PSR\,1855+09 to place limits on the allowed chirp mass and eccentricity of this system, effectively ruling out the purported binary parameters. More general applications of ruling out binaries in specific systems involve targeted campaigns to place limits on binaries in nearby, massive systems \citep{2001ApJ...562..297L,schutzma}.

These types of studies are constantly improving, as the PTA horizon grows with time, and with the addition of more pulsars. Based on the non-detection of continuous waves in PTA data thus far, we can already en masse rule out nanohertz-frequency binaries with 
$\mathcal{M}_c>0.5\times10^{8}\,\msun$ in the Virgo cluster \citep{nanograv_11yr_cw}.


\subsubsection{Constraining SMBHB Populations}~\\\vspace{-3mm}

\noindent
The previous sections discussed electromagnetic signatures of SMBHB systems that are primarily already in the microhertz to nanohertz gravitational waveband. 
However, we note that there is also a direct benefit to systematically search for ``GW precursor'' binaries---systems easily resolved as two AGN but well within the $\sim$few kpc virial radius of a typical merger. In part because they likely have a higher residence time at very large separations (Eq.~\ref{eq:dynamicalfriction}), and in part because they are directly resolvable, these systems may be more readily accessible to large electromagnetic surveys.
These can provide direct statistics on the systems that will make up our binary population \citep[\eg][]{johnny}. These systems also complement GW-based studies of massive merger environments and of the efficiency of SMBHB evolution (Sec.~\ref{sec:binaryinfluence}). So far, there have been dozens of such discoveries; a few examples include the resolved dual AGN in NGC6240 \citep{komossa+03}, the spatially resolved broad line regions in the galaxies of \citet{comerford+13}, and the record-holding binary at a separation of 7\,pc in 0402+379 \citep{rodriguez+06}. Recently, astrometric tracking of 0402+379 led \citet{bansal+17} to suggest an orbital period for this binary of $\sim10^4$\,yr.
For a summary on the detections of dual AGN, we refer the reader to \citet{Rubinur2018}

Statistics on the population of dual AGN can directly constrain some of the covariant parameters contributing to the GW background that are discussed in Section~\ref{sec:binaryinfluence} and summarized in Fig.~\ref{fig:gwspectrum}. This includes eccentricity distributions, inferred inspiral rates, and observed evidence of ongoing interaction with the galactic environment. However, the predictive power of these discoveries currently suffer from small-number statistics. Future large-scale systematic surveys, such as LSST, or the Square Kilometer Array, coupled with detailed observations using instruments with specialized capabilities, such as high resolution, rapid spectrum acquisition, and/or time-resolved imaging will significantly enhance our understanding of the binary population.

\subsubsection{Tests of Gravity}\label{sec:gr.propagation}~\\\vspace{-3mm}

\noindent As mentioned in Section~\ref{sec:grtests.dispersion}, in general relativity electromagnetic and GW signals propagate at the same speed.  This agreement can be tested if a resolvable continuous wave source also happens to be an eclipsing electromagnetic binary \citep{Will:1997bb,Cutler:2002ef}. Such a system allows for a differential test of the propagation speed of GWs and light, and the mass of the graviton. Using the order-of-magnitude limit on the graviton Compton wavelength ($\lambda_{g}=h/(m_gc)$) of such a system \citep{Will:1997bb}: \begin{equation}
\lambda_{g} > 3 \times 10^{12} {\rm km} \left(\frac{D}{200 {\rm Mpc}} \frac{100 {\rm Hz}}{f} \right)^{\frac{1}{2}}\left(\frac{1}{f \Delta t} \right)^{\frac{1}{2}}
\end{equation}
and choosing nominal values for the parameters of an \hbox{SMBHB}, of $f=1\times10^{-9}$ Hz, $D=1$~Gpc, and an observing time of $\Delta t=5$ years, one obtains a lower limit of $\lambda_{g} \sim 10^{19}$ km. This limit is three orders of magnitude less precise than the current Particle Data Group limit of~$10^{22}$\,km, suggesting that today's PTAs might not be able to produce competitive limits without more sophisticated GW data analysis schemes \citep{Choudhury:2002pu}.

\begin{figure*}
\centering
\includegraphics[width=0.9\textwidth]{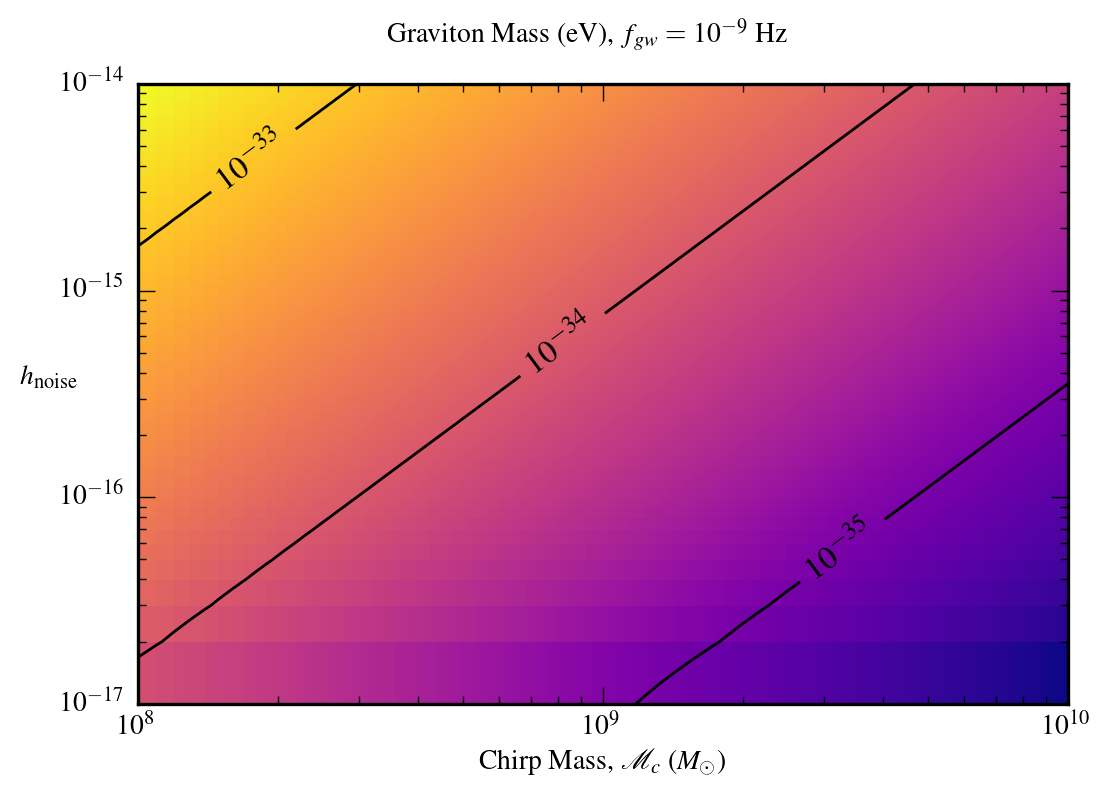}
\caption{Limits on the mass of the graviton, given a characteristic noise strain amplitude for a pulsar timing array with a continuous wave signal of given chirp mass \citep{Hazboun:2013pea}. The lines and coloring represent contours of the graviton mass limit. A fiducial frequency of $10^{-9}$~Hz is used for this illustration. }
\label{fig:graviton_mass}
\end{figure*}

The approaches in \citet{Will:1997bb}, \citet{Cutler:2002ef}, and \citet{Hazboun:2013pea} are framed in the context of a null test, in which the assumed observation is the simultaneous arrival, within the margin of error, of some fiducial feature of both the electromagnetic and GW signal. While the error in arrival time is that of both the electromagnetic and GW signals, in practice the electromagnetic arrival time is much more precise, and is often ignored in terms of these calculations. The error, related to the signal-to-noise ratio in general, and to the strain sensitivity for a specific detector, is then used to set a limit on the difference in propagation speed between the electromagnetic and GW signals. This velocity limit can then be used to calculate a lower limit on the Compton wavelength and an upper limit on the mass of the graviton~$m_g$, if one assumes a Yukawa-type potential ($V_g\sim e^{-km_gr}/r$) for the graviton. This type of potential is a phenomenological model which acts to restrict a interaction's ability to act at a distance.
Figure~\ref{fig:graviton_mass} shows a contour plot of the limits on $m_{g}$ for a generic PTA at $f_{gw}=10^{-9}$~Hz.

\citet{2017ApJ...835..103T} considers an alternate approach to testing the concordance of electromagnetic and GW propagation using gravitational lensing.  Standard analyses assume that gravitational lensing can be treated with geometric optics.  This assumption is no longer applicable if the mass of the lens is too small compared to the wavelength of the radiation; less than $10^5 M_\odot(f/Hz)^{-1}$.

Both of these tests of gravitational/electromagnetic wave propagation concordance would require the detection of an individual SMBH binary with a high enough signal-to-noise ratio in GWs and some electromagnetic measure of the orbiting SMBHs (e.g., optical or X-ray light curve).  \citet{2017ApJ...835..103T} shows that, for signal-to-noise ratios of~10--100 and GW frequencies $f \sim 10$~nHz, there is a reasonable probability of obtaining a few lenses, for which the time difference between the GW and electromagnetic signals would be of order 1--10~days, if of order 100 individual SMBH binaries could be detected as (continuous wave) GW sources.  (Longer time delays are required for detection at lower signal-to-noise ratio.)  At higher (lower) GW frequencies, shorter (longer) time delays result in similar predictions for comparable signal-to-noise ratios.

\section{Current Outlook}
\label{sec:conclude}
\vspace{-3mm}
The focus of this paper has been on the possible sources that could generate nanohertz GW signals.  We now consider briefly the observational status of pulsar timing arrays, both current and near future, as a means of providing some measure of the likelihood of detecting these signals. As indicated in the first two sections above and referenced throughout, the sensitivity of pulsar timing arrays depends on length of data set, RMS timing precision of pulsars, and the total number of contributing pulsars.

The three large national and regional PTAs are \revision{the European Pulsar Timing Array \citep[\hbox{EPTA},][]{2016MNRAS.458.3341D,2015MNRAS.453.2576L,2016MNRAS.455.1665B},} the North American Nanohertz Observatory for Gravitational Waves \citep[NANOGrav,][]{2015ApJ...813...65T,2016ApJ...821...13A}, and the Parkes Pulsar Timing Array \citep[\hbox{PPTA}, ][]{ppta-manchester13}.  These efforts, in addition to other emerging timing arrays in other nations (China, India) are currently working to combine their data into an International Pulsar Timing Array (IPTA; \eg\ \citealt{2010CQGra..27h4013H}). \revision{In the future, it is expected that new facilities such as MeerKAT and the Square Kilometre Array will increase the sensitivity of PTAs manifold through extensive pulsar searching and timing programs \citep[\eg][]{ska-era}.}

As an approximate illustration of the performance of these PTAs, \revision{recent published work has reported the timing of} 20--40 millisecond pulsars with sub-microsecond precision, while some PTAs have begun to time $>70$ pulsars to this precision. Currently, the total number of pulsars timed world-wide does not \revision{yet}  exceed 100, because there is overlap between the PTAs.  Some of this overlap is intentional, for the purposes of calibration between the different telescopes, in addition to providing the ability to independently verify any reported GW detection.

Millisecond pulsars continue to be discovered \citep[e.g.,][]{2016ApJ...819...34C,2017ApJ...846L..19P}, and, while not all are suitable for inclusion into a \hbox{PTA}, many are, so these PTAs can and are still expanding.  As a specific example, the NANOGrav~9~Year Data Release contained 37 pulsars \citep{2015ApJ...813...65T}, \revision{its 11~Year Data Release contained 45 pulsars \citep{nano-11yr-data}, 
and that collaboration is now timing more than 70 pulsars.} In addition to new pulsar discoveries, improvements to radio observing systems will continue to improve mitigation of \revision{intrinsic pulse-to-pulse variations (``jitter''; \citealt{jitter}), which dominates the noise budget of many PTA pulsars. We also have an ever-increasing understanding of other contributions to the PTA noise budget, including long-term timing noise (``red noise''); some of this is now commonly mitigated by tracking radio-frequency-dependent variations in the interstellar medium, although the origin of long-term noise in some pulsars remains yet unclear \citep[\eg][]{keith-ISM}.} These improvements can provide sudden \revision{improvements} in PTA sensitivity, as the data are re-processed using novel noise suppression techniques \citep[\eg][]{jones+17}.
Given these ongoing improvements, we consider it plausible that together, worldwide PTAs will include approximately 100 pulsars at typical precisions of approximately 500~ns by 2020. \revision{With the use of MeerKAT and SKA, these numbers will increase even more rapidly. Of course, with all PTAs, simply keeping continuous timing programs for ongoing pulsars to obtain a longer time baseline will also afford increased sensitivity.}


Many of the possibilities discussed in later sections are somewhat more speculative (\S\S\ref{sec:relic}, \ref{sec:grtests}, and~\ref{sec:beyond}). In many cases, these are likely to be ``observationally driven,'' with models that are updated to evade new limits as they are placed. Most of these models are capable of producing data over many orders of magnitude across many parameters, thus updated models will continually be available, however, they will likely cover smaller and smaller allowed parameter space.

Current PTAs are already placing astrophysical limits on the existence of SMBH binaries and their interactions with their environments (\S\ref{sec:smbh}) and the existence of cosmic strings (\S\ref{sec:strings}). Based on projections of SMBHB populations, we expect a GW background detection to be imminent from that population \citep{vs16}, with the detection of continuous-wave SMBHBs soon to follow \citep{2017arXiv170803491M}. The detection of these systems will be occurring concurrently with a number of planned synoptic sky surveys, enabling a much greater scope for multi-messenger studies (\S\ref{sec:em}) in addition to potential multi-band GW science with LISA when it flies in the mid-2030s (\S\ref{sec:lisa}).
For the forseeable future, PTAs will maintain unique access to exploring the inspiral of high-mass SMBHB systems, and will uniquely probe the physical properties of strings.



\ack 
We thank Andrew Kaiser for providing Fig.~\ref{fig:pta-lisa}; he can be contacted at {\tt ark0015@mix.wvu.edu}.
NANOGrav is supported by NSF Physics Frontier Center award \#1430284. SBS is supported by NSF award \#1458952 and by the CIFAR Azrieli Global Scholars program. Portions of this research were carried out at the Jet Propulsion Laboratory, California Institute of Technology, under a contract with the National Aeronautics and Space Administration. DRM was, until recently, a Jansky Fellow of the National Radio Astronomy Observatory, a facility of the NSF operated under cooperative agreement by Associated Universities, Inc. The Flatiron Institute is supported by the Simons Foundation. AR received funding from the European Research Council (ERC) under the European Union's Horizon 2020 Programme for Research and Innovation ERC-2014-STG under grant agreement No. 638435 (GalNUC).


\bibliographystyle{./yahapj}
\bibliography{references}


\end{document}